\pdfoutput=1
\documentclass[a4paper,11pt,english,DIV=12,parskip=half-,headsepline,footsepline=false,
  bibliography=totoc,listof=totoc,listof=nochaptergap,numbers=noenddot,captions=tableheading]{scrreprt}

\usepackage[T1]{fontenc}
\usepackage{lmodern}
\usepackage[ngerman,main=english]{babel}
\usepackage{microtype}
\usepackage{amsmath,amssymb}
\usepackage{upgreek}
\usepackage{textcomp}
\usepackage{newunicodechar}
\newunicodechar{−}{\ensuremath{-}}
\newunicodechar{×}{\ensuremath{\times}}
\newunicodechar{≈}{\ensuremath{\approx}}
\newunicodechar{≥}{\ensuremath{\geq}}
\newunicodechar{≤}{\ensuremath{\leq}}
\newunicodechar{→}{\ensuremath{\rightarrow}}
\newunicodechar{η}{\ensuremath{\eta}}
\newunicodechar{µ}{\textmu}
\newunicodechar{…}{\dots}
\usepackage{siunitx}
\usepackage{graphicx}
\usepackage{wrapfig}
\newcommand{\wraptop}{\vspace*{-\intextsep}}
\usepackage{etoolbox}
\AtBeginEnvironment{wrapfigure}{\setlength{\parskip}{0pt}}
\AtBeginEnvironment{wraptable}{\setlength{\parskip}{0pt}}
\AtBeginEnvironment{figure}{\setlength{\parskip}{0pt}}
\AtBeginEnvironment{table}{\setlength{\parskip}{0pt}}
\usepackage{float}%
\usepackage{needspace}%
\usepackage{placeins}%
\usepackage{array,booktabs,longtable,multirow,tabularx}
\usepackage{enumitem}
\usepackage[dvipsnames]{xcolor}
\usepackage{tikz}
\usetikzlibrary{arrows.meta,calc,positioning,shapes.misc,shapes.geometric,fit,backgrounds,patterns}
\usepackage[european,straightvoltages]{circuitikz}
\usepackage{xstring}%
\usepackage{rotating}%
\tikzset{
  schem/.style={font=\small, line width=0.6pt, line cap=round, line join=round,
    >={Latex[length=2mm,width=1.4mm]}},
  part/.style={line width=0.9pt},%
  leader/.style={line width=0.35pt},%
  ext/.style={line width=0.35pt, densely dashed},%
  dim/.style={line width=0.4pt, <->},%
  signal/.style={densely dashed, ->},%
  block/.style={draw, part, align=center, minimum height=1cm, inner sep=3pt},
  lbl/.style={inner sep=1.5pt, align=left},
}
\usepackage{pgfplots}
\pgfplotsset{compat=1.18}
\usepgfplotslibrary{groupplots,fillbetween}
\pgfplotsset{thesisplot/.style={
  font=\small, tick label style={font=\footnotesize}, label style={font=\small},
  grid=major, major grid style={black!15}, axis line style={black!70},
  tick align=outside, tick pos=left, every axis plot/.append style={line width=0.6pt}}}
\colorlet{plotline}{black}%
\colorlet{plotfill}{black!45}%
\colorlet{plothighlight}{black!10}%
\colorlet{thgrey}{black!50}%
\usepackage[format=plain,font=small,labelfont=bf,labelsep=colon,justification=centering,skip=6pt]{caption}
\usepackage[headsepline]{scrlayer-scrpage}
\usepackage{cite}
\usepackage{xurl}
\usepackage[hidelinks,pdftitle={Electric Racing Kart with BLDC Drive, LiFePO4 Battery System, Traction Control and Regenerative Braking},
  pdfauthor={Johannes Stockhammer, Tom Rettenwander, Philipp Huber},
  pdfsubject={76.8 V HV system, 20 kW peak, 60 Nm, DC/DC converter instead of an LV battery. Diploma thesis HTL Braunau 2016, third edition 2026},
  pdfkeywords={electric kart, racing kart, traction control, regenerative braking, LiFePO4, BLDC motor, battery management system, DC/DC converter, low-voltage supply without auxiliary battery, electric vehicle, HTL Braunau}]{hyperref}

\newcommand{\LFP}{\texorpdfstring{LiFePO\textsubscript{4}}{LiFePO4}}
\DeclareRobustCommand{\nocc}{\\{\itshape Not covered by the CC licence}}
\DeclareRobustCommand{\noccpart}[1]{\\{\itshape #1 not covered by the CC licence}}

\newcommand{\runonchapter}[1]{\begingroup\let\clearpage\relax\let\cleardoublepage\relax
  \RedeclareSectionCommand[beforeskip=2.2\baselineskip, afterskip=1.2\baselineskip]{chapter}%
  \chapter*{#1}\endgroup\addcontentsline{toc}{chapter}{#1}}

\counterwithout{figure}{chapter}
\counterwithout{table}{chapter}
\counterwithout{equation}{chapter}
\counterwithout{footnote}{chapter}
\RedeclareSectionCommand[tocbeforeskip=0.6em]{chapter}%
\renewcaptionname{english}{\bibname}{References}
\renewcaptionname{english}{\contentsname}{Table of Contents}

\clearpairofpagestyles
\ohead{\small Electric Racing Kart \textperiodcentered{} HTL Braunau}
\ofoot*{\pagemark}

\KOMAoptions{headsepline}

\setlist{itemsep=1pt,topsep=3pt}

\areaset{448.13pt}{686.55pt}%
\begin{document}

\hypersetup{pageanchor=false}%
\begin{titlepage}
\begin{center}
{\large Höhere Technische Bundeslehranstalt Braunau, Austria\par}
\vspace{1mm}
{\large Diploma Thesis (Diplomarbeit) \textperiodcentered{} 2015/16\par}
\vspace{12mm}
{\LARGE\bfseries\sffamily Electric Racing Kart with BLDC Drive,\\ \LFP{} Battery System, Traction Control\\ and Regenerative Braking\par}
\vspace{5mm}
{\large \qty{76.8}{\volt} HV system, \qty{20}{\kilo\watt} peak, \qty{60}{\newton\metre},\\ DC/DC converter instead of an LV battery\par}
\vspace{9mm}
\includegraphics[width=0.6\textwidth]{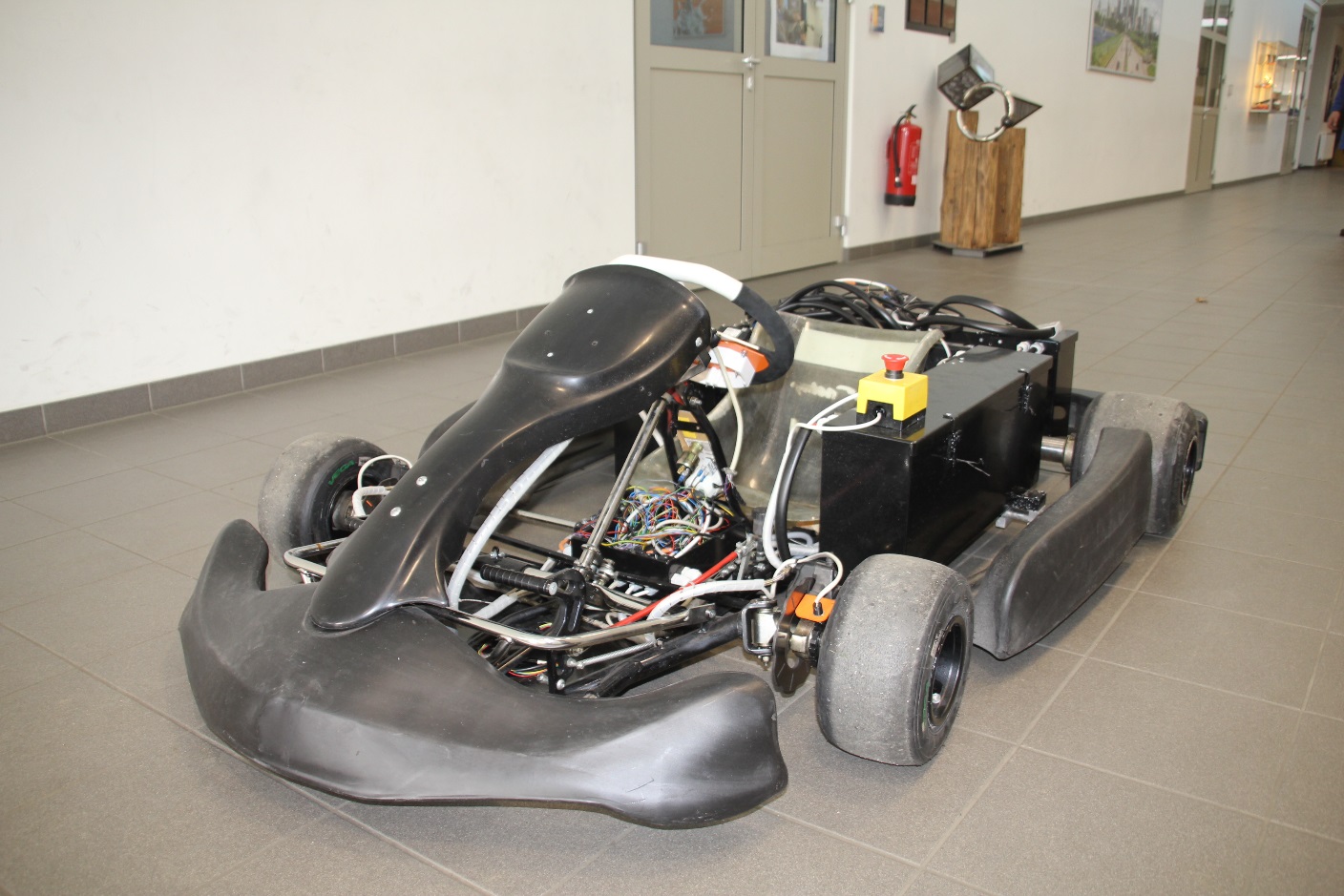}\par
\vspace{9mm}
{\Large\bfseries Johannes Stockhammer \textperiodcentered{} Tom Rettenwander \textperiodcentered{} Philipp Huber\par}
\vspace{4mm}
{As a team, we reached the national final of Jugend Innovativ 2016\\ and received a recognition prize in the special category Sustainability.\\
\href{https://www.jugendinnovativ.at/projekt/htl-elektro-kart}{\small jugendinnovativ.at/projekt/htl-elektro-kart}\par}
\vfill
{\textbf{Supervisors:} Dipl.-Ing. Thomas Gruber (theory) \textperiodcentered{} Anton Deschberger (practice)\par}
\vspace{2mm}
{Submitted in Braunau, 4 April 2016\\ \textit{Third edition, 2026}\par}
\end{center}
\end{titlepage}

\thispagestyle{empty}
\begingroup\small
\textbf{Edition.} This document is the third edition (2026, revised English edition) of the diploma thesis ``Elektrokart der HTL Braunau'', submitted in April 2016 at the Höhere Technische Bundeslehranstalt (HTL) Braunau, Austria, as part of the Reife- und Diplomprüfung.

\textbf{Citation.} Stockhammer, J.; Rettenwander, T.; Huber, P.: Electric Racing Kart with BLDC Drive, \LFP{} Battery System, Traction Control and Regenerative Braking. Diploma thesis, HTL Braunau, 2016. Third edition, 2026.

\textbf{Authors.} Johannes Stockhammer, Tom Rettenwander and Philipp Huber. Each author was responsible for the sections of his own work package; the personal reflections in Chapter~\ref{sec:personal-reflections} are the individual views of the respective author.

\textbf{Licence.} This publication is licensed under the Creative Commons licence Attribution 4.0 International (CC BY 4.0): \url{https://creativecommons.org/licenses/by/4.0/}

Excluded from this licence are third-party contents, in particular data from manufacturer datasheets and standards, and the drawings and diagrams marked ``not covered by the CC licence'' in their captions.

\textbf{Figures.} Unless stated otherwise, all photographs, tables and original engineering documents are the authors' own work. Some original engineering documents (circuit diagrams, CAD drawings and outputs of the calculation software KISSsoft) are reproduced unchanged and therefore contain German labels.

\textbf{Data and code.} Firmware, schematics, bills of material and calculations of the kart are published in the repository ``Electric Race Kart'' (\url{https://github.com/josto-me/electric_race_kart}, archived at \url{https://doi.org/10.5281/zenodo.22990337}) under the licences stated there.

\textbf{Trademarks.} Brand and product names are used for identification only and may be trademarks of their owners.

\textbf{Tools.} AI tools were used for the translation, for language editing and for preparing the schematic drawings and charts of this edition, based on the authors' own data and specifications.

\textbf{Disclaimer.} This work documents a student project of 2015/16 and is not a construction manual. Work on the \qty{76.8}{\volt} traction battery is only permissible for qualified persons in compliance with the applicable safety regulations. The operating manual in Chapter~\ref{sec:operating-manual} applies only to the vehicle built in 2016, and statements on products and suppliers reflect the experience of 2015/16.

\begin{center}\itshape Keep science moving forward.\end{center}

\endgroup
\clearpage

\pagenumbering{roman}
\hypersetup{pageanchor=true}
\chapter*{Preface}\addcontentsline{toc}{chapter}{Preface}
Electrical energy and ways of saving energy are a constant topic in the media. For this reason, we wanted to develop an electrical innovation of our own as our diploma project. When the diploma projects were chosen in the fourth year, we decided to build an electric go-kart. Our teachers, and Mr Gruber in particular, introduced us to the project.

Our motivation was not only the joy of the finished vehicle, but also the chance to create something that hardly existed at the time. At the start of the project, the market for electric go-karts was still small. The aim was therefore an energy-efficient kart built from components that a student project can handle.

We also discussed alternatives, such as an off-road kart or building and converting a petrol quad. In the end, the choice fell on a racing kart, because it looks the sportiest and is well suited as an exhibition piece.

As a team, we reached the final of the Austrian youth innovation competition Jugend Innovativ 2015/16 in the special category Sustainability and received a recognition award. Thanks to the active support of our teachers, several companies and friends, we had the best possible conditions. This cooperation made it possible to develop and realise a well-thought-out project.

\chapter*{Acknowledgements}\addcontentsline{toc}{chapter}{Acknowledgements}
First of all, we would like to thank everyone who actively supported our diploma project.

Our sincere thanks go to our two supervisors, Dipl.-Ing. Thomas Gruber and Anton Deschberger. We are especially grateful to Dipl.-Ing. Thomas Gruber, who made this great project possible. He placed a great deal of trust in us and allowed us to take on this challenge despite the associated risks.

We also thank our head of department, Dipl.-Ing. Paul Dirnberger, and our headmaster, Mag. Hans Blocher, who supported the project significantly. Special thanks go to all teachers involved, from both the practical and the theoretical side. Furthermore, we thank all participating companies. In particular, we thank Kartshop Ampfing for providing the racing kart chassis. We are particularly grateful to Dawo in Mattighofen for the powder coating of the aluminium parts. We thank Emicon Engineering, Salzburg, for creating a 3D model.

We also thank Smartemotions GmbH and Faktor GmbH.

Special mention goes to Harald Bartol. Thank you very much for your technical support and for putting us in contact with Kartshop Ampfing.

Finally, we thank our parents, who actively supported us with questions about the project not only during term time but also during the holidays.

\chapter*{Abstract}\addcontentsline{toc}{chapter}{Abstract}
The aim of this project was to convert a petrol-powered go-kart into an electric kart. The kart was to be efficient and at the same time simple in design. In addition, it should not become too heavy, since additional weight impairs the driving behaviour. Reaching these goals required attention to several aspects.

The first challenge was the weight distribution. To solve it, the battery cells were divided into three groups, each housed in its own battery box. In this way, the boxes could be placed at suitable positions on the chassis. The weight of the motor and the electronics also had to be taken into account.

For high overall efficiency, a motor with a high rated efficiency and battery cells with low self-discharge were installed. A regenerative brake feeds part of the kinetic energy back into the battery. Simple energy recovery was comparatively easy to implement, since the selected motor controller offers a recuperation function. The planned three-stage recuperation, however, required an additional microcontroller (\textmu C).

To increase the range further, the kart is equipped with a traction control system developed specifically for this vehicle. It prevents the rear wheels from spinning during acceleration by reducing the motor torque. Spinning wheels waste energy and do not contribute to acceleration, so the traction control saves energy and reduces tyre wear.

Chapter~\ref{sec:design-and-implementation} describes the design in detail. Chapter~\ref{sec:results-and-discussion} evaluates the calculations and measurements. The main results are as follows. The calculated top speed is approx.\ \qty{65}{\kilo\metre\per\hour}. The target driving time of one hour is missed. With \qty{80}{\percent} of the nominal battery energy, approx.\ \qty{44}{\minute} result at \qty{50}{\percent} motor load. With rated torque, the drive force at standstill exceeds the transmissible traction force by a factor of 1.7 to 2.0. The kart is therefore traction-limited up to approx.\ \qty{27}{\kilo\metre\per\hour} to \qty{43}{\kilo\metre\per\hour}, and traction control is essential. Regenerative braking reaches approx.\ $0.45\,g$. On a winding track, it could cover up to about \qty{30}{\percent} of the energy demand. The initial charging measurement quantifies the flat \LFP{} voltage plateau at approx.\ \qty{0.7}{\milli\volt} to \qty{1.8}{\milli\volt} per percent state of charge. During the first test runs, the cold cells showed a voltage sag under load that triggered the low-voltage alarm. The kart has no auxiliary \qty{12}{\volt} battery. The entire low-voltage system is supplied from the traction battery through an isolated DC/DC converter. Validation of these results by driving tests is the main open point.

\begin{otherlanguage}{ngerman}
\chapter*{Kurzfassung}\addcontentsline{toc}{chapter}{Kurzfassung}
{\bfseries Elektro-Rennkart mit BLDC-Antrieb, \LFP{}-Batteriesystem,\\ Antischlupfregelung und Rekuperation\par}

Im Rahmen dieses Projekts wurde ein Kart mit Benzinmotor zu einem Elektrokart umgebaut. Ziel war ein effizientes Fahrzeug mit einfachem Aufbau. Zudem sollte das Kart möglichst leicht bleiben, da ein höheres Gewicht das Fahrverhalten verschlechtert.

Ein wesentlicher Faktor ist die Gewichtsverteilung. Diese wird durch eine gezielte Aufteilung der Akkus auf drei Akkukästen erreicht. Für eine hohe Effizienz wurden ein Motor mit hohem Wirkungsgrad und Akkus mit geringer Selbstentladung verbaut. Weiters verfügt das Kart über eine Nutzbremse und eine eigens entwickelte Antischlupfregelung (ASR). Die dreistufige Rekuperation erfordert einen zusätzlichen Mikrocontroller. Die ASR verhindert durchdrehende Räder, die unnötig Energie verbrauchen würden.

Die Auswertung in Kapitel~\ref{sec:results-and-discussion} zeigt: Die berechnete Höchstgeschwindigkeit liegt bei ca.\ \qty{65}{\kilo\metre\per\hour}. Die Zielfahrzeit von einer Stunde wird verfehlt. Bei Nennmoment übersteigt die Antriebskraft im Stand die übertragbare Reifenkraft um den Faktor 1,7 bis 2,0. Die ASR ist daher ein notwendiger Teil des Antriebskonzepts. Die Rekuperation könnte auf einer kurvenreichen Strecke bis zu etwa \qty{30}{\percent} des Energiebedarfs decken. Bei kalten Zellen löste der Spannungseinbruch unter Last den Unterspannungsalarm aus, obwohl der Akku nicht leer war. Das Kart hat keine 12-V-Batterie. Das gesamte Niederspannungssystem wird über einen DC/DC-Wandler aus dem Traktionsakku versorgt. Die Bestätigung durch Messfahrten steht noch aus.

Diese Veröffentlichung ist die dritte, überarbeitete englische Ausgabe der 2016 an der HTL Braunau eingereichten Diplomarbeit.

\end{otherlanguage}

\tableofcontents
\clearpage
\pagenumbering{arabic}

\chapter{Introduction and Objectives}\label{sec:introduction-and-objectives}
This chapter presents the objectives of the project (Section~\ref{sec:objectives}) and describes its innovative aspects (Section~\ref{sec:innovation}).

\section{Objectives}\label{sec:objectives}
Electric mobility is currently the subject of broad discussion. In motor racing, electric boost systems have been used for some time to increase the torque for short periods. So far, however, the available battery capacity and the resulting short driving time severely limit their use in racing and in road traffic.

Energy storage systems that considerably increase range and efficiency are already available. The objective of the project is to adapt these systems specifically to racing. Limited installation space and the lowest possible weight are the boundary conditions. Specifically, a multi-stage recuperation system (recovery of braking energy) is developed. The recovered energy extends the driving time. For short periods, it can also be used for a higher motor torque and thus for higher driving performance. In Formula 1, comparable systems are used under the name Energy Recovery System (ERS).

Due to the high torque of the electric motor, large slip occurs at the drive axle during acceleration. To avoid this waste of energy, a traction control system with control algorithms tailored to this application is developed. As in road vehicles, the system is switched off in the event of a fault to ensure safety.

\section{Innovation}\label{sec:innovation}
The challenge of the project is to achieve a high drive efficiency and a long battery runtime in order to keep up with conventional karts powered by combustion engines. This is achieved by controlled feedback of kinetic energy into the battery and by a traction control system developed in-house.

The division of the battery pack into 24 single cells allows an optimal weight distribution. Furthermore, a lightweight chassis from kart racing contributes significantly to weight reduction and permits high lateral accelerations.

\textbf{Traction control:} The system prevents the drive wheels from spinning under strong acceleration. Large wheel slip consumes a lot of energy and does not contribute to acceleration. Traction control ensures maximum force transmission in the optimal slip range of the tyres. Speed sensors at the rear axle and at both front wheels measure the wheel speeds. Based on these signals, a control loop intervenes in the motor management. A protective function switches the system off in the event of a fault.

\textbf{Battery and battery management system:} The energy recovered during braking must be stored within a very short time. The selected \LFP{} battery technology permits very high charging currents. The use of 24 cells helps to absorb short charging current peaks. Each individual cell is monitored by a battery management system (BMS) tailored to the cell type. The BMS controls the charging and discharging process at the same time. Optionally, it allows visualisation on mobile phones and tablets.

\textbf{Energy recovery by three-stage recuperation:} During braking, part of the kinetic energy is converted into electrical energy and fed into the battery. To integrate the recuperation as efficiently as possible into normal driving, it is divided into three stages:

\begin{itemize}
  \item Stage 1: Recuperation when the throttle is released. Energy feedback starts without any action by the driver.
  \item Stage 2: At the push of a button on the steering wheel, kinetic energy is recovered and the speed is reduced without using the mechanical brake.
  \item Stage 3: If more braking force is required, the hydraulic brake acts on the rear axle in addition to the motor brake. This gives the maximum deceleration, while part of the energy is still recovered.
\end{itemize}

The innovation of the project lies above all in the combination of traction control, battery management and energy recovery by three-stage recuperation. This combination is expected to result in an electric racing kart that can keep up well with conventional petrol karts.

\chapter{Design and Implementation}\label{sec:design-and-implementation}
This chapter describes the topics the project team worked on during the project year 2015/16. Knowledge from mechatronics, electrical engineering, electronics and software development is presented where the respective topic requires it. Section~\ref{sec:chassis} gives a short overview of the chassis. Section~\ref{sec:selection-of-the-electrical-components} explains the considerations behind the selection of the motor and the battery. Section~\ref{sec:energy-storage} deals with the energy storage. Sections~\ref{sec:electric-drive} and~\ref{sec:mechanical-drive} describe the electric and the mechanical drive. Section~\ref{sec:traction-control} explains how an electronic traction control makes the kart more efficient during acceleration. Sections~\ref{sec:braking-system} and~\ref{sec:safety-systems} cover the braking system and the safety systems. Finally, Section~\ref{sec:mechanical-add-on-parts} describes the remaining mechanical add-on parts.

\section{Chassis}\label{sec:chassis}
\begin{wrapfigure}[13]{r}{0.5\textwidth}%
\vspace*{-11.6pt}%
\centering
\begin{tikzpicture}[schem, x=0.5cm, y=0.5cm,
  tube/.style={double, double distance=1.3pt, line width=0.55pt}]
  \draw[tube] (0,-1.45) -- (0,1.45);
  \draw[tube] (0,1.45) -- (1.66,2.58) -- (9.87,2.58) -- (11.0,1.92);
  \draw[tube] (0,-1.45) -- (1.66,-2.58) -- (9.87,-2.58) -- (11.0,-1.92);
  \draw[tube] (1.66,2.58) -- (1.66,-2.58);
  \draw[tube] (4.54,2.58) -- (4.54,-2.58);
  \draw[tube] (9.22,2.58) -- (9.22,-2.58);
  \draw[tube] (4.54,2.58) -- (1.66,-2.58);
  \draw[dashed, rounded corners=4mm] (5.87,-1.21) rectangle (8.32,1.21);
  \node at (7.10,0) {Seat};
  \draw[part, fill=white] (9.43,3.23) rectangle (11.21,4.35);
  \draw[part, fill=white] (9.43,-3.23) rectangle (11.21,-4.35);
  \draw[line width=1.6pt] (10.32,3.95) -- (10.32,-3.95);
  \draw[part] (0.10,3.17) rectangle (1.44,3.94);
  \draw[part] (0.10,-3.17) rectangle (1.44,-3.94);
  \draw (0.77,3.17) -- (0.77,2.62);
  \draw (0.77,-3.17) -- (0.77,-2.62);
  \node[above] at (0.77,3.94) {Front wheel};
  \node[above] at (10.32,4.35) {Rear wheel};
  \draw[leader] (11.0,1.0) -- (10.32,1.0);
  \node[lbl, right] at (11.0,1.0) {Rear\\ axle};
  \draw[leader] (5.6,2.66) -- (5.6,3.25);
  \node[lbl, above, align=center] at (5.6,3.25) {Frame tube\\ \O\,\qty{30}{\milli\metre}\\ (CrMo steel)};
  \draw[->] (6.9,-3.8) -- (4.3,-3.8);
  \node[above, inner sep=1.5pt] at (5.6,-3.8) {Driving direction};
  \draw[ext] (0.77,-3.94) -- (0.77,-4.95);
  \draw[ext] (10.32,-4.35) -- (10.32,-4.95);
  \draw[dim] (0.77,-4.75) -- node[fill=white, inner sep=2pt] {Wheelbase \qty{1040}{\milli\metre}} (10.32,-4.75);
\end{tikzpicture}
\caption{Top view of the kart frame (schematic)\nocc}\label{fig:rahmen}
\end{wrapfigure}
We bought the chassis from Kartshop Ampfing. It was built in 2012 by the Italian racing kart manufacturer CRG. Before delivery, the frame and the brake system were overhauled and checked for safety.

The kart frame is made of a chromium-molybdenum alloyed steel. The exact composition differs between manufacturers and is not disclosed. Molybdenum and chromium mainly increase the strength, toughness and hardenability of the steel. The frame tubes have a diameter of \qty{30}{\milli\metre}, and the wheelbase of our kart is \qty{1040}{\milli\metre} (Figure~\ref{fig:rahmen}).

\section{Selection of the Electrical Components}\label{sec:selection-of-the-electrical-components}
The components have to be dimensioned to match each other, and no component may overload another electrically. Service life and quality are important to withstand the high demands of racing. Mechanical robustness and electromagnetic immunity are essential above all. Since the kart is operated outdoors, connectors and electrical components must be protected against moisture. For robust connections, the Superseal connectors common in the automotive sector are used. Outdoor use also requires a temperature resistance from \qtyrange{-10}{+40}{\degreeCelsius}. Seals and housing parts must therefore be weatherproof. Contamination is a further problem, since it can quickly impair the cooling of air-cooled motors. All sensors must withstand dust and road dirt without loss of function. Not every component can meet all requirements equally, so compromises are often necessary. For this reason, various tests and calculations were carried out.

\begin{table}[!ht]
\centering
\caption{Comparison of the considered motor types~\cite{Loh15,VEM15,GM15,Ens15}}\label{tab:motors}
{\small\renewcommand{\arraystretch}{1.15}%
\begin{tabularx}{\linewidth}{@{}>{\raggedright\arraybackslash}p{3.3cm}*{4}{>{\centering\arraybackslash}X}@{}}
\toprule
 & \textbf{Asynchronous machine} & \textbf{Synchronous machine} & \textbf{Brushless DC motor} & \textbf{PM axial-flux motor} \\
\midrule
Manufacturer & Loher AG & VEM & Golden Motor & Enstroj \\
Type & \mbox{ANGA-160MB-02} & \mbox{PE1R 132 M4} & \mbox{HPM-10kW} & \mbox{EMRAX 208} \\
Rated power in \unit{\kilo\watt} & 11 & 11 & 10 & 20 \\
Rated voltage in \unit{\volt} & 400 & 370 & 72 & 115 \\
Supply frequency in \unit{\hertz} & 50 & 100 & DC (via controller) & DC (via controller) \\
Rated speed in \unit{\per\minute} & 2920 & 3000 & 4500 & 6000 \\
Cooling & air & air & air & air \\
Maintenance-free & yes & yes & yes & yes \\
Weight in \unit{\kilo\gram} & 104 & 70 & 17 & 9.1 \\
Rotor inertia & high & very high & low & low \\
Power density in \unit{\kilo\watt\per\kilo\gram} & 0.11 & 0.16 & 0.59 & 2.20 \\
Rated efficiency & 0.885 & 0.92 & 0.91 & 0.93--0.98 \\
\bottomrule
\end{tabularx}}
\end{table}

\subsection{Selection of the Motor}\label{sec:selection-of-the-motor}
The aim is a motor type with low weight and high efficiency. The research identified four possible motor types. Electric vehicles mainly use synchronous and asynchronous machines. Smaller vehicles, such as golf carts, more often use DC motors, which are almost exclusively permanent-magnet excited. Brushless DC motors are used when a long service life at moderate power is required. The highest power densities are achieved by permanent-magnet machines with liquid cooling.

Initially, a liquid-cooled motor from Enstroj~\cite{Ens15} was considered because of its impressive performance data. However, such a drive concept was beyond the scope of a student project. The additional effort and installation space of a liquid cooling system also spoke against this solution. With the revised power requirement of approx.\ \qty{10}{\kilo\watt}, several motor concepts are possible, since this power range can still be realised with air cooling. A far more important aspect is a high operating voltage to keep the currents low.

At the beginning of the project, the team had hardly any experience with motors for electric mobility. Table~\ref{tab:motors} therefore compares the main data of the motor concepts considered. On the basis of this comparison and after technical discussions with the supervisors, the model HPM-10kW was selected. The operating principles of the machine types are summarised below.

\needspace{14\baselineskip}
\subsection{The Asynchronous Machine}\label{sec:the-asynchronous-machine}
\begin{wrapfigure}[12]{r}{0.5\textwidth}%
\wraptop
\centering
\begin{tikzpicture}[schem, scale=0.62,
  cdot/.style={circle, draw, line width=0.5pt, minimum size=2.6mm, inner sep=0pt,
    path picture={\fill (path picture bounding box.center) circle (0.35mm);}},
  ccross/.style={circle, draw, line width=0.5pt, minimum size=2.6mm, inner sep=0pt,
    path picture={\draw[line width=0.5pt] (path picture bounding box.south west) -- (path picture bounding box.north east)
      (path picture bounding box.north west) -- (path picture bounding box.south east);}}]
  \draw[part] (0,0) circle (2.5);
  \draw[part] (0,0) circle (1.82);
  \draw[part] (0,0) circle (1.47);
  \foreach \a in {105,135,165,195,225,255} \node[cdot] at (\a:2.13) {};
  \foreach \a in {75,45,15,-15,-45,-75} \node[ccross] at (\a:2.13) {};
  \foreach \i in {0,...,19} \fill (\i*18:1.22) circle (0.9mm);
  \draw[part, fill=white] (0,0) circle (0.2);
  \draw[->] (30:2.78) arc[start angle=30, end angle=118, radius=2.78];
  \node[above] at (90:2.85) {Rotating field $n_{\mathrm{s}}$};
  \draw[->] (60:0.85) arc[start angle=60, end angle=150, radius=0.85];
  \node at (0,-0.62) {$n<n_{\mathrm{s}}$};
  \draw[leader] (33:2.3) -- (2.95,1.9) node[lbl, right] {Stator with\\ three-phase winding};
  \draw[leader] (10:1.65) -- (2.95,0.55) node[lbl, right] {Air gap};
  \draw[leader] (-9:1.36) -- (2.95,-0.55) node[lbl, right] {Squirrel-cage rotor\\ (short-circuited bars)};
  \draw[leader] (-35:0.2) -- (2.95,-1.9) node[lbl, right] {Shaft};
\end{tikzpicture}
\caption{Operating principle of the asynchronous machine with squirrel-cage rotor (schematic)\nocc}\label{fig:asm}
\end{wrapfigure}
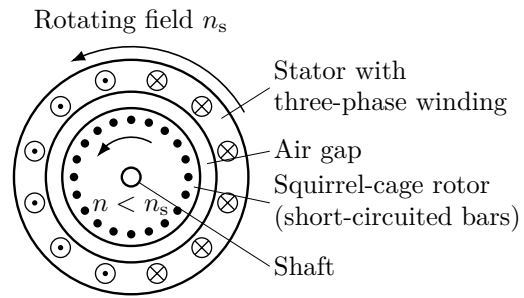
Compared with the DC machine, the asynchronous motor, especially with a squirrel-cage rotor, has a considerably simpler and more robust design (Figure~\ref{fig:asm}). It therefore requires little maintenance. A disadvantage is the close link between operating speed and stator frequency~\cite{Fis04}.

An asynchronous machine requires neither permanent magnets nor electromagnets in the rotor. In mains operation, no starting aid is required. A major disadvantage is the lower power density compared with other motor types. An advantage is the high efficiency at high rated power and the correspondingly low losses.

\subsection{The Synchronous Machine}\label{sec:the-synchronous-machine}
\begin{figure}[!ht]
\centering
\begin{tikzpicture}[schem, scale=0.82,
  slot/.style={circle, draw, line width=0.45pt, minimum size=2.2mm, inner sep=0pt},
  coil/.style={draw, line width=0.6pt, pattern=north east lines},
  cdot/.style={circle, draw, line width=0.45pt, minimum size=2.4mm, inner sep=0pt,
    path picture={\fill (path picture bounding box.center) circle (0.3mm);}},
  ccross/.style={circle, draw, line width=0.45pt, minimum size=2.4mm, inner sep=0pt,
    path picture={\draw[line width=0.45pt] (path picture bounding box.south west) -- (path picture bounding box.north east)
      (path picture bounding box.north west) -- (path picture bounding box.south east);}},
  pole/.style={font=\small\bfseries}]
  \newcommand{\uvw}[1]{%
    \foreach \t/\y in {U/0.38,V/0,W/-0.38} {
      \node[circle, draw, line width=0.45pt, minimum size=1.6mm, inner sep=0pt] (t\t) at (#1,\y) {};
      \node[right=0.5mm of t\t, font=\small] {\t};}}
  \begin{scope}
    \draw[part] (0,0) circle (2.0);
    \draw[part] (0,0) circle (1.39);
    \draw[part] (0,0) circle (1.20);
    \foreach \i in {0,...,15} \node[slot] at (11.25+\i*22.5:1.62) {};
    \foreach \a in {112,150,210,248} \node[cdot] at (\a:0.95) {};
    \foreach \a in {68,30,-30,-68} \node[ccross] at (\a:0.95) {};
    \node[pole] at (0,0.62) {N};
    \node[pole] at (0,-0.62) {S};
    \draw[part, fill=white] (0,0) circle (0.17);
    \uvw{2.45}
    \draw (10.95:2.0) -- (tU.west) (2.0,0) -- (tV.west) (-10.95:2.0) -- (tW.west);
    \node[align=left, font=\small, anchor=south west] at (-2.0,2.15) {\textbf{a)} Cylindrical rotor,\\ three-phase stator};
  \end{scope}
  \begin{scope}[xshift=6.0cm]
    \draw[part] (0,0) circle (2.0);
    \draw[part] (0,0) circle (1.75);
    \foreach \s/\l in {1/N,-1/S} {
      \draw[part, fill=white, yscale=\s] (-0.45,1.74) -- (0.45,1.74) -- (0.45,1.30) -- (0.71,1.03)
        arc[start angle=55.42, end angle=124.58, radius=1.251] -- (-0.45,1.30) -- cycle;
      \draw[coil, yscale=\s] (-0.72,1.32) rectangle (-0.50,1.62);
      \draw[coil, yscale=\s] (0.50,1.32) rectangle (0.72,1.62);
      \node[pole] at (0,1.42*\s) {\l};}
    \draw[part] (0,0) circle (0.95);
    \foreach \i in {0,...,9} \node[slot] at (18+\i*36:0.74) {};
    \draw[part, fill=white] (0,0) circle (0.17);
    \foreach \y/\t in {1.65/U,1.30/V,0.95/W} {
      \draw[line width=0.5pt] (2.40,\y-0.13) rectangle (2.52,\y+0.13);
      \draw (2.52,\y) -- (2.86,\y);
      \node[circle, draw, line width=0.45pt, minimum size=1.6mm, inner sep=0pt, fill=white] at (2.94,\y) {};
      \node[font=\small, anchor=west] at (3.0,\y) {\t};}
    \node[align=center, font=\small, above] at (2.46,1.82) {Slip\\ rings};
    \node[align=left, font=\small, anchor=south west] at (-2.0,2.15) {\textbf{b)} Field on stator,\\ three-phase rotor};
  \end{scope}
  \begin{scope}[xshift=12.0cm]
    \draw[part] (0,0) circle (2.0);
    \draw[part] (0,0) circle (1.39);
    \foreach \i in {0,...,15} \node[slot] at (11.25+\i*22.5:1.62) {};
    \draw[part, fill=white] (-0.62,1.19) arc[start angle=117.52, end angle=62.48, radius=1.342] -- (0.30,0.93)
      -- (0.30,-0.93) -- (0.62,-1.19) arc[start angle=-62.48, end angle=-117.52, radius=1.342]
      -- (-0.30,-0.93) -- (-0.30,0.93) -- cycle;
    \foreach \s in {1,-1} {
      \draw[coil, yscale=\s] (-0.52,0.38) rectangle (-0.33,0.74);
      \draw[coil, yscale=\s] (0.33,0.38) rectangle (0.52,0.74);}
    \node[pole] at (0,0.58) {N};
    \node[pole] at (0,-0.58) {S};
    \draw[part, fill=white] (0,0) circle (0.17);
    \uvw{2.45}
    \draw (10.95:2.0) -- (tU.west) (2.0,0) -- (tV.west) (-10.95:2.0) -- (tW.west);
    \node[align=left, font=\small, anchor=south west] at (-2.0,2.15) {\textbf{c)} Salient-pole rotor,\\ three-phase stator};
  \end{scope}
\end{tikzpicture}
\caption{Designs of the synchronous machine: a) internal-pole machine with cylindrical rotor, b) external-pole machine, c) internal-pole machine with salient-pole rotor (schematic)\nocc}\label{fig:sm_combined}
\end{figure}
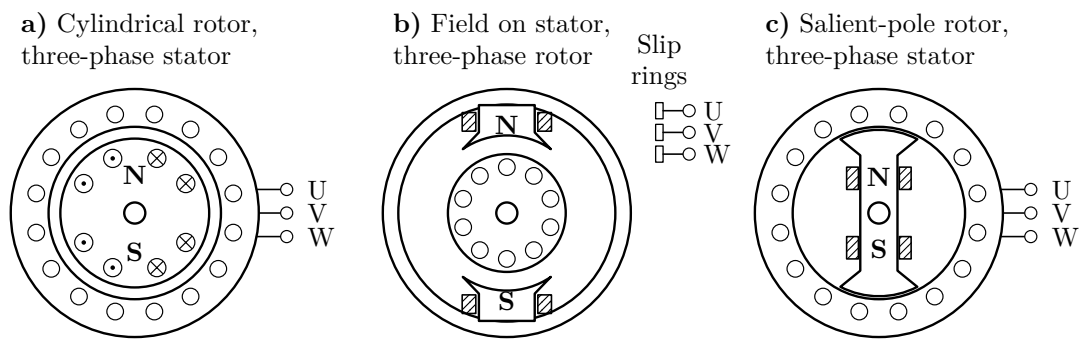

The synchronous machine exists in two basic designs, the external-pole and the internal-pole machine. Internal-pole machines are further built as salient-pole or cylindrical-rotor (non-salient pole) machines (Figure~\ref{fig:sm_combined}).

\textbf{External-pole machine:} In its basic design, it corresponds to a DC machine without commutating poles. The field winding is located on poles in the stator, and the three-phase winding is in the rotor. Since the three-phase power has to be transmitted via slip rings, this design is problematic for high power.

\textbf{Internal-pole machine:} The three-phase winding is located in the stator and the field winding in the rotor. This design is therefore suitable for high power. High speeds require a low number of pole pairs. In salient-pole machines, the individual poles take up a lot of space on the circumference, so the rotors have a large diameter and are only suitable for low speeds. High speeds are realised with cylindrical rotors. Their outer diameter is relatively small, which limits the centrifugal forces. With permanent-magnet excitation in the rotor, these motors are suitable for high speeds and power levels in electric vehicles. A further advantage is the lower moment of inertia of the rotor compared with salient-pole machines.

As an industrial drive, the synchronous machine has gained considerable importance through the development of frequency converters. It is now available as a variable-speed drive from servo motors up to the highest power ratings. Examples are a \qty{30}{\mega\watt} blast-furnace blower with synchronous motor and starting converter, and drives for cement mills, conveyor systems and rolling stands up to the megawatt range~\cite{Fis04}.

\needspace{16\baselineskip}
\subsection{The Brushless DC Motor}\label{sec:the-brushless-dc-motor}
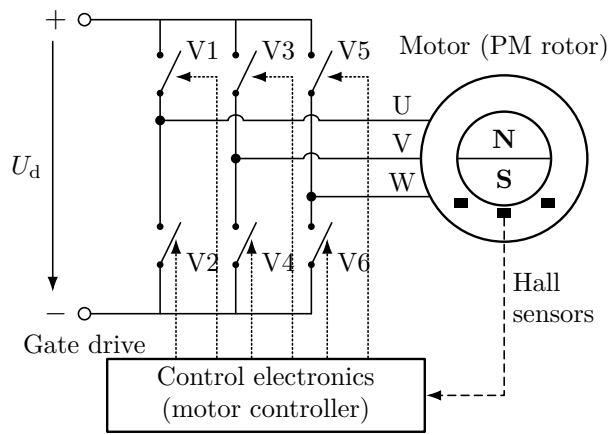
\begin{wrapfigure}[15]{r}{0.56\textwidth}%
\vspace*{-13.3pt}%
\centering
\begin{tikzpicture}[schem, y=0.72cm,
  term/.style={circle, draw, line width=0.6pt, fill=white, minimum size=1.6mm, inner sep=0pt},
  jnc/.style={circle, fill, minimum size=1.3mm, inner sep=0pt}]
  \draw (0,5.4) -- (3.0,5.4);
  \draw (0,0) -- (3.0,0);
  \node[term, label={[font=\normalsize]left:$+$}] at (0,5.4) {};
  \node[term, label={[font=\normalsize]left:$-$}] at (0,0) {};
  \draw[->] (-0.42,5.0) -- node[left] {$U_{\mathrm{d}}$} (-0.42,0.4);
  \foreach \x/\hi/\lo in {1.0/V1/V2, 2.0/V3/V4, 3.0/V5/V6} {
    \draw (\x,5.4) -- (\x,4.75);
    \draw (\x,4.05) -- (\x+0.27,4.82);
    \draw (\x,4.05) -- (\x,1.6);
    \draw (\x,0.9) -- (\x+0.27,1.67);
    \draw (\x,0.9) -- (\x,0);
    \foreach \y in {4.75,4.05,1.6,0.9} \fill (\x,\y) circle (0.45mm);
    \node[right, font=\small, inner sep=1pt] at (\x+0.3,4.85) {\hi};
    \node[right, font=\small, inner sep=1pt] at (\x+0.3,0.95) {\lo};}
  \draw (1.0,3.55) -- (1.9,3.55) arc[start angle=180, end angle=0, radius=0.1] -- (2.9,3.55)
    arc[start angle=180, end angle=0, radius=0.1] -- (4.57,3.55);
  \draw (2.0,2.85) -- (2.9,2.85) arc[start angle=180, end angle=0, radius=0.1] -- (4.45,2.85);
  \draw (3.0,2.15) -- (4.57,2.15);
  \node[jnc] at (1.0,3.55) {};
  \node[jnc] at (2.0,2.85) {};
  \node[jnc] at (3.0,2.15) {};
  \foreach \t/\y in {U/3.55,V/2.85,W/2.15} \node[above, font=\small, inner sep=2pt] at (4.2,\y) {\t};
  \draw[part] (5.55,2.85) circle[radius=1.1cm];
  \draw[part] (5.55,2.85) circle[radius=0.62cm];
  \draw (4.93,2.85) -- (6.17,2.85);
  \node[font=\small\bfseries] at (5.55,3.22) {N};
  \node[font=\small\bfseries] at (5.55,2.48) {S};
  \foreach \a in {235,270,305} \fill ($(5.55,2.85)+(\a:1.0)$) +(-0.09,-0.09) rectangle +(0.09,0.09);
  \node[above] at (5.55,4.45) {Motor (PM rotor)};
  \node[block, minimum width=4.2cm, minimum height=0.95cm] (ce) at (2.4,-1.5) {Control electronics\\ (motor controller)};
  \foreach \x in {1.0,2.0,3.0} {
    \draw[densely dotted, ->] (\x+0.21,0 |- ce.north) -- (\x+0.21,1.38);
    \draw[densely dotted, ->] (\x+0.75,0 |- ce.north) -- (\x+0.75,4.35) -- (\x+0.2,4.35);}
  \node[left, font=\small] at (0.95,-0.55) {Gate drive};
  \draw[signal] (5.55,1.76) -- (5.55,-1.5) -- (ce.east);
  \node[lbl, right, font=\small] at (5.6,0.3) {Hall\\ sensors};
\end{tikzpicture}
\caption{Basic circuit of a BLDC drive with B6 bridge and Hall position sensors (schematic)\nocc}\label{fig:bldc}
\end{wrapfigure}
The brushless DC (BLDC) motor is similar in design to a synchronous machine with permanent-magnet excitation in the rotor. A three-phase winding generates a rotating field that drives the rotor equipped with permanent magnets. In this respect, the motor does not work like a conventional DC machine. To obtain an operating behaviour like that of a DC machine, control and power electronics are required (Figure~\ref{fig:bldc}). In the kart, this task is performed by the motor controller.

Permanent-magnet excitation allows high power, and the design, which is almost identical to that of a synchronous machine, allows simple production. Medium to high power densities are possible. A four-quadrant operation can be realised with relatively simple power electronics. An input rectifier is not required, since the battery voltage is applied directly to the DC link. A further advantage over conventional DC motors is freedom from maintenance, since brushes and commutator are omitted.

\subsection{Selection of the Battery}\label{sec:selection-of-the-battery}
The requirements for the battery are manifold. On the one hand, it should offer the highest possible energy density at low weight. On the other hand, it must be able to supply the required discharge current. Since the cells are located right next to the driver, safety had top priority. Specifically, the battery must not explode or catch fire when overcharged or short-circuited, and it must not overheat under overload. With proper use, overload should not occur, since the battery is operated within its specification. Under extreme conditions, such as racing, it can occur very quickly.

A further aspect is the weight distribution on the kart. Large battery blocks are difficult to distribute, whereas many single cells allow the available space to be used optimally. With many single cells, however, a battery management system is required to balance unequal states of charge. The three technologies considered are compared below; their structure is shown in Figure~\ref{fig:batteries}.

\subsection{Lead-Acid Battery}\label{sec:lead-acid-battery}
Lead-acid batteries are robust and reliable (Figure~\ref{fig:batteries}a). A major disadvantage is the low energy density and the resulting high weight at large capacities. Very high currents can be drawn for short periods. Charging should be done with suitable chargers. With adequate care, high-quality batteries reach a service life of more than eight years. Due to its long use as a starter battery in vehicles, this technology is very mature.

Overcharging is less critical than with other battery types. It does, however, cause the electrolyte to gas and produces oxyhydrogen, which can be ignited by a small spark. Fires or explosions caused by short circuits are nevertheless very rare. In unsealed batteries, electrolyte losses can be compensated easily by adding distilled water. A battery consists of several cells connected in series and usually has a voltage of \qty{12}{\volt}. In the charged state, each cell has an open-circuit voltage of approx.\ \qty{2.1}{\volt}. Deep discharge damages a lead-acid battery permanently. If a long service life is required, the cells should not be discharged below \qty{25}{\percent} state of charge. The state of charge can be estimated relatively easily by measuring the cell voltage.

\subsection{\LFP{} Technology}\label{sec:lifepo4-technology}
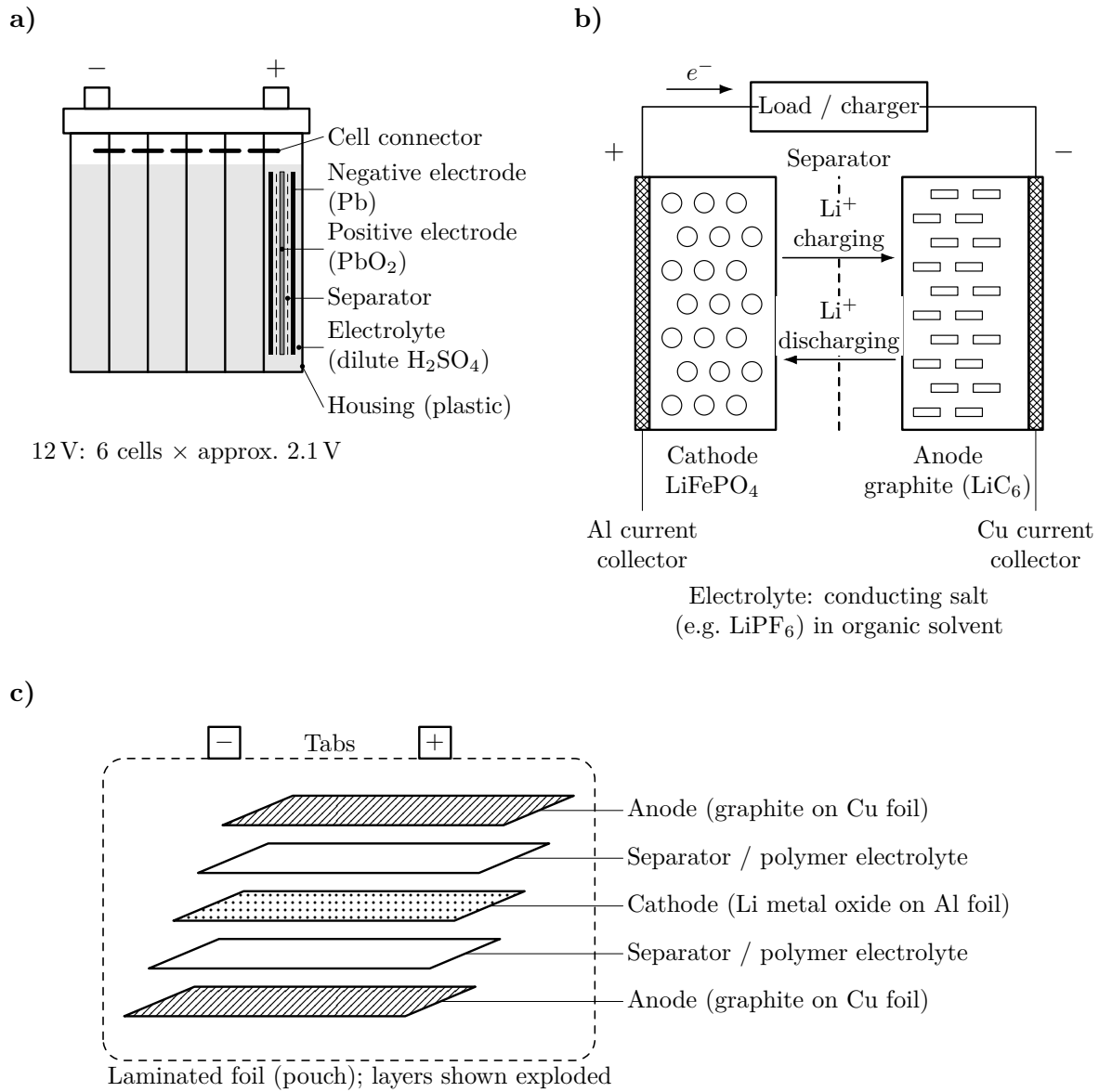
\begin{figure}[!t]
\centering
\begin{minipage}[t]{0.49\textwidth}
\textbf{a)}\par\medskip
\centering
\begin{tikzpicture}[schem]
  \fill[black!10] (0,0) rectangle (3.3,2.95);
  \draw[part] (0,0) rectangle (3.3,3.4);
  \draw[part] (-0.1,3.4) rectangle (3.4,3.75);
  \draw[part] (0.2,3.75) rectangle (0.55,4.05);
  \draw[part] (2.75,3.75) rectangle (3.1,4.05);
  \node[above, font=\normalsize] at (0.375,4.05) {$-$};
  \node[above, font=\normalsize] at (2.925,4.05) {$+$};
  \foreach \w in {0.55,1.1,1.65,2.2,2.75} {
    \draw[part] (\w,0) -- (\w,3.4);
    \draw[line width=1.8pt] (\w-0.2,3.15) -- (\w+0.2,3.15);}
  \fill (2.82,0.25) rectangle (2.88,2.85);
  \fill (3.14,0.25) rectangle (3.20,2.85);
  \draw[line width=0.4pt, fill=black!45] (2.98,0.25) rectangle (3.04,2.85);
  \draw[line width=0.4pt, densely dashed] (2.93,0.25) -- (2.93,2.85) (3.09,0.25) -- (3.09,2.85);
  \node[below] at (1.65,-0.85) {\qty{12}{\volt}: 6 cells $\times$ approx.\ \qty{2.1}{\volt}};
  \foreach \px/\py/\ly/\t in {2.95/3.15/3.35/Cell connector, 3.17/2.6/2.6/{Negative electrode\\ (Pb)},
      3.01/1.75/1.75/{Positive electrode\\ (PbO$_2$)}, 3.09/1.05/1.05/Separator,
      3.25/0.35/0.35/{Electrolyte\\ (dilute H$_2$SO$_4$)}, 3.3/0.08/-0.5/Housing (plastic)} {
    \fill (\px,\py) circle (0.35mm);
    \draw[leader] (\px,\py) -- (3.6,\ly) node[lbl, right] {\t};}
\end{tikzpicture}
\end{minipage}\hfill
\begin{minipage}[t]{0.49\textwidth}
\textbf{b)}\par\medskip
\centering
\begin{tikzpicture}[schem]
  \draw[part] (0.2,0) rectangle (2.0,3.6);
  \draw[part] (3.8,0) rectangle (5.6,3.6);
  \foreach \r in {0,...,6} {
    \pgfmathsetmacro{\y}{0.35+\r*0.48}
    \pgfmathsetmacro{\o}{mod(\r,2)*0.22}
    \foreach \x in {0.52,0.98,1.44} \draw[line width=0.45pt] (\x+\o,\y) circle (0.14);}
  \foreach \r in {0,...,9} {
    \pgfmathsetmacro{\y}{0.25+\r*0.345}
    \pgfmathsetmacro{\o}{mod(\r,2)*0.25}
    \foreach \x in {4.15,4.75} \draw[line width=0.45pt] (\x+\o-0.19,\y-0.06) rectangle (\x+\o+0.19,\y+0.06);}
  \draw[part, pattern=crosshatch] (0,0) rectangle (0.2,3.6);
  \draw[part, pattern=crosshatch] (5.6,0) rectangle (5.8,3.6);
  \draw[part, dashed] (2.9,0) -- (2.9,3.6);
  \node[above, inner sep=2pt] at (2.9,3.6) {Separator};
  \draw (0.1,3.6) -- (0.1,4.6) -- (1.75,4.6);
  \draw (4.05,4.6) -- (5.7,4.6) -- (5.7,3.6);
  \node[block, minimum width=2.3cm, minimum height=0.7cm] at (2.9,4.6) {Load / charger};
  \node[font=\normalsize] at (-0.3,3.95) {$+$};
  \node[font=\normalsize] at (6.1,3.95) {$-$};
  \draw[->] (0.45,4.85) -- node[above] {$e^-$} (1.4,4.85);
  \draw[->] (2.1,2.45) -- (3.7,2.45);
  \node[fill=white, inner sep=1pt, align=center, font=\small, above] at (2.9,2.5) {Li$^+$\\ charging};
  \draw[->] (3.7,1.0) -- (2.1,1.0);
  \node[fill=white, inner sep=1pt, align=center, font=\small, above] at (2.9,1.05) {Li$^+$\\ discharging};
  \node[align=center, below] at (1.1,-0.1) {Cathode\\ LiFePO$_4$};
  \node[align=center, below] at (4.45,-0.1) {Anode\\ graphite (LiC$_6$)};
  \draw[leader] (0.1,0) -- (0.1,-1.15);
  \draw[leader] (5.7,0) -- (5.7,-1.15);
  \node[align=center, below] at (0.1,-1.15) {Al current\\ collector};
  \node[align=center, below] at (5.7,-1.15) {Cu current\\ collector};
  \node[align=center, below] at (2.9,-2.1) {Electrolyte: conducting salt\\ (e.g.\ LiPF$_6$) in organic solvent};
\end{tikzpicture}
\end{minipage}\par\bigskip
\begin{minipage}[t]{\textwidth}
\textbf{c)}\par\medskip
\centering
\begin{tikzpicture}[schem]
  \foreach \i/\fillstyle/\t in {0/{pattern=north east lines}/{Anode (graphite on Cu foil)},
      1/{fill=white}/{Separator / polymer electrolyte},
      2/{pattern=dots}/{Cathode (Li metal oxide on Al foil)},
      3/{fill=white}/{Separator / polymer electrolyte},
      4/{pattern=north east lines}/{Anode (graphite on Cu foil)}} {
    \pgfmathsetmacro{\y}{3.0-\i*0.68}
    \pgfmathsetmacro{\s}{(4-\i)*0.35}
    \tikzset{layer/.style/.expand once=\fillstyle}
    \draw[part, layer] (\s,\y) -- (\s+4.0,\y) -- (\s+5.0,\y+0.42) -- (\s+1.0,\y+0.42) -- cycle;
    \draw[leader] (\s+4.5,\y+0.21) -- (7.1,\y+0.21) node[lbl, right] {\t};}
  \draw[dashed, rounded corners=3mm] (-0.3,-0.35) rectangle (6.7,3.95);
  \draw[part, fill=white] (1.2,3.95) rectangle (1.65,4.4);
  \draw[part, fill=white] (4.2,3.95) rectangle (4.65,4.4);
  \node at (1.425,4.175) {$-$};
  \node at (4.425,4.175) {$+$};
  \node at (2.925,4.175) {Tabs};
  \node[lbl, below right] at (-0.3,-0.35) {Laminated foil (pouch); layers shown exploded};
\end{tikzpicture}
\end{minipage}
\caption{Battery technologies: a) structure of a lead-acid battery, b) structure and operating principle of a \LFP{} cell, c) layer structure of a lithium polymer pouch cell (schematic)\nocc}\label{fig:batteries}
\end{figure}

This technology is also known as the lithium iron phosphate battery (Figure~\ref{fig:batteries}b). Its main advantages are the high short-term load capacity of the cells and the high cycle life. Owing to the thermally and chemically very stable cathode material (iron phosphate), these batteries are comparatively safe against short circuit and overcharging. However, the cells contain flammable materials. With regard to explosions, lithium iron phosphate cells are safer than other lithium-ion cells. A single deep discharge or overcharge can permanently damage a cell, but does not lead to explosion or fire. A short circuit leads to gas release only after a long time. A sustained short circuit, however, can ignite the cell.

A disadvantage is the relatively low cell voltage of approx.\ \qty{3.2}{\volt}, which requires many cells to be connected in series. The use of a BMS is therefore essential. The energy density is lower than that of other lithium-ion technologies, but sufficient for this application. Using the full capacity reduces the service life. Due to the very flat discharge curve, estimating the state of charge from the terminal voltage is difficult. Precise values can only be obtained by measuring the discharge current with a current sensor and calculating the charge drawn.

\subsection{Lithium Polymer Technology}\label{sec:lithium-polymer-technology}
LiPo batteries are mainly used in model making, where low weight is the top priority (Figure~\ref{fig:batteries}c). Lithium polymer cells react much more sensitively to overstress than most comparable battery types. Harmful influences are above all overcharging and deep discharge, as well as operation with excessive currents or outside the specified temperature range. Damage or excessive heating results in a considerable fire risk. Special chargers are required, and during discharge the current and the cell temperature must be monitored. An advantage is the slightly higher cell voltage compared with \LFP{} cells. With large capacities and high charging currents, however, the fire risk is considerable. In addition, ageing processes can unpredictably cause an internal short circuit. For safety reasons, this technology is not suitable for use in the kart.

\subsection{Selection of Charger and Battery Management System}\label{sec:selection-of-charger-and-battery-managem}
The requirements for the battery management system result from the selected \LFP{} cells. These cells can very easily be overcharged or deeply discharged and must therefore be operated strictly within their specification. For a long service life, the operating parameters have to be monitored by a battery management system. Since several cells are connected in series to reach a higher voltage, charge balancing between the cells and monitoring of the cell voltages are also necessary.

\needspace{9\baselineskip}%
\begin{wrapfigure}[9]{r}{0.45\textwidth}%
\vspace*{\dimexpr-\intextsep+11.4pt}%
\centering
\includegraphics[width=0.45\textwidth]{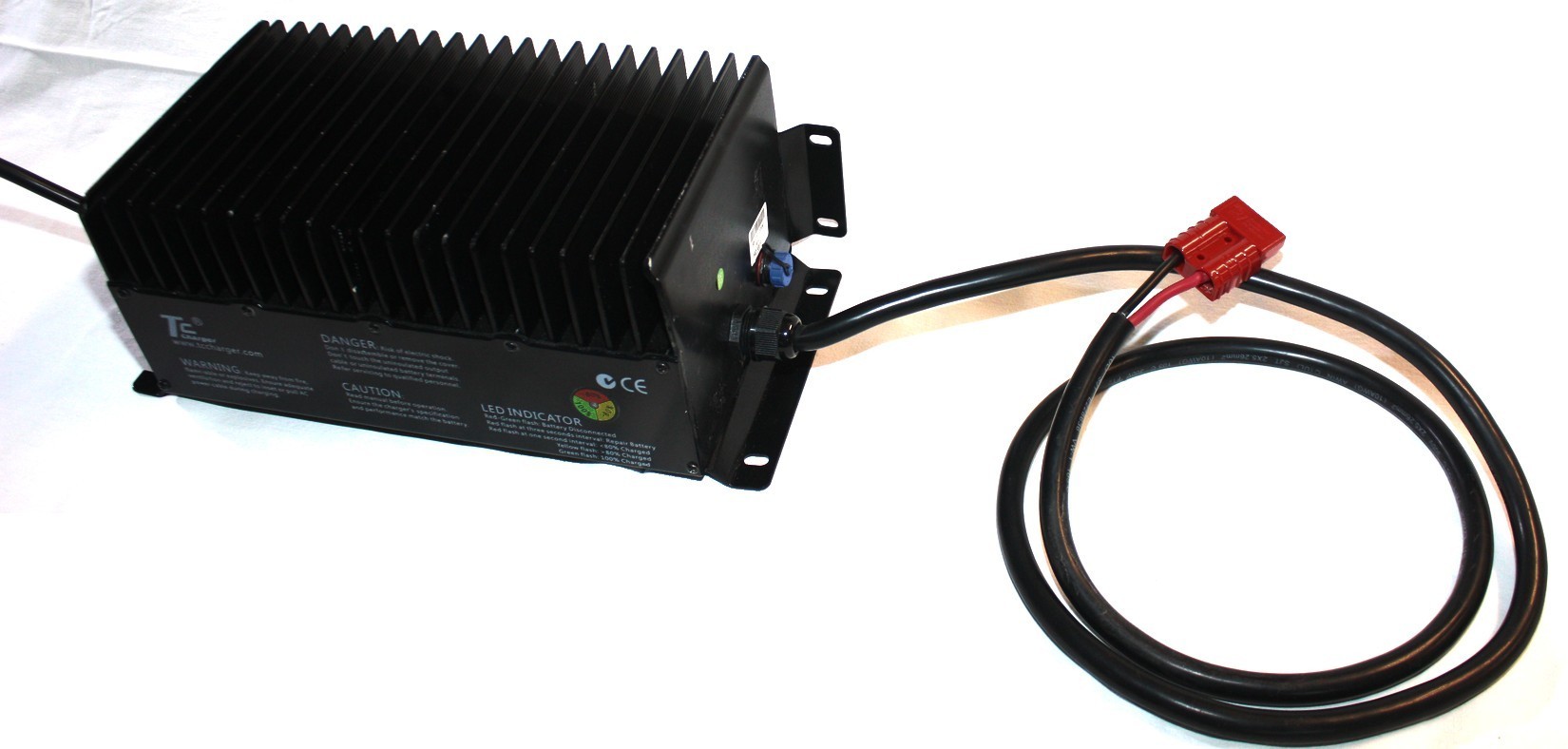}
\caption{Charger of the kart}\label{fig:image66}
\end{wrapfigure}
Many battery management systems are available with different configuration options. After thorough research, the 123\textbackslash BMS system by 123electric~\cite{BMS15} was selected. The BMS is optimised for chargers of the TC/Elcon series. After technical advice from Faktor.de, the charger CL4Y-72V-16A, adapted to \LFP{} cells, was chosen (Figure~\ref{fig:image66}). The BMS releases or stops charging via an enable interface.

A PC software shows all battery values at a glance. Gauges display the state of charge, cell voltage, charging or discharging current, total voltage and cell temperature. Three indicator LEDs show the active charging method. Red LEDs indicate a value outside the set limits: ``E'' stands for a communication error with the BMS boards, ``H'' for overvoltage or overtemperature, and ``L'' for a cell temperature below the set range or a cell voltage that is too low.

Three charging methods can be selected in the software:

\begin{itemize}
  \item \textbf{Emergency Charge} is used to recharge a deeply discharged cell or when the BMS can no longer communicate with the individual cells. All protective functions are disabled.
  \item \textbf{Quick Charge} charges the cells with maximum current until one cell reaches the maximum voltage or all cells reach the bypass voltage. Charge balancing does not take place.
  \item \textbf{Balanced Charge} charges the cells with maximum current until one cell reaches the bypass voltage. The charging current is then reduced to the set minimum and the cell voltages are balanced. Charging ends when one cell reaches the maximum voltage or all cells reach the bypass voltage.
\end{itemize}

\section{Energy Storage}\label{sec:energy-storage}
A major disadvantage of electric vehicles is the limited range or, in racing, the limited driving time. This limitation can be reduced above all by modern battery systems and the corresponding know-how. The target is a driving time of at least one hour at \qty{50}{\percent} motor load. \LFP{} single cells were chosen, as they are well suited for this application (Section~\ref{sec:selection-of-the-battery}). Each cell has a nominal voltage of \qty{3.2}{\volt}. A higher voltage is obtained by series connection. Due to the limited installation space, 24 cells can be mounted. This results in a nominal pack voltage of $24\times\qty{3.2}{\volt}=\qty{76.8}{\volt}$; the electrical components of the drive are designed for the \qty{72}{\volt} voltage class.

According to the manufacturer, only \qty{80}{\percent} of the maximum capacity should be used in normal operation to achieve a long service life. A few full charging cycles to \qty{90}{\percent} to \qty{95}{\percent} of the maximum capacity reduce the service life only slightly. The kart is mainly intended as a demonstration vehicle and not for daily racing use. To ensure the greatest possible safety for the driver, the cells are operated strictly within the specification of the manufacturer (Table~\ref{tab:cells} in Appendix~\ref{app:technical-data}). Remarkable is the high short-term discharge current of \qty{1200}{\ampere} for 5 seconds per minute.

\subsection{Calculation of the Driving Time}\label{sec:calculation-of-the-driving-time}
As a first approximation, \qty{50}{\percent} of full load is assumed in driving operation. The driving time follows from the stored energy and the average power (Table~\ref{tab:drivetime}). The stored energy is $E=24\times\qty{3.2}{\volt}\times\qty{60}{\ampere\hour}=\qty{4608}{\watt\hour}$. At \qty{50}{\percent} of the motor power of \qty{10}{\kilo\watt}, the average power is $P=\qty{5}{\kilo\watt}$, and the driving time is $t=E/P\approx\qty{55}{\minute}$.

\begin{wraptable}{r}{0.5\textwidth}
\vspace*{\dimexpr-\intextsep-1.5pt}%
\centering
\caption{Calculation of the driving time}\label{tab:drivetime}
{\small\renewcommand{\arraystretch}{1.15}%
\begin{tabular}{@{}lc@{}}
\toprule
\textbf{Parameter} & \textbf{Value} \\
\midrule
Nominal voltage of one cell in \unit{\volt} & 3.2 \\
Number of cells & 24 \\
Capacity of one cell in \unit{\ampere\hour} & 60 \\
Motor power in \unit{\watt} & \num{10000} \\
Total energy in \unit{\watt\hour} & 4608 \\
Driving time at \qty{50}{\percent} motor load in \unit{\minute} & 55 \\
\bottomrule
\end{tabular}}
\end{wraptable}
The driving time at \qty{50}{\percent} motor load is \qty{55}{\minute}. The target of one hour is therefore narrowly missed. This value refers to the full nominal energy; if only the recommended \qty{80}{\percent} are used, approx.\ \qty{44}{\minute} result (Section~\ref{sec:energy-demand-and-driving-time}). How this translates into real racing or driving operation is not yet known. Experience from the previous year's electric kart project shows that this value is realistic for racing on an indoor kart track. For visitor rides at events of HTL Braunau, the possible driving time will roughly double. The driving times achievable in outdoor racing will only become apparent in test drives, as no empirical values are available for this purpose.

\subsection{Initial Charging}\label{sec:initial-charging}
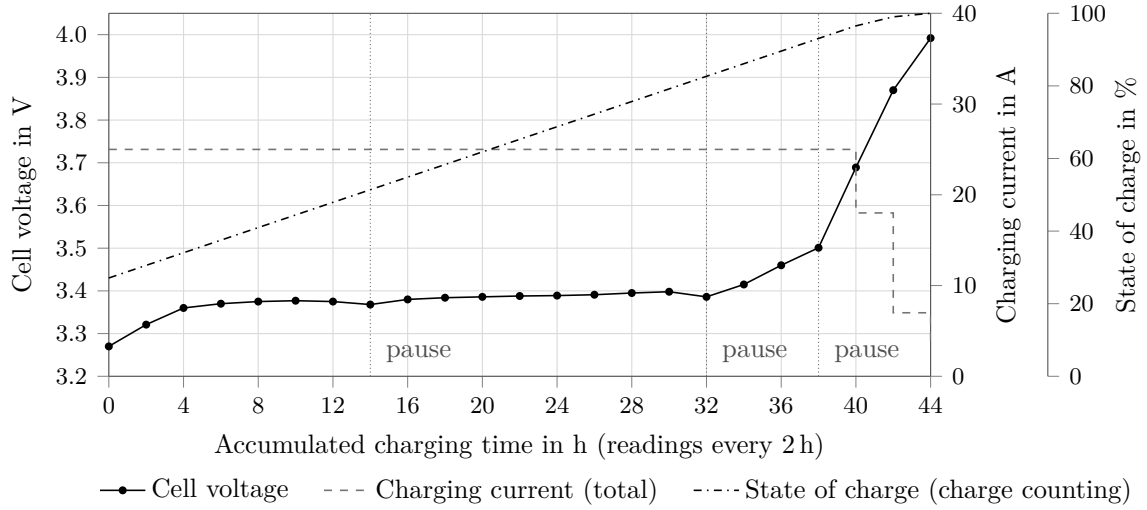
\begin{figure}[!ht]
\centering
\pgfplotsset{chargeaxis/.style={thesisplot, scale only axis, width=0.69\linewidth, height=4.8cm,
  xmin=0, xmax=44, xtick={0,4,...,44}}}
\begin{tikzpicture}
  \begin{axis}[chargeaxis, ymin=3.2, ymax=4.05, ytick={3.2,3.3,...,4.0},
      yticklabel style={/pgf/number format/.cd, fixed, fixed zerofill, precision=1},
      xlabel={Accumulated charging time in \unit{\hour} (readings every \qty{2}{\hour})},
      ylabel={Cell voltage in \unit{\volt}}, axis y line*=left,
      legend cell align=left, legend columns=3, legend style={draw=none, fill=none,
        font=\small, at={(0.62,-0.25)}, anchor=north, /tikz/every even column/.append style={column sep=1em}}]
    \foreach \x/\t in {14/{overnight\\ pause}, 32/pause, 38/pause}
      \edef\temp{\noexpand\draw[black!60, densely dotted] (axis cs:\x,3.2) -- (axis cs:\x,4.05);}\temp;
    \node[font=\small, anchor=south west, black!70] at (axis cs:14.3,3.21) {pause};
    \node[font=\small, anchor=south west, black!70] at (axis cs:32.3,3.21) {pause};
    \node[font=\small, anchor=south west, black!70] at (axis cs:38.3,3.21) {pause};
    \addplot[black, mark=*, mark size=1.2pt] table[x=t, y=u] {figures/data/charge_curve.dat};
    \addlegendentry{Cell voltage}
    \addlegendimage{black!55, dashed}
    \addlegendentry{Charging current (total)}
    \addlegendimage{black, dashdotted}
    \addlegendentry{State of charge (charge counting)}
  \end{axis}
  \begin{axis}[chargeaxis, ymin=0, ymax=40, axis y line*=right, axis x line=none, grid=none,
      ytick pos=right, yticklabel pos=right, ylabel near ticks, ylabel={Charging current in \unit{\ampere}}]
    \addplot[black!55, dashed, const plot] table[x=t, y=i] {figures/data/charge_curve.dat};
  \end{axis}
  \begin{axis}[chargeaxis, ymin=0, ymax=100, axis y line*=right, axis x line=none, grid=none, clip=false,
      ytick pos=right, yticklabel pos=right,
      every outer y axis line/.append style={xshift=1.55cm}, every y tick/.append style={xshift=1.55cm},
      every y tick label/.append style={xshift=1.55cm}]
    \addplot[black, dashdotted] table[x=t, y=soc] {figures/data/charge_curve.dat};
    \node[rotate=90, anchor=north, font=\small] at ([xshift=2.3cm]rel axis cs:1,0.5) {State of charge in \unit{\percent}};
  \end{axis}
\end{tikzpicture}
\caption{Measured cell voltage and total charging current during initial charging of the 24 parallel-connected cells, read every \qty{2}{\hour} over four days, with the state of charge from charge counting; the dotted lines mark the overnight pauses\nocc}\label{fig:charge}
\end{figure}

The initialisation process, also called initial charging, is necessary to bring all cells of a system to a common state of charge before first use. For the cell type used (\qty{60}{\ampere\hour}), the manufacturer specifies an initial charging current of 0.5\,C, i.e.\ \qty{30}{\ampere} per cell. The end-of-charge voltage for this process is \qty{4.0}{\volt}. For initial charging, all 24 cells are connected in parallel and charged together. After reaching the end-of-charge voltage, the cells should remain connected in parallel for a further 24 hours without the charger to reach a common voltage level. The charger available for this purpose delivered a total current of \qty{25}{\ampere}, i.e.\ approx.\ \qty{1}{\ampere} per cell. Charging was carried out on four consecutive days and only during the day, so that it could be supervised at all times; the pack rested overnight. During initialisation, the voltage was read every \qty{2}{\hour}, and the charging curve was derived from these readings (Figure~\ref{fig:charge}). The charging current was read from the display of the laboratory power supply and noted by hand. After \qty{44}{\hour} of charging, the current had decreased to \qty{4}{\ampere} at the end-of-charge voltage of \qty{4.0}{\volt}. In total, approx.\ \qty{1050}{\ampere\hour} were charged, i.e.\ approx.\ \qty{44}{\ampere\hour} per cell (Section~\ref{sec:battery-behaviour-during-initial-chargin}).

The small dips in Figure~\ref{fig:charge} after \qty{14}{\hour} and \qty{32}{\hour} of charging are not measurement errors. They are the points where charging was resumed after an overnight pause, and the cell voltage had relaxed towards its open-circuit value during the rest. The third pause, after \qty{38}{\hour}, falls into the steep rise at the end of charging, where the effect is masked. The measured voltages follow the typical course of a \LFP{} charging curve, as sketched schematically in the BMS documentation~\cite{BMS15}. The measurement also shows how difficult it is to determine the state of charge from the terminal voltage. The cell voltage remains almost constant over a wide range, so that small differences in terminal voltage correspond to large changes in the state of charge. The state of charge can therefore only be determined exactly by a current–time measurement. In this project, this task is performed by the battery management system.

\FloatBarrier%
\subsection{Battery Management System}\label{sec:battery-management-system}
\begin{wrapfigure}[19]{r}{0.52\textwidth}%
\wraptop
\centering
\begin{tikzpicture}[schem, y=0.74cm,
  cell/.style={block, text width=1.7cm, align=center, text height=1.7ex, text depth=0.5ex,
    minimum height=0.65cm, inner ysep=1pt, inner xsep=2.5pt},
  module/.style={block, text width=2.05cm, align=center, text height=1.7ex, text depth=0.5ex,
    minimum height=0.65cm, inner ysep=1pt, inner xsep=2.5pt, font=\small},
  bus/.style={densely dashed}]
  \draw[line width=1.6pt] (0.8,6.3) -- (0.8,3.6) (0.8,2.5) -- (0.8,-1.3);
  \draw[part, fill=white] (0.8,-0.5) circle[radius=0.28cm];
  \node[above right, inner sep=1.5pt, xshift=-0.25cm] at (0.8,6.3) {$+$ to drive / charger};
  \node[below, font=\normalsize] at (0.8,-1.3) {$-$};
  \foreach \i/\y/\c/\m in {1/5.2/Cell 1/IN module, 2/4.2/Cell 2/Cell module, 3/1.6/Cell $n-1$/Cell module, 4/0.6/Cell $n$/OUT module} {
    \node[cell, fill=white] (c\i) at (0.8,\y) {\c};
    \node[module] (m\i) at (3.4,\y) {\m};
    \draw (c\i.east) -- (m\i.west);}
  \foreach \x in {0.8,3.4} \foreach \y in {3.35,3.05,2.75} \fill (\x,\y) circle (0.45mm);
  \draw[bus] (m1.south) -- (m2.north) (m2.south) -- (3.4,3.6) (3.4,2.5) -- (m3.north) (m3.south) -- (m4.north);
  \draw[bus] (m1.east) -| (4.7,3.4) -- (4.85,3.4);
  \draw[bus] (m4.east) -| (4.7,2.6) -- (4.85,2.6);
  \node[font=\small] at (4.0,3.05) {Bus};
  \node[block, minimum width=2.4cm, minimum height=1.6cm] at (6.05,3.0) {BMS main\\ controller};
  \draw[->] (5.1,3.8) -- (5.1,6.85) node[above, inner sep=2pt] {USB (PC software)};
  \draw[->] (5.95,3.8) -- (5.95,4.5) node[above, align=center, inner sep=2pt] {Enable to\\ charger};
  \draw[->] (7.05,3.8) -- (7.05,5.75) node[above, align=center, inner sep=2pt] {Relays\\ Ry2 / Ry3};
  \draw[->] (6.05,2.2) -- (6.05,1.3) node[below, align=center, inner sep=2pt] {State-of-charge\\ gauge};
  \draw[signal] (1.08,-0.5) -- (7.85,-0.5) -- (7.85,2.6) -- (7.25,2.6);
  \node[below right, inner sep=2pt] at (1.15,-0.5) {Current sensor};
\end{tikzpicture}
\caption{Structure of the battery management system (schematic)\nocc}\label{fig:bms}
\end{wrapfigure}
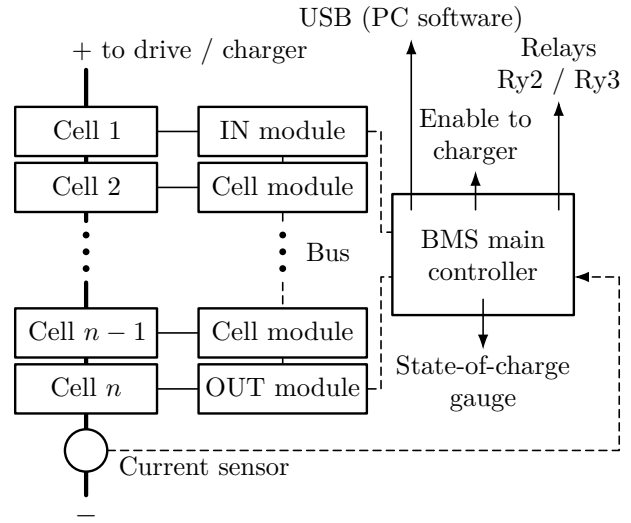
The 123\textbackslash BMS is designed for many cells connected in series. It is primarily optimised for \LFP{} cells, but can also be used with similar types such as Li-ion or LiPo. The system monitors the state of charge, the voltage and the charging and discharging process. Each cell is equipped with a cell module that monitors its electrical parameters. An IN module is mounted on the first cell and an OUT module on the last cell. If the charge difference between the cells becomes too large, the system equalises the state of charge by balancing. The BMS also monitors the maximum permissible discharge current and protects the cells against overcharging and deep discharge.

Via a USB connection, the battery parameters can be configured and the measured values read out and visualised. The cell modules transmit their data via a bus to the main controller (Figure~\ref{fig:bms}). Communication with the charger takes place via an enable interface. Unlike comparable systems, no CAN interface is required on the charger. Two relay outputs are available for fault signalling: output Ry2 signals a connected charger, and output Ry3 signals a communication problem between the modules or the violation of certain limits. Further details are given in the datasheet~\cite{BMS15}, the main data in Table~\ref{tab:bms} (Appendix~\ref{app:technical-data}).

\section{Electric Drive}\label{sec:electric-drive}

Electric drives are often dimensioned solely for the required acceleration and top speed. Due to the required feedback of braking energy, the maximum braking torque and the resulting deceleration are also of interest. As a first approximation, the rated torque of the motor can be assumed as the maximum braking torque. First, the motor data are calculated for the required top speed of approx.\ \qty{60}{\kilo\metre\per\hour} with sufficiently good acceleration.

At the beginning, there were doubts as to whether the selected motor by Golden Motor could meet the quality and performance requirements. Experience from the previous year's electric kart and extensive research, however, spoke in favour of the product. The only disadvantages are the sparse technical information on the product website and the often difficult communication with Golden Motor. Import into Europe is also difficult. The motor was therefore purchased from a German distributor.

\subsection{Calculation of the Required Motor Power}\label{sec:calculation-of-the-required-motor-power}
Due to its favourable performance data and low weight, a brushless DC motor of the type HPM-10kW by Golden Motor~\cite{GM15} was selected (Figure~\ref{fig:drive}b). The motor controller is the model VEC500, also by Golden Motor. Table~\ref{tab:motor} in Appendix~\ref{app:technical-data} lists the technical data of the motor according to the manufacturer.

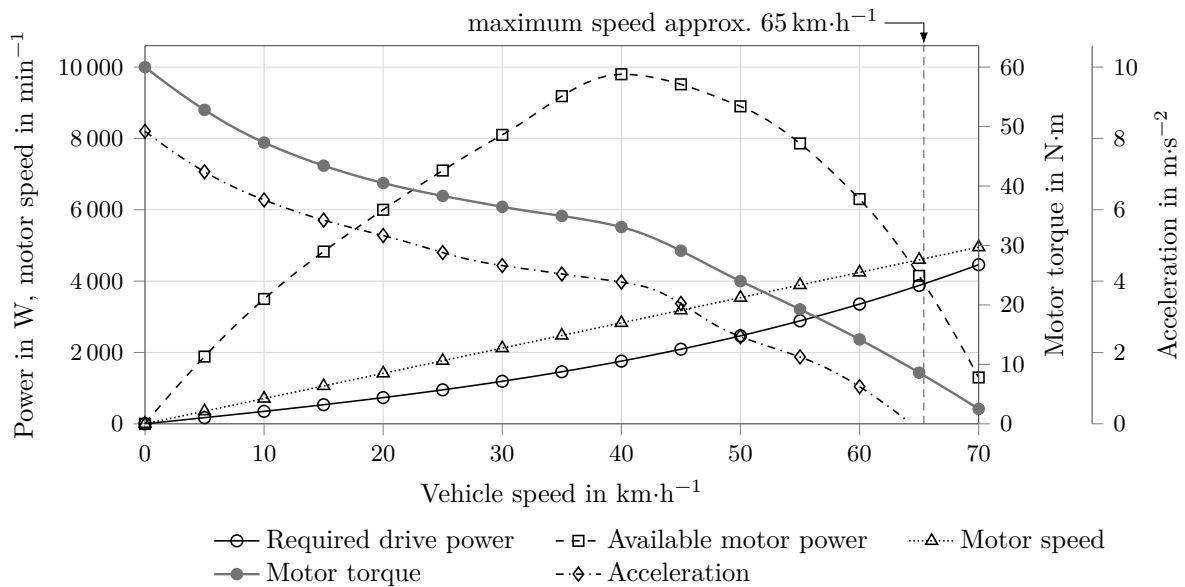
\begin{figure}[!ht]
\centering
\pgfplotsset{
  driveaxis/.style={width=0.7\linewidth, height=5.0cm, scale only axis, xmin=0, xmax=70},
  every axis plot/.append style={mark repeat=5},
  c-preq/.style={black, solid, mark=o, mark size=2.2pt},
  c-pmot/.style={black, dashed, mark=square, mark size=2pt},
  c-nmot/.style={black, densely dotted, mark=triangle, mark size=2.6pt},
  c-acc/.style={black, dashdotted, mark=diamond, mark size=2.6pt},
  c-torque/.style={black!55, solid, line width=0.9pt, mark=*, mark size=1.9pt, mark options={solid, fill=black!55}},
  c-preq/.append style={mark options={solid}}, c-pmot/.append style={mark options={solid}},
  c-nmot/.append style={mark options={solid}}, c-acc/.append style={mark options={solid}},
}
\begin{tikzpicture}
  \begin{axis}[thesisplot, driveaxis, ymin=0, ymax=10600, xtick={0,10,...,70}, ytick={0,2000,...,10000},
      restrict x to domain=0:70,
      xlabel={Vehicle speed in \unit{\kilo\metre\per\hour}},
      ylabel={Power in \unit{\watt}, motor speed in \unit{\per\minute}},
      scaled y ticks=false, yticklabel style={/pgf/number format/.cd, 1000 sep={\,}},
      axis y line*=left, legend cell align=left,
      legend columns=3, legend style={draw=none, fill=none, font=\small, at={(0.62,-0.25)}, anchor=north,
        /tikz/every even column/.append style={column sep=1.2em}}]
    \addplot[c-preq] table[x=v, y=preq] {figures/data/drive_power.dat};
    \addlegendentry{Required drive power}
    \addplot[c-pmot] table[x=v, y=pmot] {figures/data/motor_power.dat};
    \addlegendentry{Available motor power}
    \addplot[c-nmot] table[x=v, y=nmot] {figures/data/drive_power.dat};
    \addlegendentry{Motor speed}
    \addlegendimage{c-torque}
    \addlegendentry{Motor torque}
    \addlegendimage{c-acc}
    \addlegendentry{Acceleration}
    \draw[ext, black!70] (axis cs:65.4,0) -- (axis cs:65.4,10600);
  \end{axis}
  \begin{axis}[driveaxis, ymin=0, ymax=10600, hide axis, clip=false]
    \node[font=\small, anchor=east, inner sep=2pt] (vmax) at ([yshift=3mm]axis cs:62.1,10600)
      {maximum speed approx.\ \qty{65}{\kilo\metre\per\hour}};
    \draw[line width=0.4pt, -{Latex[length=1.5mm,width=1.1mm]}] (vmax.east) -| (axis cs:65.4,10650);
  \end{axis}
  \begin{axis}[thesisplot, driveaxis, ymin=0, ymax=63.6, ytick={0,10,...,60}, axis y line*=right,
      axis x line=none, grid=none, ytick pos=right, yticklabel pos=right, ylabel near ticks,
      ylabel={Motor torque in \unit{\newton\metre}}]
    \addplot[c-torque] table[x=v, y=T] {figures/data/motor_torque.dat};
  \end{axis}
  \begin{axis}[thesisplot, driveaxis, ymin=0, ymax=10.6, ytick={0,2,...,10}, axis y line*=right,
      axis x line=none, grid=none, ytick pos=right, yticklabel pos=right, clip=false,
      every outer y axis line/.append style={xshift=1.5cm}, every y tick/.append style={xshift=1.5cm},
      every y tick label/.append style={xshift=1.5cm}]
    \addplot[c-acc] table[x=v, y=a] {figures/data/acceleration.dat};
    \node[rotate=90, anchor=north, font=\small] at ([xshift=2.2cm]rel axis cs:1,0.5)
      {Acceleration in \unit{\metre\per\second\squared}};
  \end{axis}
\end{tikzpicture}
\caption{Required drive power, available motor power, motor speed, motor torque and calculated acceleration as a function of vehicle speed\nocc}\label{fig:drivepower}
\end{figure}

The driving resistance was calculated from rolling and air resistance~\cite{Pis16,Sch13} using the vehicle data in Table~\ref{tab:vehdata} (Appendix~\ref{app:technical-data}). From this, the required drive power follows as a function of the vehicle speed. Figure~\ref{fig:drivepower} compares it with the available motor power.

The required and the available power intersect at approx.\ \qty{65}{\kilo\metre\per\hour}. The calculated top speed is therefore approx.\ \qty{65}{\kilo\metre\per\hour}, and the target of approx.\ \qty{60}{\kilo\metre\per\hour} is reached. The acceleration was calculated from the maximum wheel force and the drive force for a gear ratio of 3.2 (Figure~\ref{fig:drivepower}).

The maximum acceleration at standstill is \qty{8.2}{\metre\per\second\squared}. Note that this value assumes that the full wheel force can be transmitted; in practice, it is limited by the tyre grip and by the traction control. In addition, the calculation uses a rated speed of \qty{4000}{\per\minute} and a gear ratio of 3.2, while the final design uses a rated speed of \qty{4500}{\per\minute} (manufacturer data) and a ratio of 1:3.3. The results are therefore to be understood as an estimate.

\subsection{Technical Data of the Motor Controller}\label{sec:technical-data-of-the-motor-controller}
Selecting the motor controller proved particularly difficult due to the requirements. Special attention was paid to the efficiency and to support for regenerative braking. The search for a BLDC motor controller with the desired properties was initially unsuccessful. Golden Motor, however, offers a controller matched to the motor that also supports regenerative braking. Like the motor, the controller is manufactured by Golden Motor and was also purchased from the German distributor Smartemotions. The controller operates the motor with field-oriented control, also known as vector control (Table~\ref{tab:controller} in Appendix~\ref{app:technical-data}).

\section{Mechanical Drive}\label{sec:mechanical-drive}
Power is transmitted from the motor shaft to the rear axle via two sprockets and a chain. A chain was chosen, as is usual in kart racing. Since the kart is only driven on paved surfaces, little dirt is expected. The gear ratio of 1:3.3 reduces the speed, while the torque acting on the drive axle increases accordingly. With the selected motor, a starting torque that is too low is not to be expected. An excessive starting torque can be limited by the microcontroller.

\subsection{Dimensioning of the Rear Axle}\label{sec:dimensioning-of-the-rear-axle}
\begin{figure}[tbp]
\centering
\includegraphics[width=0.8\textwidth]{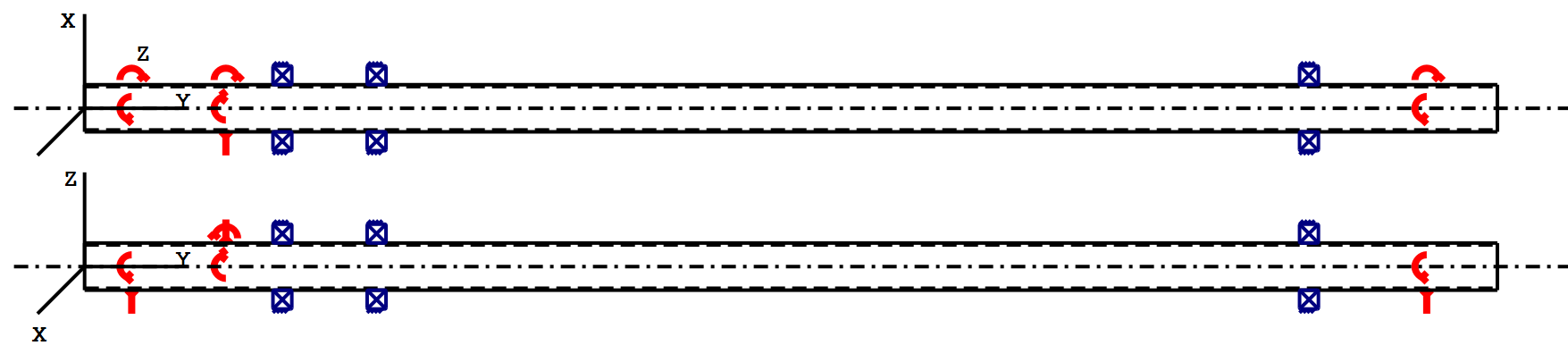}
\caption{Calculation model of the rear axle in KISSsoft with bearings and load introduction (software output)}\label{fig:image39}
\end{figure}

Kartshop Ampfing advised paying particular attention to the load on the rear axle caused by the motor torque. In racing, axles with a diameter of \qty{50}{\milli\metre} are at their load limit at a motor torque of \qty{100}{\newton\metre}. To ensure safety, the axle was recalculated with the calculation software KISSsoft~\cite{KIS15}. The calculation model with bearings and load introduction is shown in Figure~\ref{fig:image39}, the input data and main results are summarised in Table~\ref{tab:axle}.

\begin{table}[tbp]
\centering
\caption{Input data and results of the rear axle calculation with KISSsoft}\label{tab:axle}
{\renewcommand{\arraystretch}{1.15}%
\begin{tabular}{@{}lc@{}}
\toprule
\textbf{Parameter} & \textbf{Value} \\
\midrule
Material & C45 \\
Model length in \unit{\milli\metre} & 1500 \\
Outer / inner diameter in \unit{\milli\metre} & 50 / 46 (tube) \\
Speed in \unit{\per\minute} & 1000 \\
Torque at the sprocket in \unit{\newton\metre} & 330 ($\qty{100}{\newton\metre}\times3.3$) \\
Torque per wheel in \unit{\newton\metre} & 165 \\
Bearings & 3 × deep-groove ball bearing 50 × \qty{90}{\milli\metre} \\
Maximum deflection in \unit{\milli\metre} & 0.40 \\
Torsion under torque in \unit{\degree} & 0.81 \\
Max. equivalent stress (von Mises) in \unit{\newton\per\square\milli\metre} & approx.\ 110 \\
Static safety factor of bearings S0 (B1 / B2 / B3) & 4.22 / 9.85 / 33.16 \\
Nominal bearing life (B1 / B2 / B3) in \unit{\hour} & 4290 / \num{54460} / > \num{1000000} \\
\bottomrule
\end{tabular}}
\end{table}

\begin{figure}[tbp]
\pgfplotsset{axleplot/.style={thesisplot, scale only axis, width=0.66\linewidth, height=4.4cm, xmin=0, xmax=1500,
  ylabel style={align=center}, yticklabel style={text width=1.8em, align=right},
  xticklabel style={/pgf/number format/1000 sep={\,}},
  xtick={0,300,...,1500}, xlabel={Longitudinal axis $y$ in \unit{\milli\metre}},
  legend cell align=left, legend style={draw=none, fill=none, font=\small, at={(0.5,-0.3)}, anchor=north}}}
\begin{minipage}[t]{0.49\textwidth}
\textbf{a)}\par\medskip
\begin{tikzpicture}
  \begin{axis}[axleplot, ymin=-0.3, ymax=0.42, ytick={-0.2,-0.1,...,0.4},
      yticklabel style={/pgf/number format/.cd, fixed, fixed zerofill, precision=1},
      ylabel={Displacement\\ in \unit{\milli\metre}}]
    \addplot[black] table[x=y, y=w] {figures/data/axle_deflection.dat};
    \addlegendentry{Component in an arbitrary plane}
    \addplot[black!55, dashed, domain=0:1500, samples=2] {0};
    \addlegendentry{$y$ component (axial)}
  \end{axis}
\end{tikzpicture}
\end{minipage}\hfill
\begin{minipage}[t]{0.49\textwidth}
\textbf{b)}\par\medskip
\begin{tikzpicture}
  \begin{axis}[axleplot, ymin=0, ymax=125, ytick={0,20,...,120},
      ylabel={Equivalent stress\\ in \unit{\newton\per\square\milli\metre}}]
    \addplot[black] table[x=y, y=s] {figures/data/axle_stress_geh.dat};
    \addlegendentry{von Mises (distortion energy)}
    \addplot[black!55, dashed] table[x=y, y=s] {figures/data/axle_stress_ssh.dat};
    \addlegendentry{Tresca (maximum shear stress)}
  \end{axis}
\end{tikzpicture}
\end{minipage}
\caption{Results of the rear axle calculation with KISSsoft: a) calculated deflection, b) calculated equivalent stress along the rear axle}\label{fig:stress}
\end{figure}
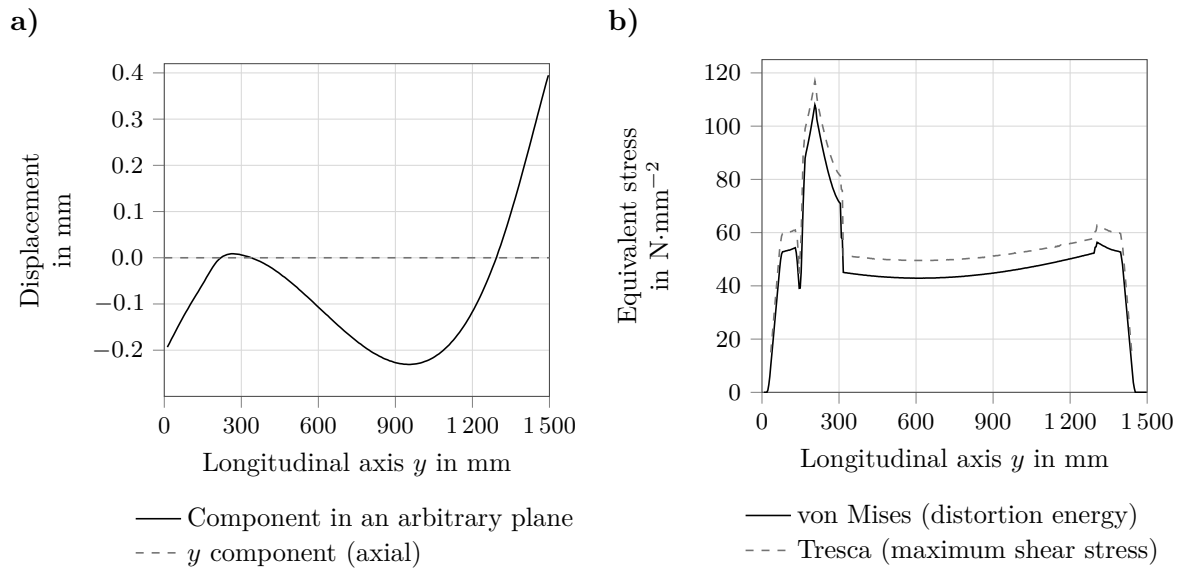

The maximum deflection is \qty{0.40}{\milli\metre}, and the torsion under load is \qty{0.81}{\degree}. The maximum equivalent stress (von Mises) of approx.\ \qty{110}{\newton\per\square\milli\metre} occurs at the load introduction of the sprocket (Figure~\ref{fig:stress}). The static safety against yielding is approx.\ 2.8 (Section~\ref{sec:mechanical-reserves-of-the-drivetrain}). For the intended use of the kart, this margin is sufficient. For racing, a fatigue assessment with measured load spectra would still be required, and a reinforced axle may be necessary. Note that the calculated nominal life of bearing B1 (\qty{4290}{\hour}) is below the target of \qty{20000}{\hour}. For continuous racing use, this bearing position would have to be reviewed.

\subsection{Dimensioning of the Chain}\label{sec:dimensioning-of-the-chain}
The BLDC motor HPM-10kW is assumed to deliver a short-term peak torque of approx.\ \qty{100}{\newton\metre} at the motor shaft. The chain has to be selected according to the resulting tensile load. With a pitch diameter of the motor sprocket of $d=\qty{55}{\milli\metre}$, the chain force follows as:
\[
  F=\frac{M}{d/2}=\frac{\qty{100}{\newton\metre}}{\qty{0.0275}{\metre}}\approx\qty{3640}{\newton}
\]
The tensile load on the chain is approx.\ \qty{3.6}{\kilo\newton}. Note that the torque values used in the individual calculations differ: the drive calculation uses a rated torque of \qty{30}{\newton\metre} and a starting torque of \qty{60}{\newton\metre}, whereas chain and axle were dimensioned conservatively with \qty{100}{\newton\metre}. To achieve the longest possible service life of the drive, a D.I.D 428HD chain was selected. In operation, the pinion on the motor shaft nevertheless failed under the torque steps of the electric motor and had to be replaced together with the chain by parts made of higher-grade materials (Section~\ref{sec:mechanical-reserves-of-the-drivetrain}).

\begin{figure}[!ht]
\centering
\begin{minipage}[t]{0.49\textwidth}
\textbf{a)}\par\medskip
\includegraphics[width=\linewidth]{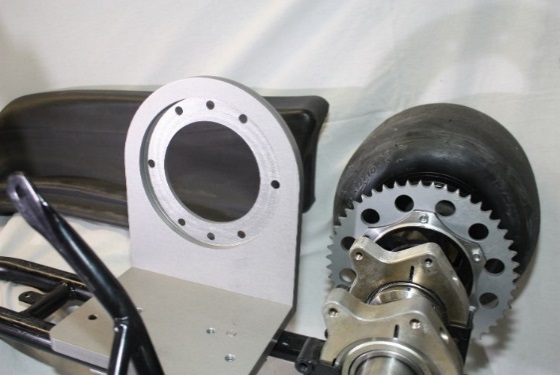}
\end{minipage}\hfill
\begin{minipage}[t]{0.49\textwidth}
\textbf{b)}\par\medskip
\includegraphics[width=\linewidth]{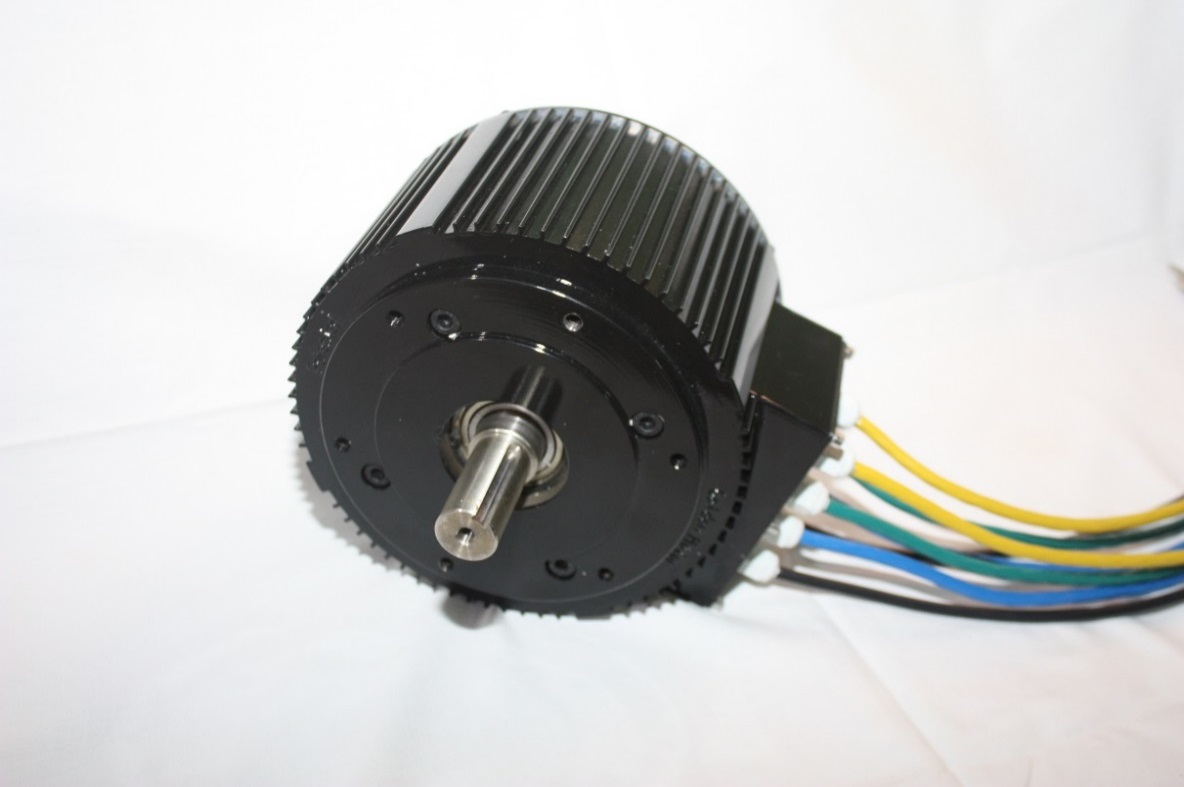}
\end{minipage}
\caption{a) Motor mount and chain drive, b) BLDC motor HPM-10kW}\label{fig:drive}
\end{figure}

\subsection{Dimensioning of the Sprockets}\label{sec:dimensioning-of-the-sprockets}
Weight saving is an important goal to realise a drive that is as energy-efficient as possible. According to practical experience from racing, a steel sprocket is required because of the torque load on the rear sprocket. For weight reasons, however, a steel sprocket was not to be used. After a technical discussion, Kartshop Ampfing finally recommended a sprocket made of a special aluminium alloy (Figure~\ref{fig:drive}a).

\section{Traction Control}\label{sec:traction-control}
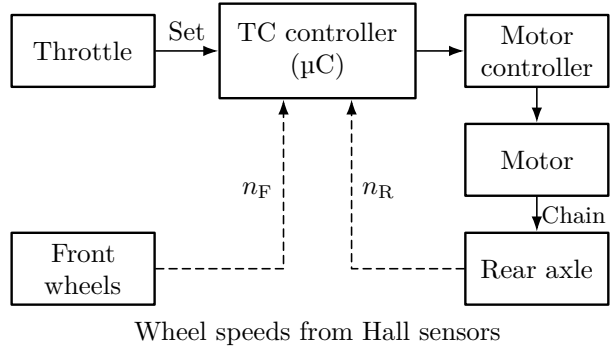
\begin{wrapfigure}[11]{r}{0.55\textwidth}%
\vspace*{\dimexpr-\intextsep-41.8pt+2\baselineskip}%
\centering
\begin{tikzpicture}[schem, node distance=0.8cm, y=0.8cm,
  b/.style={block, text width=1.75cm, minimum height=0.95cm, inner sep=2pt}]
  \node[b] (thr) at (0.55,4.0) {Throttle};
  \node[b, minimum height=1.25cm, text width=2.45cm] (tc) at (3.65,4.0) {TC controller\\ (\textmu C)};
  \node[b] (mc) at (6.55,4.0) {Motor\\ controller};
  \node[b] (mot) at (6.55,2.2) {Motor};
  \node[b] (ra) at (6.55,0.4) {Rear axle};
  \node[b] (fw) at (0.55,0.4) {Front wheels};
  \draw[->] (thr) -- node[above, font=\small] {Set} (tc);
  \draw[->] (tc) -- (mc);
  \draw[->] (mc) -- (mot);
  \draw[->] (mot) -- node[right, font=\footnotesize, inner sep=1.5pt] {Chain} (ra);
  \draw[signal] (fw.east) -- (3.2,0.4) -- (3.2,0.4 |- tc.south);
  \draw[signal] (ra.west) -- (4.1,0.4) -- (4.1,0.4 |- tc.south);
  \node[left] at (3.2,1.7) {$n_{\mathrm{F}}$};
  \node[right] at (4.1,1.7) {$n_{\mathrm{R}}$};
  \node[below, font=\small] at (3.65,-0.3) {Wheel speeds from Hall sensors};
\end{tikzpicture}
\caption{Control loop of the traction control (schematic)\nocc}\label{fig:tcloop}
\end{wrapfigure}

Traction control (TC) prevents the drive wheels from spinning under strong acceleration. Large wheel slip consumes a lot of energy and does not contribute to acceleration. Traction control ensures maximum force transmission in the optimal slip range of the tyres~\cite{Pis16}. Speed sensors at the rear axle and at both front wheels measure the wheel speeds. Based on these signals, a control loop intervenes in the motor management (Figure~\ref{fig:tcloop}). The slip of the drive axle follows from the speed of the rear axle $v_{\mathrm{R}}$ and the speed of a front wheel $v_{\mathrm{F}}$ as $\lambda=(v_{\mathrm{R}}-v_{\mathrm{F}})/v_{\mathrm{R}}$. A protective function switches the system off in the event of a fault.

\subsection{Implementation}\label{sec:implementation}
Processing the wheel speeds and the developed control algorithm require high computing power and low latency. Simple 8-bit microcontrollers are only partly capable of this task. The choice therefore fell on the 32-bit microcontroller SAM3X8E (ARM Cortex-M3) by Atmel, which is used on the Arduino Due platform. It offers sufficient computing power for this application, but requires sound programming skills.

\subsection{Electronic and Mechanical Components}\label{sec:tc-components}
The following components were used to realise the traction control. Their properties were selected with particular care, and it was also important that the parts are robust and resistant to moisture. This is of course not the case for the installed \textmu C, so it was mounted in a box made especially for the electronic components. For all parts mounted outside, moisture resistance was given particular attention.

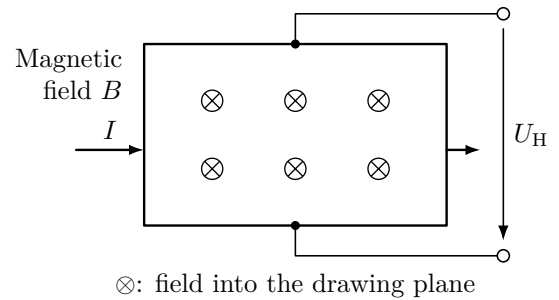
\begin{wrapfigure}[13]{R}{0.45\textwidth}%
\vspace*{\dimexpr-\intextsep+2\baselineskip-15.9pt}%
\centering
\begin{tikzpicture}[schem,
  ccross/.style={circle, draw, line width=0.5pt, minimum size=2.8mm, inner sep=0pt,
    path picture={\draw[line width=0.5pt] (path picture bounding box.south west) -- (path picture bounding box.north east)
      (path picture bounding box.north west) -- (path picture bounding box.south east);}},
  term/.style={circle, draw, line width=0.6pt, fill=white, minimum size=1.6mm, inner sep=0pt}]
  \draw[part] (0,0) rectangle (4.0,2.4);
  \foreach \x in {0.9,2.0,3.1} \foreach \y in {0.75,1.65} \node[ccross] at (\x,\y) {};
  \node[align=right, left] at (-0.15,2.0) {Magnetic\\ field $B$};
  \node[align=center, below] at (2.0,-0.5) {$\otimes$: field into the drawing plane};
  \draw[->, line width=0.9pt] (-0.9,1.0) -- node[above] {$I$} (0,1.0);
  \draw[->, line width=0.9pt] (4.0,1.0) -- (4.45,1.0);
  \fill (2.0,2.4) circle (0.6mm);
  \fill (2.0,0) circle (0.6mm);
  \draw (2.0,2.4) -- (2.0,2.8) -- (4.75,2.8);
  \draw (2.0,0) -- (2.0,-0.4) -- (4.75,-0.4);
  \node[term] at (4.75,2.8) {};
  \node[term] at (4.75,-0.4) {};
  \draw[->] (4.75,2.6) -- node[right] {$U_{\mathrm{H}}$} (4.75,-0.2);
\end{tikzpicture}
\caption{Hall effect (schematic)\nocc}\label{fig:hall}
\end{wrapfigure}
\textbf{Hall sensor:} The Hall sensor uses the Hall effect to measure magnetic fields. If a current-carrying conductor is penetrated by a magnetic field, the Lorentz force deflects the charge carriers. As a result, a voltage (Hall voltage) arises perpendicular to both the direction of current and the magnetic field (Figure~\ref{fig:hall}).

\begin{figure}[!ht]
\begin{minipage}[t]{0.49\textwidth}
\textbf{a)}\par\medskip
\includegraphics[width=\linewidth]{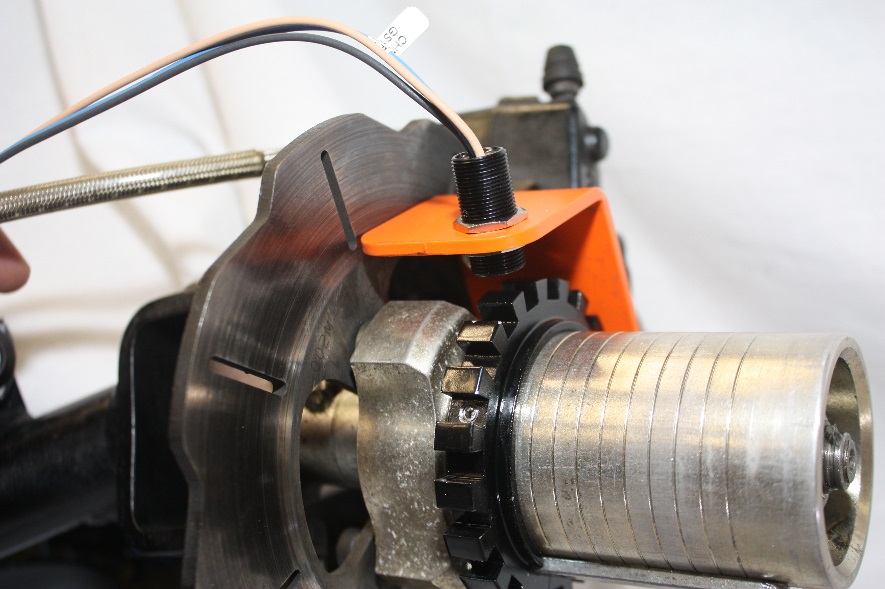}
\end{minipage}\hfill
\begin{minipage}[t]{0.49\textwidth}
\textbf{b)}\par\medskip
\centering
\begin{tikzpicture}[schem, scale=1.45]%
  \draw[line width=1.6pt] (55:3.0) ++(0,-2.7) arc[start angle=55, end angle=125, radius=3.0];
  \foreach \a in {72,90,108}
    \draw[part, fill=white, rotate around={\a-90:(0,-2.7)}] (-0.17,0.32) -- (-0.17,0.0) -- (0.17,0.0) -- (0.17,0.32);
  \draw[part] (-0.3,0.55) rectangle (0.3,1.85);
  \node[lbl, right] at (0.4,1.15) {Hall\\ sensor};
  \draw[->] (0.3,1.75) -- (0.7,1.97);
  \draw (0.8,1.9) -- (1.1,1.9) -- (1.1,2.35) -- (1.45,2.35) -- (1.45,1.9) -- (1.8,1.9) -- (1.8,2.35) -- (2.15,2.35) -- (2.15,1.9) -- (2.45,1.9);
  \node[align=center, below] at (0,-0.75) {Sensor ring with recesses};
\end{tikzpicture}
\end{minipage}
\caption{a) Speed measurement at the front axle with sensor ring and Hall sensor, b) principle of the speed measurement with Hall sensor and sensor ring (schematic)\noccpart{b)}}\label{fig:speed}
\end{figure}

Three Hall sensors are mounted on the kart as speed sensors. They measure the speed of the rear axle and of the two front wheels. The Cherry GS100701 sensors stand out for their robust aluminium housing and high accuracy. Since the sensor already contains a magnet, no additional magnetic field is required. Together with the self-made sensor rings and the \textmu C, the sensor can thus measure the wheel speed (Figure~\ref{fig:speed}).

\textbf{Microcontroller:} The \textmu C is the heart of the traction control. Since all three Hall sensors have to be evaluated very quickly, a high clock frequency is required. The SAM3X8E runs at \qty{84}{\mega\hertz} and has 512 KB of flash memory. This is large enough to store all data as well as the program. The flash memory keeps its content without power and needs little energy.

Since the \textmu C does more than just control the traction, it needs enough inputs and outputs (I/O pins). The selected \textmu C offers 54 digital I/O pins, 12 analogue inputs and 2 analogue outputs (DAC). The permissible total current of all I/O pins is \qty{130}{\milli\ampere}, which is sufficient for the connected signal lines.

\textbf{Mechanical components:} Some mechanical parts were developed and manufactured in the HTL workshops; others were produced by an external company.

\begin{wrapfigure}[10]{r}{0.4\textwidth}%
\wraptop
\centering
\includegraphics[width=0.4\textwidth]{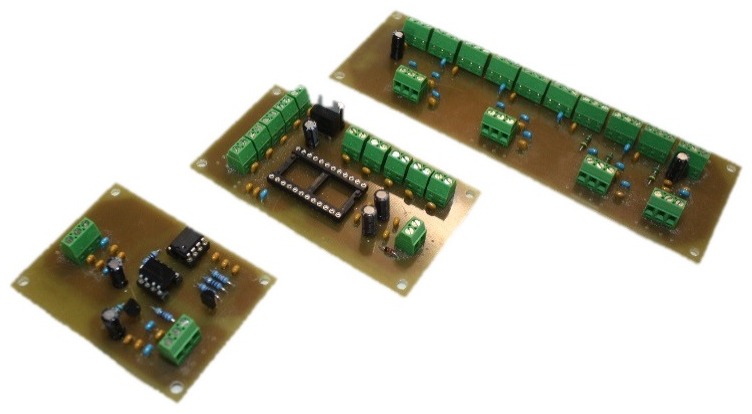}
\caption{Circuit boards of the traction control}\label{fig:tchw}
\end{wrapfigure}
\textbf{Sensor rings:} The sensor rings were developed especially for the kart (Figure~\ref{fig:speed}a). They are located at the positions where the Hall sensors are mounted. Each ring has several identical recesses that are needed for the measurement. The Hall sensor detects every single recess and passes the signal on to the \textmu C. From the number of pulses per unit of time and the known number of recesses, the speed can be calculated. So that the rings can be positioned exactly below the sensors, they can be moved along the axles and are fixed with four grub screws each after positioning. A ring shifting during the drive would lead to incorrect measurements and thus to a malfunction of the traction control.

\textbf{Sensor holders:} The sensor holders are a simple construction that positions the Hall sensors exactly above the sensor rings. The two holders at the front wheels are screwed to the brake mount. In this way, they always stay in the intended position above the sensor rings, even when steering. At the rear axle, the holder is screwed to one of the axle bearing brackets, so that here, too, the sensor cannot be moved out of position.

\textbf{Circuit boards:} The hardware of the traction control must meet special requirements for use in the kart. First, the electronics must be mechanically robust. Second, electromagnetic interference from the motor must not cause malfunctions. The traction control intervenes in the signal path between throttle pedal and motor controller and reduces the motor torque depending on the slip. A fault in the circuit could send wrong information to the motor controller. In the worst case, this could result in an unintended full-throttle position. Since a defect cannot be ruled out even with interference-proof hardware, an emergency stop switch was mounted within reach of the driver.

To ensure electromagnetic compatibility, only shielded cables were used, and ceramic suppression capacitors were provided at every I/O of the circuit boards. Each board has ground planes as large as possible. Vehicle ground and battery voltage are galvanically isolated by optocouplers (Figure~\ref{fig:tchw}). The block diagrams and circuit diagrams are given in Appendix~\ref{app:block-diagrams-and-circuit-diagrams}. Schematics, bills of material and the firmware are publicly available in the project repository~\cite{Sto26}.

\needspace{22\baselineskip}%
\subsection{Software}\label{sec:software}
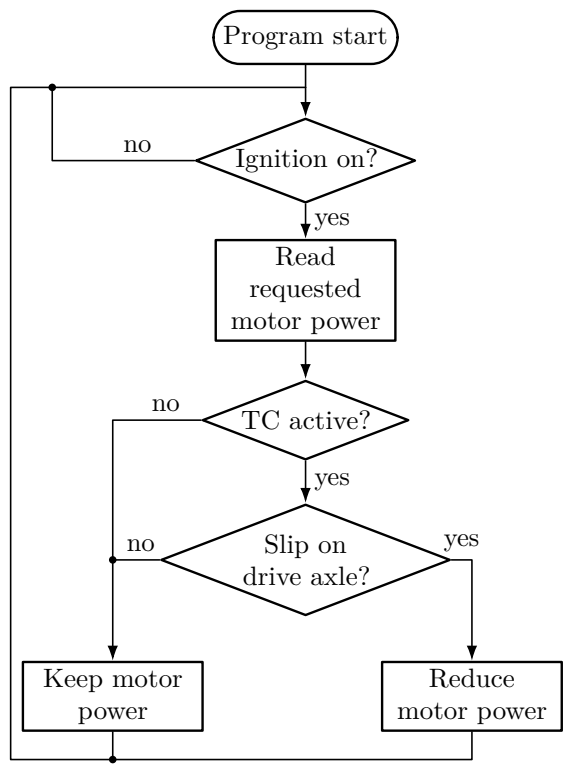
\begin{wrapfigure}[23]{l}{0.5\textwidth}%
\wraptop
\centering
\begin{tikzpicture}[schem, y=0.88cm,
  proc/.style={block, text width=2.2cm, minimum height=0.85cm, inner sep=2.5pt},
  dec/.style={draw, part, diamond, aspect=2.6, text width=2.1cm, align=center, inner sep=0pt},
  term/.style={draw, part, rounded corners=3.5mm, minimum width=2.4cm, minimum height=0.7cm},
  yn/.style={inner sep=3pt, font=\small}]
  \node[term] (start) at (3.4,11.6) {Program start};
  \node[dec] (ign) at (3.4,9.75) {Ignition on?};
  \node[proc] (read) at (3.4,7.8) {Read requested motor power};
  \node[dec] (tc) at (3.4,5.85) {TC active?};
  \node[dec] (slip) at (3.4,3.75) {Slip on drive axle?};
  \node[proc] (keep) at (0.85,1.7) {Keep motor power};
  \node[proc] (red) at (5.6,1.7) {Reduce motor power};
  \draw (start) -- (3.4,10.85);
  \draw[->] (3.4,10.85) -- (ign);
  \draw[->] (ign) -- node[right, yn] {yes} (read);
  \draw[->] (read) -- (tc);
  \draw[->] (tc) -- node[right, yn] {yes} (slip);
  \draw (ign.west) -- node[above, yn, pos=0.4] {no} (0.05,9.75) -- (0.05,10.85) -- (3.4,10.85);
  \draw[->] (tc.west) -- node[above, yn, pos=0.4] {no} (0.85,5.85) -- (keep.north);
  \draw (slip.west) -- node[above, yn, pos=0.4] {no} (0.85,3.75);
  \fill (0.85,3.75) circle (0.5mm);
  \draw[->] (slip.east) -- node[above, yn] {yes} (5.6,3.75) -- (red.north);
  \draw (keep.south) -- (0.85,0.75);
  \draw (red.south) -- (5.6,0.75) -- (-0.5,0.75) -- (-0.5,10.85) -- (0.05,10.85);
  \fill (0.85,0.75) circle (0.5mm);
  \fill (0.05,10.85) circle (0.5mm);
\end{tikzpicture}
\caption{Flowchart of the traction control program\nocc}\label{fig:flow}
\end{wrapfigure}
The program code, i.e.\ the control algorithm, can best be described with a flowchart (Figure~\ref{fig:flow}). Such a chart is often drawn before a program is written. It helps not only to explain the program code, but also during programming itself. In a way, the chart runs through the program like a common thread.

The structure of the chart is comparable to the behaviour of a state machine. With such a program structure, the program, viewed from the outside, always runs through the same cycle. A certain state is queried and, after evaluation, a suitable action is taken. The program runs in an endless loop and repeats itself continuously. The chart shows, for example, that the program starts again and queries the ignition once more if the ignition is switched off.

The individual steps are explained in more detail below:

\textbf{Step 1:} After the program start, the program checks whether the ignition is switched on. If it is switched off, the query is repeated. If it is switched on, the program continues.

\textbf{Step 2:} In the next step, the currently requested motor power is read and stored. The \textmu C performs no further queries in this state and then executes the next part of the program.

\textbf{Step 3:} In the third step, the \textmu C checks whether the traction control is already active, i.e.\ whether the corresponding variable is set, and whether it is to be activated or deactivated. For this purpose, the values relevant to the traction control are calculated. If the traction control is deactivated, the previously read motor power is kept unchanged. If the traction control is to be activated, the \textmu C switches to the next state.

\textbf{Step 4:} In this, probably the most important state, the slip of the drive axle is calculated. The program then decides whether the motor power has to be reduced or, if the axle is not slipping, can be kept.

\section{Braking System}\label{sec:braking-system}
The braking system is the most important safety device on the kart. In operation, it must always allow the kart to stop safely. In addition, the braking energy is to be converted back into electrical energy and fed into the battery.

\subsection{Concept}\label{sec:concept}
The braking concept comprises not only the race-proven hydraulic brake, but also a regenerative brake.

In regenerative braking, the kinetic energy of the kart is converted into electrical energy. For this purpose, the electric drive motor is operated as a generator. In this mode, it absorbs the mechanical power of the drive axle and converts it into electrical energy, which is fed back into the battery. The special feature of the regenerative brake is that it works without wear and that energy is recovered with good efficiency.

In this way, the kart can be braked mostly with the motor brake, and the mechanical brake only has to be used for hard braking. A similar system is used in the automotive sector. In electric and hybrid vehicles with regenerative braking, the driver usually does not notice it, since it is applied automatically when the brake pedal is pressed. For light braking, only the regenerative brake is used. For harder braking, the hydraulic brake is applied in addition. This coupling was deliberately not implemented in the kart. The brakes are operated separately in order to demonstrate the effect of the regenerative brake more clearly.

\subsection{Regenerative Braking}\label{sec:recuperation}
In a technical sense, recuperation simply means energy recovery. The regenerative brake was first used in rail vehicles, where the traction motors were switched to generator operation for braking. In the early days of this technology, only about \qty{5}{\percent} of the energy used could be recovered, and the braking force was weak and uneven. Later, considerably higher recovery rates of up to \qty{50}{\percent} with reliable braking force were achieved with asynchronous motors. The technology was continuously developed further. In the automotive sector, brake energy regeneration was introduced after 2000, for example under the name ``EfficientDynamics'' (BMW). Today, it is used above all in electric and hybrid vehicles.

\subsection{Implementation}\label{sec:braking-implementation}
The regenerative brake of the kart works as described above and is divided into three recuperation levels:

\textbf{Stage 1 -- motor brake:} The first recuperation stage starts as soon as the throttle is released. It brakes only slightly, comparable to the engine brake of a vehicle with a combustion engine. Despite the weak braking effect, energy is already recovered.

\textbf{Stage 2 -- light braking:} In the second stage, the kart is actively decelerated by the regenerative brake. This stage is activated by pressing a button on the steering wheel. It is well suited for gentle deceleration, for example before a slight bend.

\textbf{Stage 3 -- hard braking:} In the third stage, the hydraulic brake is used in addition to the regenerative brake. This achieves the maximum deceleration of the kart. With the hydraulic brake, however, braking is no longer wear-free, and only a small part of the energy is recovered. Most of the kinetic energy is converted into heat in the hydraulic brakes.

The aim of the different stages is braking that is both safe and recovers energy. As described above, the button for the second stage is located next to the display on the steering wheel. Additional use of the hydraulic brake leads to the third stage. Of all stages, the second one is the most efficient.

\subsection{Components}\label{sec:components}
The most important components are the installed \qty{10}{\kilo\watt} BLDC motor and the matching motor controller. Energy recovery is realised by the motor controller, which switches the motor to generator operation at the push of a button and thus generates electrical energy. A further component of the recuperation is the \textmu C. Only the \textmu C makes it possible to start recuperation at the push of a button.

\section{Safety Systems}\label{sec:safety-systems}
The topic of safety systems was discussed intensively in the project team. There are many ways to protect the driver of a kart. The most important are a good brake, protective clothing, a helmet and a neck support.

Initially, a stable headrest was to be designed and firmly attached to the kart. This plan could not be implemented, because under the given time pressure a working kart was defined as the main goal. In addition, the frame hardly offers any space to mount a support of the size originally planned.

Because of this missing safety feature, other safety measures are used. A good brake and suitable protective clothing remain essential parts of the safety concept. In addition, a speed limiter was installed that can be set in several stages. For inexperienced drivers, the kart is limited to \qty{15}{\kilo\metre\per\hour}, for example.

The central electrical safety device is the emergency stop switch within reach of the driver (Section~\ref{sec:tc-components}). Pressing it switches off the drive, while the hydraulic brake remains fully functional (Section~\ref{sec:start-up}). In addition, the motor switches off when the throttle and the brake pedal are pressed at the same time.

\section{Mechanical Add-on Parts}\label{sec:mechanical-add-on-parts}
The mechanical part of our project covers the design of housings and their mounting on the kart frame. The focus is on protecting the electrical components. The battery boxes were drawn in AutoCAD. The other parts designed in-house were modelled in the 3D CAD system PTC Pro/ENGINEER, and their drawings are given in Appendix~\ref{app:cad-design-documentation}.

\subsection{Battery Boxes}\label{sec:battery-boxes}
The battery boxes are meant to protect the cells and the BMS boards against mechanical loads and contamination. They are made of \qty{2}{\milli\metre} aluminium sheet, which is lighter than steel and meets our requirements well. For an optimal weight distribution, the cells are divided among three boxes: 12 cells at the rear axle, 10 cells on the left of the driver and 2 cells on the right of the driver.

\begin{wrapfigure}[13]{r}{0.42\textwidth}%
\wraptop
\centering
\includegraphics[width=0.42\textwidth]{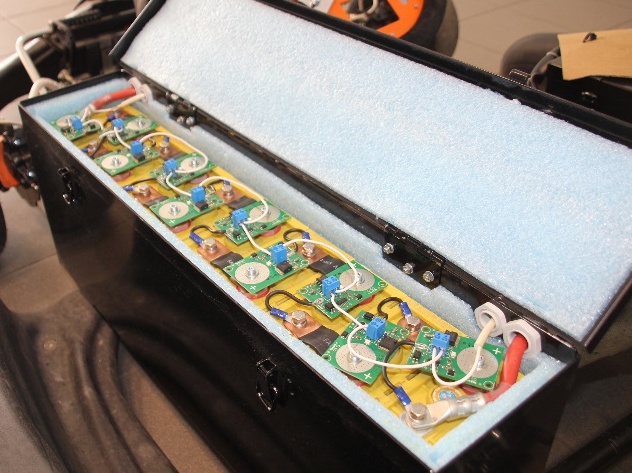}
\caption{Battery box with cells and BMS modules}\label{fig:box}
\end{wrapfigure}
Our plan was to mount the cells lying down in the boxes, which would have made the kart look more aerodynamic. However, the cells have a safety vent on the top and must not be mounted on their side. In the event of a fault, gas escapes through this vent. We therefore had to design the boxes for upright cells (Figure~\ref{fig:box}). The boxes are lined with foam, which supports the upright cells. Since a cell expands when overcharged, the foam still leaves room for the expanding cell in such a fault. The boxes are fixed to the frame with specially made tube clamps, the same system that is also used for the motor.

\subsection{Plastic Fairings}\label{sec:plastic-fairings}
The plastic fairings give the kart its appearance. The front bumper and the steering column fairing provide good aerodynamics. At the same time, the fairing parts act as a crumple zone. Their high strength and hollow design protect important components such as the motor and the battery boxes in collisions.

\subsection{Motor Mount}\label{sec:motor-mount}
A reliable and rigid mounting of the motor on the frame is particularly important. The motor must not shift or wobble while driving. The motor mount is a special design for mounting electric motors on a kart frame (Figure~\ref{fig:drive}a).

\chapter{Results and Discussion}\label{sec:results-and-discussion}
This chapter evaluates the calculations, bench measurements and design data of 2015/16 together. It was added for this edition. No new measurements were carried out. All values are either calculated or were measured during commissioning, and driving tests with recorded data are not available. Section~\ref{sec:verification-against-the-requirements} checks whether the requirements were met. Sections~\ref{sec:energy-demand-and-driving-time} to~\ref{sec:mechanical-reserves-of-the-drivetrain} show what the data reveal about energy balance, traction, battery, low-voltage supply and drivetrain of an electric kart of this class. Section~\ref{sec:limitations} names the limits of these results.

\section{Verification against the Requirements}\label{sec:verification-against-the-requirements}
The requirements were taken from the project objectives (Chapter~\ref{sec:introduction-and-objectives}) and from the project planning. Table~\ref{tab:req} compares each requirement with the result achieved and names the basis of the assessment.

\begin{table}[tbp]
\centering
\caption{Verification of the project requirements}\label{tab:req}
{\small\renewcommand{\arraystretch}{1.15}%
\begin{tabularx}{\linewidth}{@{}>{\raggedright\arraybackslash}p{3.3cm}*{4}{>{\centering\arraybackslash}X}@{}}
\toprule
\textbf{Requirement} & \textbf{Target} & \textbf{Result} & \textbf{Basis} & \textbf{Status} \\
\midrule
Top speed & approx.\ \qty{60}{\kilo\metre\per\hour} & approx.\ \qty{65}{\kilo\metre\per\hour} & calculation & met \\
Driving time at \qty{50}{\percent} motor load & ≥ \qty{60}{\minute} & \qty{55}{\minute} (\qty{100}{\percent} capacity); \qty{44}{\minute} (\qty{80}{\percent} capacity) & calculation & not met \\
Traction control & functional prototype & implemented in hardware and firmware & implementation & met; effect not measured \\
Energy recovery & three stages & three stages implemented & implementation & met; energy not measured \\
Limitation of speed and power & adjustable & multi-stage speed limiter; power demand limited to \qty{86}{\percent} & implementation & met \\
Rear axle strength & \qty{100}{\newton\metre} motor torque & static safety against yield approx.\ 2.8 & KISSsoft & met for intended use \\
Headrest and neck support & integrated & not implemented & inspection & not met \\
\bottomrule
\end{tabularx}}
\end{table}

Five of seven requirements are met. The target driving time is missed, and the planned headrest was not implemented. Two central functions, traction control and energy recovery, are implemented, but their effect has not yet been quantified by measurements.

\section{Energy Demand and Driving Time}\label{sec:energy-demand-and-driving-time}
\textbf{Capacity actually usable.} The driving time of \qty{55}{\minute} (Section~\ref{sec:energy-storage}) is based on the full nominal energy of \qty{4608}{\watt\hour}. The cell manufacturer, however, recommends using only \qty{80}{\percent} of the capacity in normal operation. With \qty{3686}{\watt\hour}, the driving time at \qty{50}{\percent} motor load (\qty{5}{\kilo\watt}) drops to approx.\ \qty{44}{\minute}.

\textbf{Steady-state demand.} A second estimate uses the driving resistance model of Section~\ref{sec:electric-drive} with the vehicle data of Table~\ref{tab:vehdata} (Appendix~\ref{app:technical-data}) and a drive efficiency of 0.9. Table~\ref{tab:consumption} shows the demand at constant speed.

\begin{table}[tbp]
\centering
\caption{Calculated energy demand at constant speed (vehicle data of the drive calculation, drive efficiency 0.9, driving time with \qty{80}{\percent} of the nominal battery energy)}\label{tab:consumption}
{\small\renewcommand{\arraystretch}{1.15}%
\begin{tabularx}{\linewidth}{@{}*{6}{>{\centering\arraybackslash}X}@{}}
\toprule
\textbf{Speed in \unit{\kilo\metre\per\hour}} & \textbf{Wheel power in \unit{\watt}} & \textbf{Battery power in \unit{\watt}} & \textbf{Energy in \unit{\watt\hour\per\kilo\metre}} & \textbf{Driving time in \unit{\minute}} & \textbf{Share of rolling resistance} \\
\midrule
30 & 1192 & 1324 & 44 & 167 & \qty{86}{\percent} \\
40 & 1757 & 1952 & 49 & 113 & \qty{78}{\percent} \\
50 & 2466 & 2741 & 55 & 81 & \qty{70}{\percent} \\
60 & 3356 & 3728 & 62 & 59 & \qty{61}{\percent} \\
\bottomrule
\end{tabularx}}
\end{table}

At a constant \qty{60}{\kilo\metre\per\hour}, the battery has to deliver approx.\ \qty{3.7}{\kilo\watt}, and the usable energy lasts approx.\ \qty{59}{\minute}. The assumption of \qty{5}{\kilo\watt} average power is therefore about 1.3 times the steady-state demand at \qty{60}{\kilo\metre\per\hour}. This margin covers acceleration phases. Whether it is sufficient in racing can only be answered by measurements on the track.

\begin{figure}[!ht]
\centering
\pgfplotsset{
  energyaxis/.style={width=0.7\linewidth, height=5.0cm, scale only axis, xmin=10, xmax=65},
  c-e070/.style={black, solid, mark=o, mark size=2.2pt, mark options={solid}, mark repeat=5},
  c-e035/.style={black, dashed, mark=square, mark size=2pt, mark options={solid}, mark repeat=5},
  c-t070/.style={black!55, solid, line width=0.9pt, mark=*, mark size=1.9pt, mark options={solid, fill=black!55}, mark repeat=5},
  c-t035/.style={black, densely dotted, mark=triangle, mark size=2.6pt, mark options={solid}, mark repeat=5},
}
\begin{tikzpicture}
  \begin{axis}[thesisplot, energyaxis, ymin=0, ymax=70, xtick={10,20,...,60}, axis y line*=left,
      xlabel={Constant vehicle speed in \unit{\kilo\metre\per\hour}},
      ylabel={Energy demand in \unit{\watt\hour\per\kilo\metre}},
      legend cell align=left, legend columns=2, legend style={draw=none, fill=none, font=\small,
        at={(0.55,-0.25)}, anchor=north, /tikz/every even column/.append style={column sep=1.2em}}]
    \addplot[c-e070] table[x=v, y=e070] {figures/data/energy.dat};
    \addlegendentry{Energy demand, $\mu_{\mathrm{R}}=0.07$ (design value)}
    \addplot[c-e035] table[x=v, y=e035] {figures/data/energy.dat};
    \addlegendentry{Energy demand, $\mu_{\mathrm{R}}=0.035$ (half)}
    \addlegendimage{c-t070}
    \addlegendentry{Driving time, $\mu_{\mathrm{R}}=0.07$}
    \addlegendimage{c-t035}
    \addlegendentry{Driving time, $\mu_{\mathrm{R}}=0.035$}
  \end{axis}
  \begin{axis}[thesisplot, energyaxis, ymin=0, ymax=700, axis y line*=right, axis x line=none, grid=none,
      ytick pos=right, yticklabel pos=right, ylabel near ticks,
      ylabel={Driving time (\qty{80}{\percent} capacity) in \unit{\minute}}]
    \addplot[c-t070] table[x=v, y=t070] {figures/data/energy.dat};
    \addplot[c-t035] table[x=v, y=t035] {figures/data/energy.dat};
  \end{axis}
\end{tikzpicture}
\caption{Calculated battery energy demand and driving time at constant speed for two rolling resistance coefficients\nocc}\label{fig:energy}
\end{figure}
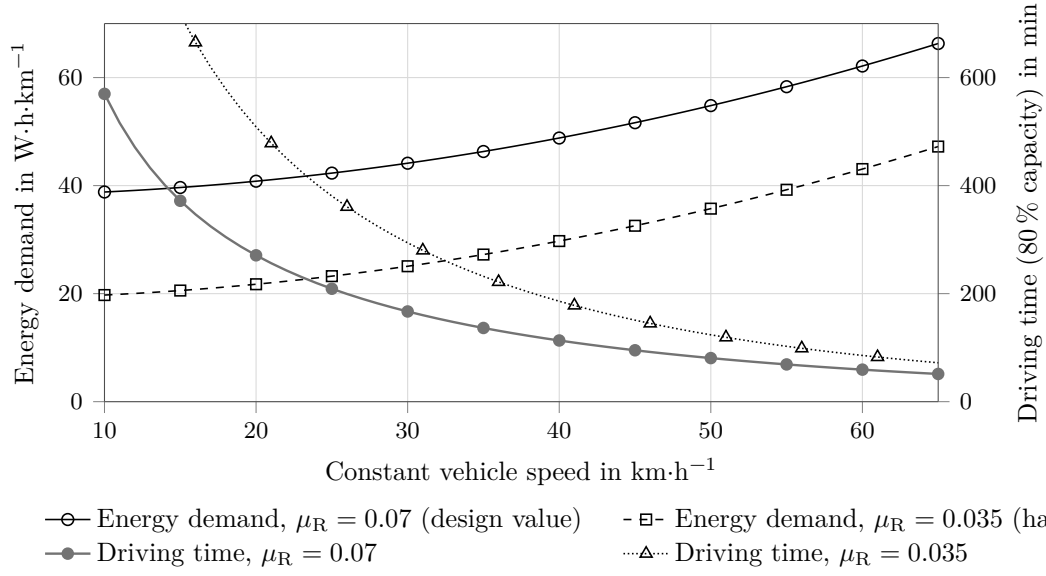

\textbf{Sensitivity.} Up to \qty{60}{\kilo\metre\per\hour}, rolling resistance accounts for \qty{61}{\percent} to \qty{86}{\percent} of the driving resistance. The rolling resistance coefficient is therefore the most sensitive parameter of the model. The value $\mu_{\mathrm{R}}$ = 0.07 used in the design is a conservative assumption and was not measured. Should the real value be only half as large, the demand at \qty{60}{\kilo\metre\per\hour} drops from \qty{62}{\watt\hour\per\kilo\metre} to \qty{43}{\watt\hour\per\kilo\metre}, i.e.\ by approx.\ \qty{31}{\percent} (Figure~\ref{fig:energy}).

\textbf{Battery size and vehicle mass.} The driving time target of one hour determined the battery size. With 24 cells of \qty{2.3}{\kilo\gram} each, the battery weighs approx.\ \qty{55}{\kilo\gram}, i.e.\ almost \qty{30}{\percent} of the total mass of \qty{187}{\kilo\gram}, and the kart became noticeably heavy. In the actual use as a demonstration vehicle with short rides, this capacity was never needed. Since the number of cells in series is fixed by the \qty{72}{\volt} voltage class of the drive, a lighter battery means cells of smaller capacity. Halving the capacity would save approx.\ \qty{28}{\kilo\gram} or \qty{15}{\percent} of the total mass, reduce rolling resistance and inertia accordingly, and still allow approx.\ \qty{30}{\minute} at a constant \qty{60}{\kilo\metre\per\hour} with \qty{80}{\percent} of the energy. In retrospect, the battery should have been sized for the real duty cycle rather than for the one-hour target.

\section{Traction Limit and Need for Traction Control}\label{sec:traction-limit-and-need-for-traction-con}
The force that the rear tyres can transmit is limited by the static friction coefficient $\mu_{\mathrm{H}}$ = 0.9, the vehicle mass and the share x of the weight that rests on the rear axle:
\[
  F_{\mathrm{max}}=\mu_{\mathrm{H}}\,m\,g\,x
\]
The mass budget of the drive calculation gives $m=\qty{187}{\kilo\gram}$ including a \qty{75}{\kilo\gram} driver. The rear axle load share was not measured. For a kart with rear-mounted motor and battery boxes, a range of \qty{50}{\percent} to \qty{60}{\percent} is assumed. The traction limit therefore lies between approx.\ \qty{830}{\newton} and \qty{990}{\newton}.

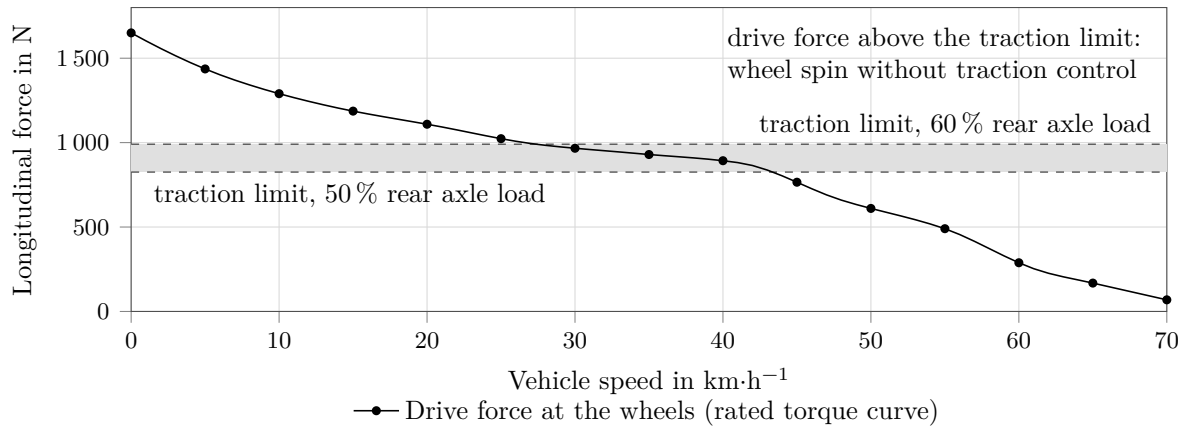
\begin{figure}[!ht]
\centering
\begin{tikzpicture}
  \begin{axis}[thesisplot, width=0.97\linewidth, height=5.6cm, xmin=0, xmax=70, ymin=0, ymax=1800,
      xtick={0,10,...,70}, xlabel={Vehicle speed in \unit{\kilo\metre\per\hour}},
      ylabel={Longitudinal force in \unit{\newton}},
      yticklabel style={/pgf/number format/.cd, 1000 sep={\,}},
      legend cell align=left, legend style={draw=none, fill=none, font=\small, at={(0.5,-0.25)}, anchor=north}]
    \fill[black!12] (axis cs:0,825.5) rectangle (axis cs:70,990.6);
    \draw[black!70, dashed, line width=0.5pt] (axis cs:0,825.5) -- (axis cs:70,825.5);
    \draw[black!70, dashed, line width=0.5pt] (axis cs:0,990.6) -- (axis cs:70,990.6);
    \node[font=\small, anchor=south east, inner sep=0pt] at (axis cs:69,1030) {traction limit, \qty{60}{\percent} rear axle load};
    \node[font=\small, anchor=north west, inner sep=0pt] at (axis cs:1.5,770) {traction limit, \qty{50}{\percent} rear axle load};
    \node[font=\small, align=left, anchor=north east] at (axis cs:69,1740) {drive force above the traction limit:\\ wheel spin without traction control};
    \addplot[black, mark=*, mark size=1.4pt, mark repeat=5] table[x=v, y=F] {figures/data/traction.dat};
    \addlegendentry{Drive force at the wheels (rated torque curve)}
  \end{axis}
\end{tikzpicture}
\caption{Drive force at the wheels compared with the traction limit of the rear axle (own calculation)\nocc}\label{fig:traction}
\end{figure}

With the rated torque curve of the motor (\qty{60}{\newton\metre} at standstill), a gear ratio of 3.3 and a wheel radius of \qty{0.12}{\metre}, the drive force at standstill is approx.\ \qty{1650}{\newton}. This is 1.7 to 2.0 times the traction limit. The drive force remains above the limit up to approx.\ \qty{27}{\kilo\metre\per\hour} (\qty{60}{\percent} rear axle load) or approx.\ \qty{43}{\kilo\metre\per\hour} (\qty{50}{\percent} rear axle load), as Figure~\ref{fig:traction} shows. In this speed range, the acceleration of the kart is limited by traction and not by the motor. Driving practice confirmed this. With the traction control switched off, the rear axle could be made to slide with the throttle at any time, and the kart could be drifted without difficulty.

Two consequences follow. First, the calculated acceleration of \qty{8.2}{\metre\per\second\squared} at standstill (Section~\ref{sec:electric-drive}) cannot be transmitted. Taking tyre grip and rolling resistance into account, the realistic launch acceleration is approx.\ \qtyrange{3.8}{4.6}{\metre\per\second\squared}. Second, a driver will inevitably spin the rear wheels at full throttle. Beyond the friction maximum, the transmitted force decreases with increasing slip, and the excess energy is converted into heat and tyre wear~\cite{Pac12}. The traction control is therefore a necessary part of the drive concept. This agrees with the estimate made during planning that traction control becomes useful from a torque of approx.\ \qty{100}{\newton\metre} at the rear axle. With $\qty{60}{\newton\metre}\times3.3\approx\qty{200}{\newton\metre}$, this value is clearly exceeded.

\section{Braking and Energy Recovery Potential}\label{sec:braking-and-energy-recovery-potential}
\textbf{Regenerative braking force.} If the motor brakes with its rated torque of \qty{30}{\newton\metre}, the braking force at the rear wheels is approx.\ \qty{825}{\newton}. This corresponds to a deceleration of \qty{4.4}{\metre\per\second\squared} ($0.45\,g$). Above approx.\ \qty{44}{\kilo\metre\per\hour}, the rated power of \qty{10}{\kilo\watt} limits the braking force further. At \qty{60}{\kilo\metre\per\hour}, only approx.\ \qty{600}{\newton} (\qty{3.2}{\metre\per\second\squared}) remain. Hard braking therefore still requires the hydraulic brake. The data thus confirm the three-stage concept of Section~\ref{sec:braking-system}, in which the hydraulic brake is added only in stage 3.

\textbf{Recoverable energy.} Table~\ref{tab:recup} shows the kinetic energy released in typical braking events. For the recoverable share, an overall efficiency of 0.7 is assumed for the chain from the wheel to the battery (drivetrain, motor in generator mode, controller and battery). This value was not measured.

\begin{table}[htp]
\centering
\caption{Kinetic energy and recoverable energy of typical braking events ($m=\qty{187}{\kilo\gram}$, assumed overall efficiency 0.7)}\label{tab:recup}
{\small\renewcommand{\arraystretch}{1.15}%
\begin{tabularx}{\linewidth}{@{}*{4}{>{\centering\arraybackslash}X}@{}}
\toprule
\textbf{Braking event} & \textbf{Kinetic energy in \unit{\watt\hour}} & \textbf{Recoverable energy in Wh ($\eta=0.7$)} & \textbf{Equivalent distance at \qty{60}{\kilo\metre\per\hour} in \unit{\metre}} \\
\midrule
60 → \qty{0}{\kilo\metre\per\hour} & 7.2 & 5.1 & 81 \\
60 → \qty{30}{\kilo\metre\per\hour} & 5.4 & 3.8 & 61 \\
50 → \qty{25}{\kilo\metre\per\hour} & 3.8 & 2.6 & 42 \\
40 → \qty{20}{\kilo\metre\per\hour} & 2.4 & 1.7 & 27 \\
\bottomrule
\end{tabularx}}
\end{table}

A single braking event from \qty{60}{\kilo\metre\per\hour} to \qty{30}{\kilo\metre\per\hour} returns approx.\ \qty{4}{\watt\hour}. This seems small, but on a winding track the kinetic energy is turned over many times. For an illustrative track with eight such braking events per kilometre, approx.\ \qty{43}{\watt\hour\per\kilo\metre} of kinetic energy are released and approx.\ \qty{30}{\watt\hour\per\kilo\metre} can be recovered. The total demand in this scenario is approx.\ \qty{100}{\watt\hour\per\kilo\metre} (driving resistance at an average of \qty{45}{\kilo\metre\per\hour} plus acceleration energy). Recuperation could therefore cover up to about \qty{30}{\percent} of the consumption. This is an upper estimate: rolling and air resistance during braking reduce the recoverable energy further. On fast tracks with few braking points, the effect is much smaller.

\textbf{Brake wear.} Beyond energy, regenerative braking reduces the use of the friction brake. Non-exhaust emissions from brake, tyre and road wear have become a relevant source of traffic-related particulate matter~\cite{OECD20}. The kart addresses this with regenerative braking and traction control. For the kart itself, this effect is a hypothesis and has not been measured.

\section{Battery Behaviour during Initial Charging}\label{sec:battery-behaviour-during-initial-chargin}
During initial charging, the 24 cells were connected in parallel and charged with a total current of \qty{25}{\ampere}, i.e.\ approx.\ \qty{1}{\ampere} per cell (0.017\,C). Charging was spread over four days with overnight pauses, and the voltage was read every \qty{2}{\hour} until, after \qty{44}{\hour} of charging, the current had decreased to \qty{4}{\ampere} at the end-of-charge voltage of \qty{4.0}{\volt} (Figure~\ref{fig:charge}). In total, approx.\ \qty{1050}{\ampere\hour} were charged, i.e.\ approx.\ \qty{44}{\ampere\hour} per cell.

\textbf{State of charge and capacity.} The voltage alone does not tell at which state of charge the cells started. The voltage curve in the BMS documentation~\cite{BMS15} is a schematic illustration of a typical \LFP{} cell, not a measured curve of the cells used, and is therefore not suitable for reading off state-of-charge values. For \LFP{}, the open-circuit voltage is almost constant between approx.\ \qty{20}{\percent} and \qty{90}{\percent} state of charge, and it shows a hysteresis between charging and discharging of the order of \qty{10}{\milli\volt} that persists even after long rest periods~\cite{Ros11,Dre10}. After storage, the cells rest on the lower branch of this hysteresis. The rest voltage of \qty{3.27}{\volt} at the start is therefore compatible with a wide range of states of charge, roughly \qty{20}{\percent} to \qty{40}{\percent}. The evaluation is therefore reversed. In total, approx.\ \qty{44}{\ampere\hour} per cell were charged. With the rated capacity of \qty{60}{\ampere\hour}, the cells started at approx.\ \qty{27}{\percent} state of charge. Charging losses shift this value slightly upwards, since not all of the charge is stored. New cells are usually delivered partially charged, so this starting point is plausible, and the measurement is consistent with the rated capacity. It does not, however, determine the capacity independently. An evaluation of the same readings by one of the authors at the time, integrating the current from \qty{3.25}{\volt} to \qty{3.5}{\volt}, likewise gave approx.\ \qty{60}{\ampere\hour}. The state-of-charge line in Figure~\ref{fig:charge} follows from this charge counting, anchored at \qty{100}{\percent} at the end of charging and based on the rated capacity. It shows how the state of charge rises linearly while the voltage stays almost constant. Two further sources of uncertainty concern the measurement itself. The charging current was read from the built-in display of the laboratory power supply instead of a calibrated current clamp, and the readings were noted by hand, so the \qty{2}{\hour} interval was kept only approximately. Together, these effects limit the integrated charge to an accuracy of roughly ±\qty{10}{\percent}. A capacity test by full discharge with calibrated instruments would remove these uncertainties.

The measurement quantifies the flat voltage plateau of \LFP{}. Between \qty{4}{\hour} (\qty{3.36}{\volt}) and \qty{36}{\hour} (\qty{3.46}{\volt}) of charging, i.e.\ between approx.\ \qty{34}{\percent} and \qty{90}{\percent} state of charge according to the charge counting, the cell voltage rose by only \qty{100}{\milli\volt}. This is approx.\ \qty{1.8}{\milli\volt} per percent state of charge. In the middle of the plateau, between \qty{6}{\hour} and \qty{30}{\hour}, the rise is only \qty{28}{\milli\volt} for approx.\ \qty{25}{\ampere\hour} or \qty{42}{\percent} of the capacity, i.e.\ approx.\ \qty{0.7}{\milli\volt} per percent. The relaxation of the voltage during the overnight pauses, visible as dips of \qty{7}{\milli\volt} and \qty{12}{\milli\volt} after the first two pauses, is of the same order as this rise. It illustrates the slow open-circuit voltage recovery of \LFP{}, which lasts for hours~\cite{Ros11}. A voltage error of only \qty{10}{\milli\volt} therefore corresponds to about \qty{6}{\percent} to \qty{15}{\percent} state of charge. In the plateau, a state-of-charge estimation based on the terminal voltage is thus unusable, and the hysteresis adds a further uncertainty that depends on the history of the cell~\cite{Dre10}. The charge has to be counted by current integration, as the BMS of the kart does~\cite{Ple15}. This property of \LFP{}, a very flat open-circuit voltage curve with pronounced hysteresis, had been described in the literature before the project~\cite{Ros11}. In 2016, however, it was hardly reflected in the products available to a student project.

\section{Cell Temperature and Voltage Sag under Load}\label{sec:cell-temperature-and-voltage-sag-under-l}
A second observation concerns the behaviour of the cells under load. During the first test runs, the cells were cool, and high discharge currents were drawn during acceleration. The terminal voltage then dropped so far that the low-voltage alarm of the battery management system was triggered, although the state of charge was not low. After the cells had warmed up in operation, the alarm no longer occurred. No current or temperature data were recorded for these events, so the observation is qualitative.

The behaviour is explained by the internal resistance of the cells. The voltage drop under load is the product of current and internal resistance, and the internal impedance of lithium-ion cells rises considerably at low temperature, while the available energy decreases~\cite{Jag16}. That the charge-transfer resistance dominates this rise had already been shown in 2003~\cite{Zha03}. The datasheet range of \qtyrange{-35}{+85}{\degreeCelsius} (Table~\ref{tab:cells}, Appendix~\ref{app:technical-data}) states where the cell may be operated. It does not state where the cell delivers its full performance. For a kart, this has three consequences. First, a low-voltage threshold based on the terminal voltage alone confuses a cold cell under load with an empty cell. The threshold has to be compensated for the load current or evaluated with a delay. Second, the cells should be warmed up with moderate load before full power is demanded. Third, the temperature has to be treated as a state variable of the battery and as more than a protection limit. A state-of-charge estimation that also uses an internal resistance model~\cite{Ple15} would resolve both problems observed in this project, the flat plateau and the load-dependent voltage sag.

\section{Low-Voltage Supply without Auxiliary Battery}\label{sec:low-voltage-supply-without-auxiliary-bat}
The kart has no separate \qty{12}{\volt} battery. The entire low-voltage system, i.e.\ microcontroller, sensors, display, throttle signal, relays and battery management, is supplied from the \qty{76.8}{\volt} traction battery through the isolated DC/DC converter TEN 8-7212WI (\qty{72}{\volt} to \qty{12}{\volt}) and a downstream \qty{5}{\volt} linear regulator (Appendix~\ref{app:block-diagrams-and-circuit-diagrams}). This saves a second battery with its own charging and maintenance, reduces weight and installation space and avoids a discharged auxiliary battery as a source of failure. As a consequence, the low-voltage system is only live while the traction battery is connected. In the kart, this is acceptable, because the safety-relevant functions, the hydraulic brake and the emergency stop switch, are purely mechanical or electromechanical and do not depend on the low-voltage supply. During storage, however, the emergency stop switch must remain released, so that the battery management system stays supplied and protects the cells against deep discharge (Section~\ref{sec:storage}). The galvanic isolation of the converter keeps the traction voltage away from the control electronics and the sensor cabling.

\textbf{Voltage fluctuations.} During commissioning, a consequence of the missing auxiliary battery was observed. The low-voltage rails showed voltage fluctuations. An auxiliary battery is a low-impedance buffer with a capacity of several ampere-hours. The DC/DC converter, by contrast, is rated at \qty{8}{\watt}, i.e.\ approx.\ \qty{0.7}{\ampere} at \qty{12}{\volt}, and its output is buffered only by a few microfarads (Appendix~\ref{app:block-diagrams-and-circuit-diagrams}). Every load step, for example the switching of a relay or the display, therefore appears as a dip on the rail.

\textbf{Coupling of the sensor signals.} The steep edges of the Hall sensor pulses could be measured on practically all lines of the low-voltage system. A fast edge contains spectral components far above the pulse frequency itself. Through the common ground return and the parasitic capacitance and mutual inductance between adjacent lines, such edges couple into every neighbouring line~\cite{Ott09}. On the kart, the sensor cables run over the whole vehicle next to the motor and battery cables, and the boards are connected by screw terminals and discrete wiring.

\textbf{Why smaller structures help.} The interference produced and picked up by a circuit scales with its geometry. The radiated field of a current loop is proportional to the loop area and to the rate of change of the current, and as a common rule of thumb, the inductance of a trace or wire is roughly \qty{1}{\nano\henry} per millimetre, so every long connection between converter, filter capacitor and load turns a current step into a voltage spike $L\,\mathrm{d}i/\mathrm{d}t$~\cite{Ott09}. The hand-made boards of the kart use through-hole components, screw terminals with a pitch of \qty{3.81}{\milli\metre} and generous trace lengths, which results in large loops. A compact layout with surface-mounted components, filter capacitors placed directly at the converter pins, a continuous ground plane and short, paired supply and return traces reduces loop area and inductance substantially and would have improved the electromagnetic compatibility of the low-voltage system. Measures with the same aim are bulk capacitance on the \qty{12}{\volt} and \qty{5}{\volt} rails, twisted or shielded sensor lines with their own return conductor, RC low-pass filters at the sensor inputs of the microcontroller and a common-mode choke at the DC/DC input. These observations were made with an oscilloscope during commissioning, but were not recorded.

\section{Parameters of the Traction Control}\label{sec:parameters-of-the-traction-control}
Table~\ref{tab:asrpar} lists the parameters of the traction control as implemented in the archived firmware~\cite{Sto26} and the resulting behaviour.

\begin{table}[tbp]
\centering
\caption{Parameters of the traction control firmware and their consequences}\label{tab:asrpar}
{\small\renewcommand{\arraystretch}{1.15}%
\begin{tabularx}{\linewidth}{@{}>{\raggedright\arraybackslash}X>{\centering\arraybackslash}p{3.0cm}>{\raggedright\arraybackslash}X@{}}
\toprule
\textbf{Parameter} & \textbf{Value} & \textbf{Consequence} \\
\midrule
Control cycle & \qty{1}{\milli\second} (\qty{1}{\kilo\hertz}) & timer interrupt of the SAM3X8E \\
Slip threshold (rear axle vs.\ front wheel) & \qty{20}{\percent} & fixed design value \\
Reduction ramp & 2 PWM steps per ms & full reduction within \qty{110}{\milli\second} \\
Recovery ramp & 1 PWM step per ms & full recovery within \qty{220}{\milli\second} \\
Maximum PWM value & 220 of 255 & power demand limited to \qty{86}{\percent} \\
Standstill detection & \qty{200}{\milli\second} without pulse & wheel is regarded as stationary \\
Missing front wheel signal & no intervention & kart behaves like without traction control \\
Ignition release & only with released pedal & no unintended start at power-up \\
\bottomrule
\end{tabularx}}
\end{table}

The threshold controller with ramps is simple, robust and needs only a few parameters. Its weak point is the fixed slip threshold of \qty{20}{\percent}. The slip at which the tyre transmits the highest force depends on tyre, surface and temperature~\cite{Pac12}. The threshold was set as a design value and still has to be identified on the track.

A second aspect concerns the reference speed. Three speed sensors are mounted on the kart (Section~\ref{sec:tc-components}), but the archived firmware compares the rear axle with one front wheel only. In a corner, the front wheels turn at different speeds. For a track width of approx.\ \qty{1}{\metre} and a corner radius of \qty{10}{\metre}, the speed of one front wheel deviates by approx.\ ±\qty{5}{\percent} from the vehicle speed. The threshold of \qty{20}{\percent} is large enough to tolerate this error. A kart has a rigid rear axle without differential, so the inner rear wheel slips in every corner anyway. An intervention in tight corners is therefore not necessarily a malfunction.

\section{Mechanical Reserves of the Drivetrain}\label{sec:mechanical-reserves-of-the-drivetrain}
\textbf{Rear axle.} The rear axle was calculated with a motor torque of \qty{100}{\newton\metre}, i.e.\ \qty{330}{\newton\metre} at the sprocket. This load case is conservative: the rated torque of the motor is \qty{30}{\newton\metre}, and the starting torque of the rated curve is \qty{60}{\newton\metre}. The maximum equivalent stress of approx.\ \qty{110}{\newton\per\square\milli\metre} is well below the minimum yield strength of C45 in the normalised condition (\qty{305}{\newton\per\square\milli\metre} for diameters from \qty{16}{\milli\metre} to \qty{100}{\milli\metre}~\cite{DIN06}). The static safety factor is therefore approx.\ 2.8. A fatigue assessment was not carried out.

\textbf{Grip as a torque limit.} In traction, the torque at the rear axle is limited by tyre grip. With the traction limit of \qty{830}{\newton} to \qty{990}{\newton}, only approx.\ \qty{100}{\newton\metre} to \qty{120}{\newton\metre} can be supported at the wheels. The design load of \qty{330}{\newton\metre} thus covers steady traction loads about three times over. Load peaks from kerbs, chain shocks or the acceleration of the drivetrain inertia are not covered by this estimate.

\textbf{Chain drive under torque steps.} The chain was dimensioned statically for a peak torque of \qty{100}{\newton\metre} (Section~\ref{sec:dimensioning-of-the-chain}). In operation, this proved insufficient for the pinion on the motor shaft. The torque step of the electric motor from standstill destroyed the original pinion completely, and a new pinion and chain had to be made from higher-grade materials. The pinion is the most highly stressed element of the chain drive because it has the smallest pitch diameter and the fewest teeth, so the full chain force of approx.\ \qty{3.6}{\kilo\newton} acts on very few teeth. An electric motor delivers its torque within milliseconds, and the play of the chain lets the drivetrain accelerate freely for a moment before the load is taken up. The pinion therefore sees an impact instead of a steady load. In chain design, such shock loads are covered by an application factor that is considerably higher for a drive with sudden torque steps than for uniform load. Two measures follow: a hardened steel pinion, and a limitation of the torque rate in the firmware. The ramp of the traction control, which raises the power demand by at most one PWM step per millisecond (Table~\ref{tab:asrpar}), also acts as such a limitation.

\textbf{Bearings.} The most highly loaded bearing B1 next to the sprocket reaches a nominal life of \qty{4290}{\hour} in the design load case. This is below the target of \qty{20000}{\hour}. For a demonstration vehicle with a few dozen operating hours per year, the value is sufficient. In continuous racing use, this bearing position would have to be reviewed.

\section{Limitations}\label{sec:limitations}
The results of this chapter are subject to the following limitations:
\begin{itemize}
  \item No driving tests with recorded data are available. Lap times, energy consumption on the track, wheel slip and recovered energy have not been measured.
  \item Several model parameters are assumptions: rolling resistance coefficient, drag area, rear axle load share and efficiencies.
  \item The individual calculations of 2015/16 use different values for torque (\qty{30}{\newton\metre}, \qty{60}{\newton\metre}, \qty{100}{\newton\metre}), gear ratio (3.2 and 3.3), rated speed (\qty{4000}{\per\minute} and \qty{4500}{\per\minute}) and vehicle mass (\qty{180}{\kilo\gram} and \qty{187}{\kilo\gram}). The results are therefore to be understood as estimates.
  \item The initial charging curve is a single measurement of the parallel-connected pack, read by hand every \qty{2}{\hour}, with the current taken from the display of the power supply rather than from a calibrated instrument. The state of charge during charging is derived from the rated capacity, and the capacity itself was not measured. The voltage sag of the cold cells was observed, but not recorded. The effects on tyre wear and particulate emissions are hypotheses.
\end{itemize}

\chapter{Operating Manual}\label{sec:operating-manual}
This chapter contains the operating manual of the kart. It covers safety instructions (Section~\ref{sec:safety-instructions}), the handling of the battery (Section~\ref{sec:handling-of-the-battery}), electrical accidents (Section~\ref{sec:instructions-for-electrical-accidents}) and the procedures for start-up, parking, charging and storage (Sections~\ref{sec:start-up} to~\ref{sec:storage}).

{\small\itshape This operating manual refers exclusively to the vehicle built in 2016 and is reproduced for documentation purposes. It is not a general guide for the construction or operation of electric vehicles.\par}

\section{Safety Instructions}\label{sec:safety-instructions}
\begin{itemize}
  \item The kart may only be driven with a helmet, protective clothing and sturdy footwear.
  \item The kart must not be driven in poor visibility (dusk, darkness, fog).
  \item The kart must not be driven in rain or snowfall.
  \item The kart may only be driven on paved surfaces in closed-off areas. It is not approved for use on public roads.
  \item The driver must not be under the influence of alcohol, drugs or medication.
\end{itemize}

\section{Handling of the Battery}\label{sec:handling-of-the-battery}
The \LFP{} cells must be handled with particular care. The cell voltage must not fall below \qty{2.8}{\volt}, as the cells are otherwise deeply discharged and damaged. The cells are therefore a potential hazard even when the kart is not in use. The terminals are the greatest source of danger. The positive and negative terminals must never be connected to each other. Otherwise, a very high short-circuit current flows, discharges the cells within a very short time and damages them.

In the event of a fault, a safety vent on the top of the cells releases a gas mixture. This gas must not be inhaled or get into the nose or eyes. Swelling, deformation, cracks in the surface and fire can also occur. The risk of explosion is low with this cell chemistry. A damaged cell must not be used any longer.

Voltage measurements may only be carried out at the terminals of a cell. Care must be taken not to touch any components of the BMS.

In summer, the kart should not be left standing in the sun for long. At temperatures above \qty{85}{\degreeCelsius}, the cells must no longer be loaded. This temperature can be reached, because the battery boxes are powder-coated black and the heat cannot escape well. Any contact with water or other liquids must be avoided. The battery boxes must not be left open unattended, since the cells could become dirty and a person could cause a short circuit.

The cells contain toxic substances. They do not belong in household waste but must be disposed of properly. For transport or disposal, the terminals should be covered with insulating tape to avoid an accidental short circuit.

\section{Instructions for Electrical Accidents}\label{sec:instructions-for-electrical-accidents}
Special care is required when working with the battery, since each cell can deliver currents of up to \qty{1200}{\ampere} for a short time. During maintenance work on the kart, the basic safety rules for work on electrical installations must be followed to avoid accidents:

\begin{itemize}
  \item Disconnect completely
  \item Secure against reconnection
  \item Verify absence of voltage
  \item Cover adjacent live parts with insulating material if necessary
\end{itemize}

The injuries caused by an electrical accident depend on several factors:

\begin{itemize}
  \item Current and voltage
  \item Frequency
  \item Current path through the body
  \item Duration of the current flow
  \item Point in time of the current flow (within the cardiac cycle)
\end{itemize}

The hazard zones for alternating and direct current differ. Alternating current is dangerous at lower currents than direct current (Figure~\ref{fig:zones}). Quantitative values are given in IEC TS 60479-1~\cite{IEC05}.

\begin{figure}[tbp]
\small
\begin{minipage}[t]{0.49\textwidth}
\textbf{a)}\par\medskip
\centering
\begin{tikzpicture}[schem]
  \draw[part] (0,0) rectangle (6.2,5.4);
  \draw[part] (0.86,0) -- (0.86,5.4);
  \draw[part] (2.48,5.4) -- (2.48,4.15) -- (4.10,0);
  \draw[part] (3.52,5.4) -- (3.52,4.46) -- (4.66,0);
  \draw[dashed] (3.97,5.4) -- (3.97,4.46) -- (5.11,0);
  \draw[dashed] (4.42,5.4) -- (4.42,4.46) -- (5.56,0);
  \foreach \x/\t in {0.86/a, 2.48/b, 3.52/c$_1$, 3.97/c$_2$, 4.42/c$_3$} \node[anchor=north west, inner sep=1pt] at (\x,5.35) {\t};
  \node at (0.43,1.45) {AC-1};
  \node at (1.75,1.45) {AC-2};
  \node at (3.2,3.75) {AC-3};
  \node at (5.75,2.6) {AC-4};
  \node[below] at (3.1,0) {Body current $I$ (logarithmic)};
  \node[above, rotate=90] at (0,2.7) {Duration of current flow $t$ (logarithmic)};
\end{tikzpicture}\par\medskip
\raggedright
\begin{tabular}{@{}>{\bfseries}l@{\hspace{0.8em}}p{0.78\linewidth}@{}}
AC-1 & perception possible, usually no startle reaction \\
AC-2 & involuntary muscular contractions likely, usually no harmful physiological effects \\
AC-3 & strong muscular contractions, difficulty in breathing, reversible disturbances of heart function \\
AC-4 & ventricular fibrillation likely, cardiac and respiratory arrest, severe burns possible \\
\end{tabular}
\end{minipage}\hfill
\begin{minipage}[t]{0.49\textwidth}
\textbf{b)}\par\medskip
\centering
\begin{tikzpicture}[schem]
  \draw[part] (0,0) rectangle (6.2,5.4);
  \draw[part] (1.61,0) -- (1.61,5.4);
  \draw[part] (2.97,5.4) -- (2.97,4.15) -- (4.11,0);
  \draw[part] (4.03,5.4) -- (4.03,4.46) -- (4.96,0);
  \draw[dashed] (4.43,5.4) -- (4.43,4.46) -- (5.36,0);
  \draw[dashed] (4.83,5.4) -- (4.83,4.46) -- (5.76,0);
  \foreach \x/\t in {1.61/a, 2.97/b, 4.03/c$_1$, 4.43/c$_2$, 4.83/c$_3$} \node[anchor=north west, inner sep=1pt] at (\x,5.35) {\t};
  \node at (0.81,1.45) {DC-1};
  \node at (2.45,1.45) {DC-2};
  \node at (3.72,3.6) {DC-3};
  \node at (5.8,3.6) {DC-4};
  \node[below] at (3.1,0) {Body current $I$ (logarithmic)};
  \node[above, rotate=90] at (0,2.7) {Duration of current flow $t$ (logarithmic)};
\end{tikzpicture}\par\medskip
\raggedright
\begin{tabular}{@{}>{\bfseries}l@{\hspace{0.8em}}p{0.78\linewidth}@{}}
DC-1 & slight pricking sensation possible when making or breaking current \\
DC-2 & involuntary muscular reactions likely, usually no harmful effects \\
DC-3 & strong involuntary muscular reactions, reversible disturbances of impulse formation and conduction in the heart \\
DC-4 & ventricular fibrillation likely, severe burns possible \\
\end{tabular}
\end{minipage}\par\medskip
c$_1$: onset of probability of ventricular fibrillation; c$_2$, c$_3$: increasing probability. Qualitative representation based on IEC TS 60479-1~\cite{IEC05}, not to scale.
\caption{Time--current zones of the effects of current on persons (qualitative): a) alternating current (15--\qty{100}{\hertz}), b) direct current\nocc}\label{fig:zones}
\end{figure}

\subsection{Rescue and First Aid}\label{sec:rescue-and-first-aid}
If an electrical accident occurs, first aid must be given. The danger has to be assessed correctly, and self-protection always has priority.

If a person is caught in the circuit, he or she must be freed as quickly as possible. If possible, the circuit must be switched off. In order not to endanger oneself, the person should be pulled out of the circuit with an insulating object (e.g.\ a wooden broom or rubber gloves). To reduce the risk of an electric shock, the rescuer should stand on an insulating surface or wear insulating footwear.

After the rescue, the general first-aid measures apply. An emergency call must be made (in Austria: ambulance 144, European emergency number 112). Consciousness and breathing must be checked. If the person is not breathing, cardiopulmonary resuscitation must be started immediately, and an automated external defibrillator must be used if available. If the person is breathing, he or she must be placed in the recovery position. Even without visible injuries, a medical examination is always required after an electrical accident, as cardiac arrhythmias can occur with a delay. Details can be found in the current first-aid guidelines of the European Resuscitation Council~\cite{ERC21}, on which the courses of the Austrian Red Cross are based.

\section{Start-up}\label{sec:start-up}
Before start-up, the following points must be checked:

\begin{itemize}
  \item Charger disconnected
  \item Throttle pedal, brake pedal and brake lever not blocked
  \item Tyre pressure between \qty{100}{\kilo\pascal} and \qty{120}{\kilo\pascal}
  \item Brake fluid level between the upper and lower mark
  \item Motor ventilation slots free
  \item Motor controller connected at the Anderson connector
  \item Brake pads checked
\end{itemize}

Once these points are fulfilled, the kart can be started:

\begin{enumerate}
  \item Release the emergency stop switch.
  \item A loud signal then sounds for 30 seconds. During this time, the battery management system is initialised.
  \item When the signal stops, the ignition key can be turned. The direction of travel is selected with the forward/reverse switch. It is recommended to set the traction control switch to ``on''. The kart is now ready to drive.
\end{enumerate}

The left pedal is the brake, and the throttle pedal is on the right. The state of charge can be read on the battery gauge. If the state of charge becomes too low, a signal sounds and the kart must be recharged. If the kart is nevertheless used further, it switches off automatically.

If the throttle and the brake pedal are pressed at the same time while driving, the motor switches off. It can be used again once the throttle has been released for 2 seconds and the brake pedal is no longer pressed.

If a fault occurs during operation, the emergency stop switch must be pressed immediately. The hydraulic brakes remain fully functional. The emergency stop switch may only be released again after the fault has been eliminated.

\textbf{Note:} Immediately after charging, the kart should be driven gently and frequent braking with the motor brake should be avoided. This prevents overcharging of the cells by recuperation.

\section{Parking}\label{sec:parking}
The kart should only be parked on level ground. Parking consists of the following two steps:

\begin{enumerate}
  \item Switch off the ignition.
  \item Disconnect the motor controller.
\end{enumerate}

\section{Charger and Charging}\label{sec:charger-and-charging}
The kart must not be in operation during charging. The following steps are required for charging with the charger of the kart (Figure~\ref{fig:image66}):

\begin{enumerate}
  \item Release the emergency stop switch.
  \item Connect the Anderson connector.
  \item Connect the control connector.
  \item Connect the charger to the mains.
\end{enumerate}

A light on the charger shows the three possible states:

\begin{itemize}
  \item Red light: charging is in progress and the cells are not yet fully charged.
  \item Flashing green light: balancing of the individual cells is not yet completed.
  \item Green light: charging is completed.
\end{itemize}

\section{Storage}\label{sec:storage}
The kart must be stored in a building or garage, as it is not weatherproof. During storage, the kart can remain connected to the charger. If it is not connected to the charger, the cells must be charged once a month. The emergency stop switch must be released, so that the battery management system protects the cells against deep discharge.

\chapter{Conclusion and Outlook}\label{sec:conclusion-and-outlook}
Within one year, a petrol racing kart was converted into an electric kart with a \qty{10}{\kilo\watt} BLDC drive, a \qty{76.8}{\volt} \LFP{} battery with cell-level monitoring, a three-stage energy recovery and a traction control developed in-house. The kart combines ecological and economic sustainability with use in racing. The evaluation in Chapter~\ref{sec:results-and-discussion} condenses the calculations and measurements of the project into the following findings.

\textbf{Key findings.}

\begin{enumerate}
  \item The kart meets five of seven requirements. The calculated top speed is approx.\ \qty{65}{\kilo\metre\per\hour} (Section~\ref{sec:verification-against-the-requirements}).
  \item The driving time target of one hour is missed. At \qty{50}{\percent} motor load, \qty{55}{\minute} result with the full nominal energy and approx.\ \qty{44}{\minute} with the recommended \qty{80}{\percent}. At a constant \qty{60}{\kilo\metre\per\hour}, approx.\ \qty{59}{\minute} are possible (Section~\ref{sec:energy-demand-and-driving-time}).
  \item Up to \qty{60}{\kilo\metre\per\hour}, rolling resistance causes \qty{61}{\percent} to \qty{86}{\percent} of the driving resistance. The rolling resistance coefficient is therefore the most sensitive model parameter. Halving it lowers the demand at \qty{60}{\kilo\metre\per\hour} by approx.\ \qty{31}{\percent}.
  \item With rated torque, the drive force at standstill of approx.\ \qty{1650}{\newton} exceeds the transmissible traction force by a factor of 1.7 to 2.0. Up to approx.\ \qty{27}{\kilo\metre\per\hour} to \qty{43}{\kilo\metre\per\hour}, the kart is limited by traction and not by the motor. Traction control is therefore a necessary part of the drive concept (Section~\ref{sec:traction-limit-and-need-for-traction-con}).
  \item The realistic launch acceleration of approx.\ \qtyrange{3.8}{4.6}{\metre\per\second\squared} is set by tyre grip.
  \item Regenerative braking alone reaches approx.\ $0.45\,g$ at rated torque. Hard braking still needs the hydraulic brake, which confirms the three-stage braking concept (Section~\ref{sec:braking-and-energy-recovery-potential}).
  \item One braking event from \qty{60}{\kilo\metre\per\hour} to \qty{30}{\kilo\metre\per\hour} returns approx.\ \qty{4}{\watt\hour}. On a winding track, recuperation could cover up to about \qty{30}{\percent} of the energy demand (upper estimate).
  \item In the voltage plateau of \LFP{}, the cell voltage changes by only approx.\ \qty{0.7}{\milli\volt} to \qty{1.8}{\milli\volt} per percent state of charge, and the voltage relaxes by about \qty{10}{\milli\volt} during a pause. The state of charge can only be determined by current integration. Because of the flat plateau and the hysteresis, the starting state of charge cannot be read from the voltage. With the rated \qty{60}{\ampere\hour}, the charge of approx.\ \qty{44}{\ampere\hour} per cell corresponds to a start at approx.\ \qty{27}{\percent}, which is plausible for new cells. The measurement is consistent with the rated capacity, but does not determine it (Section~\ref{sec:battery-behaviour-during-initial-chargin}).
  \item Tyre grip limits the torque at the rear axle in traction to approx.\ \qty{100}{\newton\metre} to \qty{120}{\newton\metre}. The axle design load of \qty{330}{\newton\metre} is therefore conservative for traction loads. The static safety against yield is approx.\ 2.8 (Section~\ref{sec:mechanical-reserves-of-the-drivetrain}).
  \item With cool cells, high discharge currents caused a voltage sag that triggered the low-voltage alarm although the battery was not empty. The internal resistance of \LFP{} cells is strongly temperature-dependent, so a voltage-based low-voltage threshold has to be compensated for load and temperature (Section~\ref{sec:cell-temperature-and-voltage-sag-under-l}).
  \item The kart has no auxiliary \qty{12}{\volt} battery. The entire low-voltage system is supplied from the traction battery through an isolated DC/DC converter (Section~\ref{sec:low-voltage-supply-without-auxiliary-bat}).
  \item The battery of approx.\ \qty{55}{\kilo\gram}, sized for the one-hour target, makes up almost \qty{30}{\percent} of the vehicle mass. For the real duty cycle of a demonstration vehicle, a battery of half the capacity would have sufficed and would have saved approx.\ \qty{15}{\percent} of the total mass.
  \item The torque step of the electric motor from standstill destroyed the original pinion on the motor shaft. A static chain design with \qty{100}{\newton\metre} does not cover such shock loads. A hardened pinion and a torque-rate limitation in the firmware are required.
  \item Without the buffer of an auxiliary battery, the low-voltage rails showed voltage fluctuations, and the edges of the Hall sensor pulses coupled into the whole low-voltage system. Buffering and a compact, EMC-oriented layout of the DC/DC and filter stage are decisive, because smaller structures mean smaller loop areas and lower inductance.
\end{enumerate}

\textbf{Added value of the project.} The kart is a fully documented reference design of an electric racing kart in the \qty{10}{\kilo\watt} class, from the requirements through calculation and component selection to the firmware. In 2016, it brought lithium iron phosphate cells into a vehicle at a time when this chemistry was still uncommon in European passenger cars. Only from about 2020 did \LFP{} gain a large share of the electric vehicle market, and by 2022 it reached just under \qty{30}{\percent}~\cite{IEA23}. The project encountered two weaknesses of a purely voltage-based battery monitoring, the flat voltage plateau and the load- and temperature-dependent voltage sag. State-of-charge estimation with cell models addresses these problems today~\cite{Ple15}. The low-voltage supply without an auxiliary battery is a further simplification that is unusual in vehicle design. The hardware of the traction control can be built with standard components, and its know-how lies in the software. An implementation in other electric vehicles is therefore possible with comparatively little effort. Firmware, schematics, bills of material and calculations are publicly available~\cite{Sto26}, so that later student teams and other institutions can reproduce and extend the work. As a team, Johannes Stockhammer, Tom Rettenwander and Philipp Huber reached the final of the Austrian youth innovation competition Jugend Innovativ 2015/16 in the special category Sustainability and received a recognition award~\cite{JI16}. Approx.\ 1300 working hours are documented in the project diaries (Appendix~\ref{app:project-diaries}).

The environmental aspect goes beyond energy consumption. Non-exhaust emissions from tyre and brake wear are a relevant source of particulate matter~\cite{OECD20}. Regenerative braking reduces the use of the friction brake, and traction control limits wheel spin. Both measures therefore address non-exhaust emissions. For the kart, this effect remains a hypothesis until it is measured.

\textbf{Outlook.} The most important open point is the validation on the track. The following test programme is proposed:

\begin{enumerate}
  \item Data logging of wheel speeds, throttle signal, motor current and battery voltage. The microcontroller already evaluates the wheel speeds at \qty{1}{\kilo\hertz}.
  \item Launch tests from 0 to \qty{40}{\kilo\metre\per\hour} with and without traction control.
  \item Lap tests at a kart track with measurement of lap time, energy per lap and recovered energy per lap from the current integration of the BMS.
  \item A capacity test of the cells by a full discharge at a defined current, to verify the rated \qty{60}{\ampere\hour} and to obtain the open-circuit voltage curve of the cells actually used.
  \item Identification of the model parameters: rolling resistance by a coast-down test, rear axle load by weighing the kart with driver.
  \item Tuning of the slip threshold and the ramps of the traction control based on the recorded data.
  \item Determination of tyre wear by weighing the tyres before and after a defined distance.
  \item Oscilloscope measurements of the \qty{12}{\volt} and \qty{5}{\volt} rails and of the sensor lines, followed by a revised, compact DC/DC and filter board.
  \item Measurement of the chain force during launches with a strain gauge on the motor mount.
  \item A smaller and lighter battery sized for the measured duty cycle, combined with fast charging between sessions. \LFP{} cells tolerate high charging currents, so short breaks can be used to recharge instead of carrying the energy for a full hour. Since the cells accept charge better when warm, charging directly after driving is favourable. The charger and the BMS limits have to be matched to the higher currents.
\end{enumerate}

New technologies that are first used in motor racing often find their way into series production vehicles later on. With the proposed measurements, the kart can make a small but measurable contribution to this.

\chapter{Personal Reflections}\label{sec:personal-reflections}
In this chapter, each of us describes his personal motivation for the diploma project and briefly reflects on his experiences and on the cooperation in the team. During the project year, the initially empty kart chassis was given a new appearance, and a working electric kart was created.

With this project, we have contributed to a long-standing project of HTL Braunau. We hope that future classes of HTL Braunau will realise further innovative ideas on and with the kart. We are firmly convinced that our kart can inspire future students for technology.

\section{Johannes Stockhammer}\label{sec:johannes-stockhammer}
I was fascinated by the possibilities of electric mobility long before the diploma project. I had already tried many racing vehicles, but none of them was the perfect vehicle for me. Towards the end of the fourth year, our teacher in the subject AT1 asked me whether I would like to build an electric kart as the showcase project of HTL Braunau the following year. I could hardly believe my luck. Now I had the chance to design a kart that was ``perfect'' for me, together with my project partners.

Working together with my project partners taught me above all to respect other points of view and to find compromises. I also learned that it is sometimes better to let a matter rest and start again later. Through my contacts in motor racing, I took on the role of contact person for companies right from the start. In technical discussions, I was often able to contribute my knowledge to the team. Our ambitious goals for motor power, torque and efficiency repeatedly led to problems that were not easy to solve.

I also learned to keep proper records right from the beginning. This saves a lot of time and trouble later on. The time frame for completing the project is short anyway and requires careful planning. For me, project work therefore often did not end with the last lesson of the day, but continued long after. I never really got used to the stress towards the end, when everything has to be finished at once.

I have many positive memories of our joint work on the final project, which I will not forget so soon.

Suddenly, one realises how quickly five years at HTL can pass.

\section{Tom Rettenwander}\label{sec:tom-rettenwander}
One of my hobbies is riding motorcycles, so this diploma project dealt with a familiar kind of transport for me. My motorcycle is powered by a combustion engine. The electric drive and the energy supply therefore changed the picture of a vehicle I was used to. The fact that I could contribute to a technology of the future with this project was the main reason for me to choose it.

Most of the time, the work in the project team was good teamwork. At the beginning, we had to discard many ideas and find compromises. Whenever mechanical problems occurred, they were solved quickly with the help of a project partner. The joint and independent work made the lessons much livelier than normal lessons.

During the different tasks, it was important to me to stay up to date with my project partners. It often helped me to set a daily goal and to work towards it in a focused way. A big challenge for me was to find a suitable place for all components on the limited area of the kart frame.

Thanks to the many different areas covered by our diploma project, I was able to learn a lot and gain new experience. We often stayed longer at HTL to finish the work we had started. The time pressure caused by the shortened final year and partly long delivery times, together with studying for the other subjects and the Matura alongside the project, made it a stressful time for me.

I am already looking forward to the day when all work on our kart is finished.

\section{Philipp Huber}\label{sec:philipp-huber}
When I first heard about the project, it was clear to me that I could identify with it well. I often feel more at home on wheels than on my own feet. Even before it was announced that such diploma projects would be offered, I had made plans to build my own kart. The biggest obstacle had always been the lack of a team, because I knew I would not manage it alone.

The project team was fully committed right from the start, and work began immediately with great enthusiasm. When a new team comes together, however, small problems are the order of the day. After a few weeks, these problems were reduced to a minimum and soon stopped altogether. We all had to learn to respect the others' opinions and ideas about the kart. Finding compromises was never a problem in the team.

Keeping my team partners up to date the whole time was very easy for me. Thanks to a daily plan that we drew up at the beginning of our work, it was never a problem to bring the others up to date.

I think a project like ours requires a lot of know-how and knowledge of the subject. Planning and dimensioning all components took a lot of time and nerves, all the more so as this was the first team project I had ever worked on. Thanks to the comprehensive education at HTL, however, almost all necessary areas of knowledge could be covered. I have to say that I learned a great deal from my team partners.

Despite the rather short period in which the kart had to be completed, we never worked under stress. Instead, we invested many hours of our free time. Given all the hard work our team put into the project, I am sure that it can be completed in the remaining time.

\appendix
\chapter{Technical Data of the Components}\label{app:technical-data}
The following tables list the data of the cells and the battery management system (Table~\ref{tab:cells} and Table~\ref{tab:bms}), the data of the motor and the motor controller (Table~\ref{tab:motor} and Table~\ref{tab:controller}), and the vehicle data used in the drive calculation of Section~\ref{sec:electric-drive} (Table~\ref{tab:vehdata}). Unless stated otherwise, the values are taken from the manufacturers' datasheets.

\begin{table}[!ht]
\centering
\caption{Technical data of the \LFP{} cells according to the manufacturer~\cite{Win15}}\label{tab:cells}
{\small\renewcommand{\arraystretch}{1.15}%
\begin{tabularx}{\textwidth}{@{}>{\raggedright\arraybackslash}X>{\centering\arraybackslash}p{2.5cm}@{\hspace{1.5em}}>{\raggedright\arraybackslash}X>{\centering\arraybackslash}p{2.5cm}@{}}
\toprule
\textbf{Parameter} & \textbf{Value} & \textbf{Parameter} & \textbf{Value} \\
\midrule
Manufacturer & Winston & Maximum discharge current (max.\ \qty{15}{\minute}) in \unit{\ampere} & 180 \\
Type & WB-LYP60AHA\newline (\LFP{}) & Short-term peak discharge current (\qty{5}{\second} per minute) in \unit{\ampere} & 1200 \\
Nominal voltage in \unit{\volt} & 3.2 & Temperature range during charging/discharging in \unit{\degreeCelsius} & $-35$ to $+85$ \\
Lower voltage limit in \unit{\volt} & 2.8 & Housing temperature resistance in \unit{\degreeCelsius} & 200 \\
Maximum voltage in \unit{\volt} & 4.0 & Weight in \unit{\kilo\gram} & 2.3 \\
Nominal capacity in \unit{\ampere\hour} & 60 & Terminal thread & M6 \\
Cycles at \qty{80}{\percent} DOD & $>3000$ & Housing material & plastic \\
Cycles at \qty{70}{\percent} DOD & $>5000$ & Dimensions in \unit{\milli\metre} & $203\times114\times61$ \\
Optimal discharge current in \unit{\ampere} & 30 & Initial charging required & yes \\
\bottomrule
\end{tabularx}}
\medskip
\caption{Technical data of the battery management system~\cite{BMS15}}\label{tab:bms}
{\small\renewcommand{\arraystretch}{1.15}%
\begin{tabularx}{\textwidth}{@{}>{\raggedright\arraybackslash}X>{\centering\arraybackslash}p{2.5cm}@{\hspace{1.5em}}>{\raggedright\arraybackslash}X>{\centering\arraybackslash}p{2.5cm}@{}}
\toprule
\textbf{Parameter} & \textbf{Value} & \textbf{Parameter} & \textbf{Value} \\
\midrule
Manufacturer & 123electric & Quiescent current of one cell board in \unit{\micro\ampere} & $<100$ \\
Type & 123\textbackslash BMS & Current sensor range in \unit{\ampere} & 100, 200 or 400 \\
Supply voltage of BMS controller in \unit{\volt} & 8 to 30 & ADC resolution in bit & 10 \\
Controller quiescent current incl.\ current sensor in \unit{\ampere} & $<0.007$ & Adjustable cell voltage (\qty{0.01}{\volt} steps) in \unit{\volt} & 2 to 5 \\
Number of cells & 2 to 255 & Adjustable cell temperature (\qty{1}{\degreeCelsius} steps) in \unit{\degreeCelsius} & $-40$ to $+99$ \\
Balancing current in \unit{\ampere} & 1 & Acquisition time per cell in \unit{\second} & 0.015 \\
\bottomrule
\end{tabularx}}
\end{table}

\begin{table}[!ht]
\small\renewcommand{\arraystretch}{1.15}%
\begin{minipage}[t]{0.485\textwidth}
\caption{Technical data of the motor HPM-10kW~\cite{GM15}}\label{tab:motor}
\begin{tabular*}{\linewidth}{@{}l@{\extracolsep{\fill}}c@{}}
\toprule
\textbf{Parameter} & \textbf{Value} \\
\midrule
Manufacturer & Golden Motor \\
Type & HPM-10kW \\
Rated power in \unit{\kilo\watt} & 10 \\
Rated voltage in \unit{\volt} & 72 \\
Rated current in \unit{\ampere} & 140 \\
Short-term peak power in \unit{\kilo\watt} & 22 \\
Short-term peak current in \unit{\ampere} & 300 \\
Rated speed in \unit{\per\minute} & 4500 \\
Maximum speed in \unit{\per\minute} & 6000 \\
Cooling & air \\
Maintenance-free & yes \\
Weight in \unit{\kilo\gram} & 17 \\
Stator housing material & aluminium \\
Winding inductance in \unit{\micro\henry} & 77 \\
Winding resistance in \unit{\milli\ohm} & 6 \\
Dimensions in \unit{\milli\metre} & \O\,$206\times217$ \\
Rotor inertia & low \\
Power density in \unit{\kilo\watt\per\kilo\gram} & 0.59 \\
Rated efficiency & 0.91 \\
\bottomrule
\end{tabular*}
\end{minipage}\hfill
\begin{minipage}[t]{0.485\textwidth}
\caption{Technical data of the motor controller VEC500-72~\cite{GM14}}\label{tab:controller}
\begin{tabular*}{\linewidth}{@{}l@{\extracolsep{\fill}}c@{}}
\toprule
\textbf{Parameter} & \textbf{Value} \\
\midrule
Manufacturer & Golden Motor \\
Type & VEC500-72 \\
Rated power in \unit{\kilo\watt} & 10 \\
Rated voltage in \unit{\volt} & 72 \\
Rated current in \unit{\ampere} & 150 \\
Short-term peak power in \unit{\kilo\watt} & 20 \\
Maximum phase current in \unit{\ampere} & 500 \\
Control method & field-oriented \\
 & (vector) control \\
Cooling & air \\
Maintenance-free & yes \\
Weight in \unit{\kilo\gram} & 3.2 \\
Housing material & plastic, aluminium \\
Dimensions in \unit{\milli\metre} & $200\times190\times58$ \\
Rated efficiency & $>0.90$ \\
 & \\
 & \\
 & \\
 & \\
\bottomrule
\end{tabular*}
\end{minipage}
\end{table}

\begin{table}[!ht]
\centering
\caption{Vehicle data used for the drive calculation}\label{tab:vehdata}
{\small\renewcommand{\arraystretch}{1.15}%
\begin{tabular}{@{}lc@{}}
\toprule
\textbf{Parameter} & \textbf{Value} \\
\midrule
Drag coefficient $c_{\mathrm{w}}$ & 0.65 \\
Frontal area in \unit{\square\metre} & 0.7 \\
Vehicle mass incl.\ driver in \unit{\kilo\gram} & 180 \\
Rolling resistance coefficient $\mu_{\mathrm{R}}$ & 0.07 \\
Static friction coefficient $\mu_{\mathrm{H}}$ & 0.9 \\
Air density in \unit{\kilo\gram\per\cubic\metre} & 1.23 \\
Wheel diameter in \unit{\metre} & 0.24 \\
Drive efficiency & 0.9 \\
Motor power in \unit{\watt} & \num{10000} \\
Rated speed in \unit{\per\minute} & 4000 \\
Rated torque in \unit{\newton\metre} & 30 \\
Gear ratio & 3.2 \\
Rolling resistance force in \unit{\newton} & 123.6 \\
\bottomrule
\end{tabular}}
\end{table}

\chapter{Project Diaries}\label{app:project-diaries}
Due to the shortened final year at HTL of just under seven months, the project time was limited. Nevertheless, many hours were invested in the diploma project. Since each team member worked on different tasks, each kept his own project diary. The diaries (Table~\ref{tab:d1} to Table~\ref{tab:d3}) show the activities and the working time spent. They are reproduced as recorded in 2015/16; obvious typing errors in the dates have been corrected.

\subsection*{Johannes Stockhammer}
{\small\renewcommand{\arraystretch}{1.1}%
\begin{longtable}{@{}p{1.9cm}>{\raggedright\arraybackslash}p{\dimexpr\textwidth-3.2cm-4\tabcolsep\relax}>{\centering\arraybackslash}p{1.3cm}@{}}
\caption{Project diary of Johannes Stockhammer}\label{tab:d1}\\
\toprule
\textbf{Date} & \textbf{Activity} & \textbf{Hours} \\
\midrule
\endfirsthead
\toprule
\textbf{Date} & \textbf{Activity} & \textbf{Hours} \\
\midrule
\endhead
\bottomrule
\endlastfoot
2015-08-03 & Reviewing products available on the market & 3 \\
2015-09-04 & Overview of possible manufacturers and suppliers & 1 \\
2015-09-05 & Checking which battery types are suitable for the project & 3 \\
2015-09-11 & Technical support by Harald Bartol & 2 \\
2015-09-18 & Introduction to project lessons / familiarisation with laboratories and workshops & 0 \\
2015-09-28 & Defining the project goals & 1 \\
2015-09-29 & Overview of available technologies & 0 \\
2015-10-02 & Project planning & 3 \\
2015-10-05 & Dimensioning the required motor power & 0 \\
2015-10-06 & Determining the required drive components & 0 \\
2015-10-09 & Selecting the frame (weight, stability, etc.) & 3 \\
2015-10-12 & Trip to Ampfing / technical discussion at Kartshop Ampfing / selecting the frame & 6 \\
2015-10-13 & Dimensioning the battery and the electrical drive components & 6 \\
2015-10-16 & Requesting information on further components / chassis / meeting with Mr Dirnberger & 2 \\
2015-10-19 & E-mail correspondence with companies / collecting the chassis in Ampfing / disassembly & 2 \\
2015-10-20 & Revising the component list & 1 \\
2015-10-23 & Searching for European Golden Motor distributors / comparing electrical components & 2 \\
2015-10-27 & Internet research on the service life of electrical components and batteries & 3 \\
2015-10-30 & Creating the project poster / discussing competition entries & 2 \\
2015-11-03 & Contacting Faktor.de / selecting the battery management system & 3 \\
2015-11-06 & Ordering the BMS, cells and accessories from Faktor.de & 4 \\
2015-11-09 & Finding ways to charge the cells / providing a sufficiently powerful power supply & 7 \\
2015-11-10 & Receiving the delivery from Faktor.de / checking for completeness & 4 \\
2015-11-13 & Setting up the charging environment / documentation & 2 \\
2015-11-16 & Charging the cells / dimensioning the mechanical drive & 7 \\
2015-11-20 & Charging the cells / technical support by Kartshop Ampfing & 4 \\
2015-11-23 & Charging the cells / trip to Ampfing / dimensioning the drive components / buying sprockets and chain & 8 \\
2015-11-24 & Searching for a suitable seat for the frame & 4 \\
2015-11-27 & Research on safety systems for racing karts & 8 \\
2015-11-30 & Project meeting on implementing traction control & 7 \\
2015-12-01 & Preparing for the open day & 6 \\
2015-12-04 & Open day / project presentation & 10 \\
2015-12-14 & Drawing the battery boxes / implementing the motor mount & 9 \\
2015-12-15 & Comparing alternative suppliers to Smartemotions & 5 \\
2015-12-18 & Order from Smartemotions Germany & 4 \\
2015-12-21 & Planning the division of work for the thesis & 6 \\
2015-12-22 & Setting up a shared working environment for the thesis document & 4 \\
2015-12-27 & Writing the thesis & 9 \\
2015-12-28 & Writing the thesis & 12 \\
2015-12-29 & Writing the thesis & 10 \\
2015-12-30 & Writing the thesis & 7 \\
2016-01-02 & Writing the thesis & 9 \\
2016-01-03 & Writing the thesis & 8 \\
2016-01-04 & Writing the thesis & 3 \\
2016-01-05 & Work on the kart frame & 13 \\
2016-01-08 & Battery boxes / electrical planning & 23 \\
2016-01-11 & Drawing the sensor ring & 4 \\
2016-01-12 & Cleaning the kart frame & 7 \\
2016-01-15 & Manufacturing the sensor ring / discussion with teacher & 4 \\
2016-01-18 & Removing the rear axle / greasing the bearings / fitting the display holder to the steering wheel & 8 \\
2016-01-19 & Component selection for the traction control & 10 \\
2016-01-22 & Making the electronics housing / battery box lid & 7 \\
2016-01-24 & Drawing the sensor rings in Pro/ENGINEER & 15 \\
2016-01-25 & Project folder / sorting documents / selecting a cable supplier / correspondence with Faktor.de & 3 \\
2016-01-26 & Project folder / derusting the kart frame & 6 \\
2016-01-29 & Submitting the project folder / priming the kart frame / painting the sensor rings & 25 \\
2016-02-01 & Painting and reassembling the kart frame & 10 \\
2016-02-02 & Completing assembly of the kart frame / mounting the sensor rings / painting the rims & 4 \\
2016-02-05 & Planning the traction control components / preparing the order list & 38 \\
2016-02-08 & Mounting the rims / selecting the voltage regulator for the \qty{12}{\volt} control voltage & 6 \\
2016-02-09 & Revising the order list & 1 \\
2016-02-12 & Information afternoon / drawing the traction control boards & 10 \\
2016-02-13 & Developing the traction control boards & 8 \\
2016-02-14 & Designing the traction control boards and revising them for EMC / ordering components & 12 \\
2016-02-15 & Project folder / documentation / display holder & 8 \\
2016-02-16 & Work on the kart in the HTL workshop / removing the steering wheel / making the speed sensor holders & 10 \\
2016-02-17 & Drawing the display cover and the speed sensor holders & 5 \\
2016-02-18 & Exposing, developing and etching the boards / quality check & 5 \\
2016-02-19 & Cutting and lacquering the boards / collecting ordered parts in Simbach & 6 \\
2016-02-20 & Soldering components and mounting the boards / testing the boards / creating the program structure & 8 \\
2016-02-21 & Writing the project report for Jugend Innovativ / work on the traction control & 6 \\
2016-02-22 & Ordering the speed sensors / soldering the capacitors / mounting the boards & 9 \\
2016-02-27 & Writing the project report for Jugend Innovativ & 6 \\
2016-02-28 & Writing the project report for Jugend Innovativ / proofreading the partners' sections & 5 \\
2016-03-02 & Mounting the sensors, work on the project presentation & 3 \\
2016-03-07 & Electrical test setup of the BMS / testing the motor & 18 \\
2016-03-08 & Project presentations & 4 \\
2016-03-11 & Programming the BMS and ordering & 2 \\
2016-03-12 & Creating the circuit diagram of the control and power electronics & 5 \\
2016-03-13 & Orders from Conrad and RS Components, alarm function for battery monitoring & 5 \\
2016-03-14 & Control electronics, drive, ordering & 9 \\
2016-03-15 & Mounting the boards, painting the hinges & 3 \\
2016-03-21 & Mounting the throttle holder and the dashboard bracket & 10 \\
2016-03-22 & Wiring, mounting the electronics & 35 \\
2016-03-23 & Connecting the electronics / mounting the connectors and the motor controller & 15 \\
2016-03-24 & Programming the microcontroller / wiring harness of the control electronics & 13 \\
2016-03-25 & Completing the wiring / testing the motor / adjusting the BMS / writing the thesis & 14 \\
2016-03-26 & Trip to Ampfing, mounting the chain, setting motor parameters & 7 \\
2016-03-27 & Drawing the chain guard and dashboard / trip to Ampfing / electrical wiring & 13 \\
2016-03-28 & Writing the thesis & 9 \\
2016-03-29 & Work on the kart & 12 \\
2016-04-04 & Writing the thesis & 6 \\
2016-04-05 & Writing the thesis / project folder & 4 \\
\midrule
 & Total & 669 \\
\end{longtable}}

\subsection*{Philipp Huber}
{\small\renewcommand{\arraystretch}{1.1}%
\begin{longtable}{@{}p{1.9cm}>{\raggedright\arraybackslash}p{\dimexpr\textwidth-3.2cm-4\tabcolsep\relax}>{\centering\arraybackslash}p{1.3cm}@{}}
\caption{Project diary of Philipp Huber}\label{tab:d2}\\
\toprule
\textbf{Date} & \textbf{Activity} & \textbf{Hours} \\
\midrule
\endfirsthead
\toprule
\textbf{Date} & \textbf{Activity} & \textbf{Hours} \\
\midrule
\endhead
\bottomrule
\endlastfoot
2015-09-21 & Introduction to project lessons, familiarisation with laboratories and workshops & 2 \\
2015-09-21 & General internet research & 8 \\
2015-09-22 & Internet research on kart frames & 6 \\
2015-09-25 & Internet research on motors & 2 \\
2015-09-28 & Project definition, diploma project registration & 7 \\
2015-09-29 & Internet research on motors & 4 \\
2015-10-02 & Internet research on batteries & 3 \\
2015-10-05 & Internet research on batteries, creating the schedule & 6 \\
2015-10-06 & Component list, meeting & 4 \\
2015-10-09 & Comparing components & 2 \\
2015-10-13 & Second component list & 4 \\
2015-10-16 & Project meeting & 4 \\
2015-10-19 & Removing the kart fairing and seat & 6 \\
2015-10-20 & Collecting the chassis, disassembly, cleaning & 6 \\
2015-10-23 & Component layout on the kart & 2 \\
2015-10-30 & Planning & 6 \\
2015-11-03 & Creating the project poster & 4 \\
2015-11-06 & Ordering the cells & 4 \\
2015-11-09 & Setting up a way to charge the cells & 6 \\
2015-11-10 & Battery delivery received & 4 \\
2015-11-13 & Preparing battery charging and finding a place for it & 2 \\
2015-11-16 & Charging the cells, kart floor panel & 6 \\
2015-11-17 & Charging the cells, kart floor panel & 4 \\
2015-11-20 & Moving the cells to the basement, charging & 4 \\
2015-11-23 & Design work in AutoCAD, measuring the kart, trip to Ampfing, buying sprockets and chain & 8 \\
2015-11-24 & Design work in AutoCAD & 5 \\
2015-11-27 & Revising and printing the poster & 4 \\
2015-11-30 & Design work in AutoCAD & 6 \\
2015-12-01 & Preparing for the open day & 4 \\
2015-12-04 & Open day & 8 \\
2015-12-14 & Planning the voltage regulation & 8 \\
2015-12-15 & Preparing the 123\textbackslash BMS for mounting on the cells & 4 \\
2015-12-18 & Making ground cables for the 123\textbackslash BMS & 2 \\
2015-12-21 & Soldering the BMS, final version for Jugend Innovativ & 6 \\
2016-01-02 & Writing the thesis & 2 \\
2016-01-05 & Planning the programming & 8 \\
2016-01-08 & Making special screws for the 123\textbackslash BMS & 5 \\
2016-01-11 & Soldering bus connections of the 123\textbackslash BMS & 6 \\
2016-01-15 & Manufacturing the rear battery box & 2 \\
2016-01-18 & Manufacturing a battery box & 6 \\
2016-01-19 & Planning and drawing the display holder plate & 4 \\
2016-01-25 & Making the display holder & 7 \\
2016-01-26 & Manufacturing the electronics box & 4 \\
2016-01-29 & Derusting and priming the kart frame & 2 \\
2016-02-01 & Painting and reassembling the kart & 6 \\
2016-02-22 & Mounting the sensor rings on the kart & 6 \\
2016-02-23 & Drilling and tapping the tube clamps & 4 \\
2016-02-26 & Drilling and tapping the tube clamps & 4 \\
2016-02-29 & Trial assembly and fitting of tube clamps, mounting blocks and battery boxes & 7 \\
2016-03-02 & Creating the project presentation & 2 \\
2016-03-04 & Preparing the battery boxes, mounting blocks and electronics box for coating & 2 \\
2016-03-05 & Revising the project presentation & 2 \\
2016-03-06 & Writing the thesis & 2 \\
2016-03-07 & Cleaning and coating the battery boxes, mounting blocks and electronics box & 4 \\
2016-03-08 & Project presentation & 4 \\
2016-03-11 & Coating the battery boxes, mounting blocks and electronics box & 5 \\
2016-03-14 & Cutting foam and gluing it into the battery boxes & 6 \\
2016-03-15 & Fitting the cells into the battery boxes & 4 \\
2016-03-18 & Mounting the battery boxes and the electronics box & 4 \\
2016-03-23 & Writing the thesis & 5 \\
2016-03-24 & Writing the thesis & 3 \\
2016-03-29 & Writing the thesis & 4 \\
2016-04-01 & Writing the thesis & 8 \\
2016-04-02 & Writing the thesis & 2 \\
2016-04-03 & Writing the thesis & 3 \\
\midrule
 & Total & 294 \\
\end{longtable}}

\subsection*{Tom Rettenwander}
{\small\renewcommand{\arraystretch}{1.1}%
\begin{longtable}{@{}p{1.9cm}>{\raggedright\arraybackslash}p{\dimexpr\textwidth-3.2cm-4\tabcolsep\relax}>{\centering\arraybackslash}p{1.3cm}@{}}
\caption{Project diary of Tom Rettenwander}\label{tab:d3}\\
\toprule
\textbf{Date} & \textbf{Activity} & \textbf{Hours} \\
\midrule
\endfirsthead
\toprule
\textbf{Date} & \textbf{Activity} & \textbf{Hours} \\
\midrule
\endhead
\bottomrule
\endlastfoot
2015-09-21 & Introduction to project lessons, familiarisation with laboratories and workshops & 2 \\
2015-09-21 & General internet research & 6 \\
2015-09-22 & Internet research on kart frames & 4 \\
2015-09-25 & Internet research on motors & 2 \\
2015-09-28 & Project definition, diploma project registration & 6 \\
2015-09-29 & Internet research on motors & 4 \\
2015-10-02 & Internet research on batteries, viewing a kart in Markl & 7 \\
2015-10-05 & Internet research on batteries, creating the schedule & 6 \\
2015-10-06 & Component list & 4 \\
2015-10-09 & Comparing components & 3 \\
2015-10-12 & Viewing a kart in Ampfing & 3 \\
2015-10-13 & Second component list & 4 \\
2015-10-16 & Creating the work breakdown structure, project meeting & 4 \\
2015-10-19 & Removing the kart fairing and seat & 6 \\
2015-10-20 & Disassembly and cleaning & 4 \\
2015-10-23 & Component layout on the kart & 2 \\
2015-10-30 & Planning, creating the schedule & 6 \\
2015-11-03 & Creating the project poster & 4 \\
2015-11-06 & Ordering the cells & 4 \\
2015-11-09 & Obtaining a power supply for charging the cells & 6 \\
2015-11-10 & Battery delivery received & 4 \\
2015-11-13 & Making a protective cover for the cells & 2 \\
2015-11-16 & Charging the cells, painting the floor panel & 6 \\
2015-11-17 & Charging the cells, maintenance of the power supply unit & 4 \\
2015-11-20 & Moving the cells to the basement, charging & 4 \\
2015-11-23 & Measuring the kart, design work in AutoCAD & 6 \\
2015-11-24 & Design work in AutoCAD & 5 \\
2015-11-27 & Revising and printing the poster & 4 \\
2015-11-30 & Design work in AutoCAD & 7 \\
2015-12-01 & Preparing for the open day & 6 \\
2015-12-04 & Open day & 9 \\
2015-12-14 & Comparing motor variants & 6 \\
2015-12-15 & Revising the component layout on the kart & 4 \\
2015-12-18 & Preparing the Jugend Innovativ entry & 6 \\
2015-12-20 & Preparing the Jugend Innovativ entry & 3 \\
2015-12-21 & Soldering the BMS, final version for Jugend Innovativ & 8 \\
2015-12-22 & Special screws for the BMS & 4 \\
2015-12-29 & Writing the thesis & 2 \\
2015-12-30 & Writing the thesis & 3 \\
2016-01-04 & Writing the thesis & 2 \\
2016-01-07 & Design work in AutoCAD & 2 \\
2016-01-08 & Design work in AutoCAD & 5 \\
2016-01-11 & Bending and welding the right battery box & 6 \\
2016-01-12 & Bending the left and rear battery boxes & 4 \\
2016-01-15 & Welding the left and rear battery boxes & 5 \\
2016-01-18 & Measuring the battery boxes, correcting AutoCAD drawings, bending battery box lids & 6 \\
2016-01-19 & Bending the battery box lids, designing the electronics box & 4 \\
2016-01-22 & Planning the headrest and battery box mounting & 4 \\
2016-01-25 & Buying hinge and latch for the battery box, measuring, drilling the battery box & 7 \\
2016-01-26 & Filing the battery box lids and fitting the latches & 4 \\
2016-01-29 & Punching the latch hooks & 2 \\
2016-02-01 & Bending and drilling the latch hooks, fitting the latches & 6 \\
2016-02-12 & 3D drawing for the CNC milling machine, milling the tube clamps & 8 \\
2016-02-22 & 3D drawing for the CNC milling machine, milling the tube clamps & 6 \\
2016-02-23 & Cutting the mounting blocks to size and drilling & 4 \\
2016-02-26 & Drilling and tapping the mounting blocks & 4 \\
2016-02-29 & Trial assembly and fitting of tube clamps, mounting blocks and battery boxes & 7 \\
2016-03-01 & Trial fitting of the boxes & 4 \\
2016-03-02 & Creating the project presentation & 2 \\
2016-03-04 & Preparing the battery boxes, mounting blocks and electronics box for coating & 2 \\
2016-03-05 & Revising the project presentation & 2 \\
2016-03-06 & Writing the thesis & 2 \\
2016-03-07 & Cleaning and coating the battery boxes, mounting blocks and electronics box, cleaning the kart & 7 \\
2016-03-08 & Project presentation & 4 \\
2016-03-11 & Coating the battery boxes, mounting blocks and electronics box & 5 \\
2016-03-14 & Cutting foam and gluing it into the battery boxes & 6 \\
2016-03-15 & Fitting the cells into the battery boxes & 4 \\
2016-03-18 & Mounting the battery boxes and the electronics box & 4 \\
2016-03-22 & Writing the thesis & 2 \\
2016-03-23 & Writing the thesis & 3 \\
2016-03-24 & Writing the thesis & 2 \\
2016-03-29 & Writing the thesis & 4 \\
2016-03-31 & Writing the thesis & 2 \\
2016-04-01 & Writing the thesis & 7 \\
2016-04-02 & Writing the thesis & 6 \\
2016-04-03 & Writing the thesis & 6 \\
2016-04-04 & Writing the operating manual & 8 \\
2016-04-05 & Adding missing documents to the thesis & 5 \\
\midrule
 & Total & 353 \\
\end{longtable}}

\chapter{Work Breakdown Structure}\label{app:work-breakdown-structure}
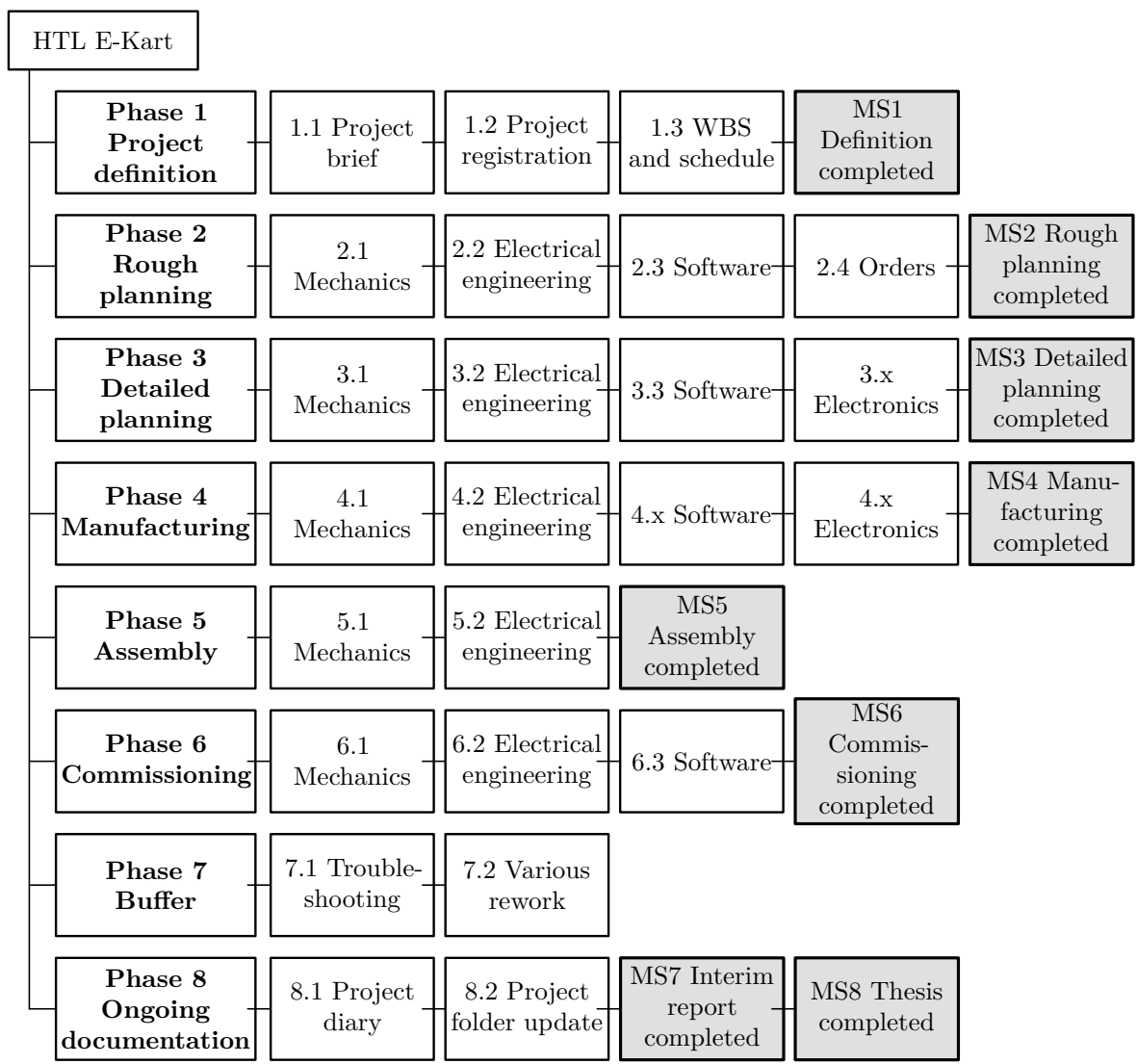
\begin{figure}[!ht]
\centering
\begin{tikzpicture}[schem, every node/.append style={execute at begin node={\hyphenpenalty=10000\exhyphenpenalty=10000}},
  phase/.style={block, text width=2.6cm, minimum height=1.4cm, font=\small\bfseries, inner sep=2pt, anchor=west, fill=white},
  wp/.style={block, text width=2.1cm, minimum height=1.4cm, inner sep=2pt, align=center, anchor=west, fill=white},
  ms/.style={wp, fill=black!12, line width=1.2pt}]
  \node[block, minimum width=2.6cm, minimum height=0.8cm, anchor=west] (root) at (0,0.2) {HTL E-Kart};
  \draw (0.3,-0.2) -- (0.3,-13.1);
  \foreach \k/\p/\items in {
      0/{Phase 1\\ Project definition}/{1.1 Project brief, 1.2 Project registration, 1.3 WBS and schedule, *MS1 Definition completed},
      1/{Phase 2\\ Rough planning}/{2.1 Mechanics, 2.2 Electrical engineering, 2.3 Software, 2.4 Orders, *MS2 Rough planning completed},
      2/{Phase 3\\ Detailed planning}/{3.1 Mechanics, 3.2 Electrical engineering, 3.3 Software, 3.x Electronics, *MS3 Detailed planning completed},
      3/{Phase 4\\ Manufacturing}/{4.1 Mechanics, 4.2 Electrical engineering, 4.x Software, 4.x Electronics, *MS4 Manu-\\facturing completed},
      4/{Phase 5\\ Assembly}/{5.1 Mechanics, 5.2 Electrical engineering, *MS5 Assembly completed},
      5/{Phase 6\\ Commissioning}/{6.1 Mechanics, 6.2 Electrical engineering, 6.3 Software, *MS6 Commis-\\sioning completed},
      6/{Phase 7\\ Buffer}/{7.1 Trouble-\\shooting, 7.2 Various rework},
      7/{Phase 8\\ Ongoing documentation}/{8.1 Project diary, 8.2 Project folder update, *MS7 Interim report completed, *MS8 Thesis completed}} {
    \pgfmathsetmacro{\y}{-1.2-\k*1.7}
    \draw (0.3,\y) -- (0.65,\y);
    \node[phase] (p) at (0.65,\y) {\p};
    \foreach \it [count=\j] in \items {
      \pgfmathsetmacro{\x}{3.6+(\j-1)*2.4}
      \draw (\x-0.3,\y) -- (\x,\y);
      \IfBeginWith{\it}{*}{\StrGobbleLeft{\it}{1}[\txt]\node[ms] at (\x,\y) {\txt};}
        {\node[wp] at (\x,\y) {\it};}}
  }
  \node[anchor=north west] at (0,-13.9) {MS = milestone (shaded)};
\end{tikzpicture}
\caption{Work breakdown structure of the project (status 16 October 2015)\nocc}\label{fig:wbs}
\end{figure}

In the work breakdown structure, the diploma project was divided into phases and work packages (Figure~\ref{fig:wbs}).

\chapter{Block Diagrams and Circuit Diagrams}\label{app:block-diagrams-and-circuit-diagrams}
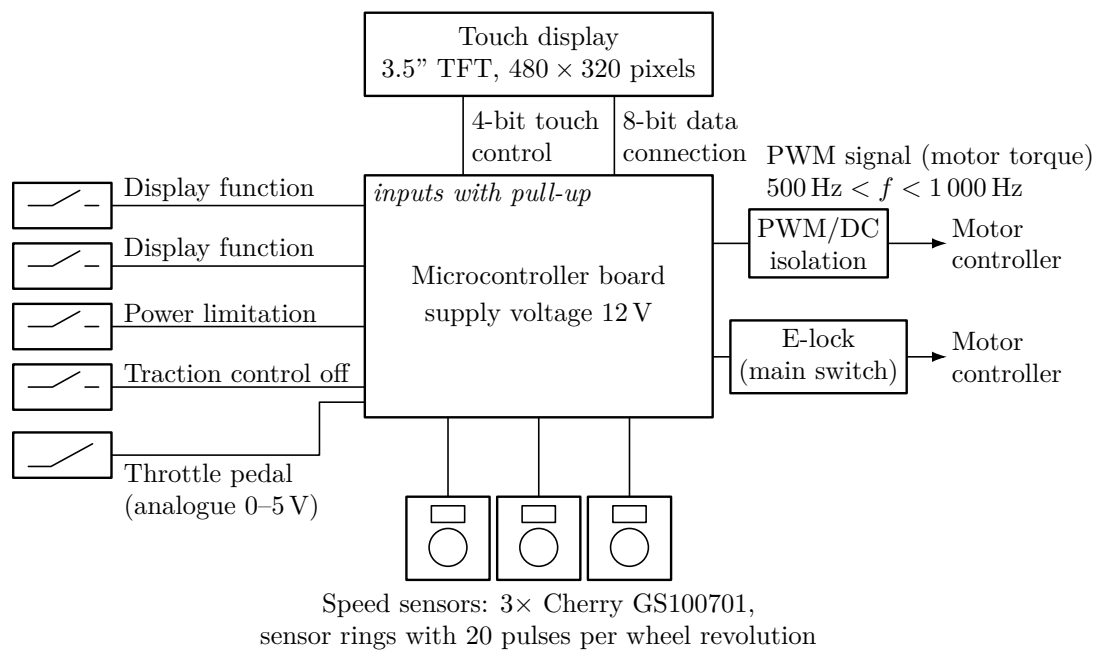
\begin{figure}[!ht]
\centering
\begin{tikzpicture}[schem,
  sw/.style={draw, part, minimum width=1.3cm, minimum height=0.6cm, inner sep=0pt,
    path picture={\draw[line width=0.6pt] ([xshift=-0.45cm]path picture bounding box.center) -- ++(0.32,0) -- ++(0.36,0.2)
      ([xshift=0.3cm]path picture bounding box.center) -- ++(0.18,0);}},
  sensor/.style={draw, part, minimum width=1.1cm, minimum height=1.1cm, inner sep=0pt,
    path picture={\draw[line width=0.6pt] ([yshift=0.32cm]path picture bounding box.center) +(-0.22,-0.11) rectangle +(0.22,0.11)
      ([yshift=-0.12cm]path picture bounding box.center) circle (0.25);}}]
  \node[block, minimum width=4.6cm, minimum height=3.2cm] (uc) at (7.5,3.0) {Microcontroller board\\[2pt] supply voltage \qty{12}{\volt}};
  \node[anchor=north west, font=\small\itshape, inner sep=3pt] at (uc.north west) {inputs with pull-up};
  \node[block, minimum width=4.6cm, minimum height=1.1cm] (disp) at (7.5,6.2) {Touch display\\ 3.5'' TFT, $480\times320$ pixels};
  \draw (6.5,5.65) -- (6.5,4.6);
  \draw (8.5,5.65) -- (8.5,4.6);
  \node[lbl, right] at (6.55,5.12) {4-bit touch\\ control};
  \node[lbl, right] at (8.55,5.12) {8-bit data\\ connection};
  \foreach \y/\t in {4.2/Display function, 3.4/Display function, 2.6/Power limitation, 1.8/Traction control off} {
    \node[sw] at (1.2,\y) {};
    \draw (1.85,\y) -- (5.2,\y);
    \node[above right, inner sep=1.5pt] at (1.95,\y) {\t};}
  \node[draw, part, minimum width=1.3cm, minimum height=0.6cm, inner sep=0pt] (thr) at (1.2,0.9) {};
  \draw[line width=0.6pt] (0.75,0.75) -- (1.05,0.75) -- (1.6,1.05);
  \draw (1.85,0.9) -- (4.6,0.9) -- (4.6,1.6) -- (5.2,1.6);
  \node[lbl, below right] at (1.95,0.85) {Throttle pedal\\ (analogue 0--\qty{5}{\volt})};
  \node[block, minimum width=1.8cm, minimum height=0.9cm] (pwm) at (11.2,3.7) {PWM/DC\\ isolation};
  \node[block, minimum width=1.8cm, minimum height=0.9cm] (el) at (11.2,2.2) {E-lock\\ (main switch)};
  \draw (9.8,3.7) -- (pwm.west);
  \draw (9.8,2.2) -- (el.west);
  \draw[->] (pwm.east) -- (12.9,3.7) node[lbl, right] {Motor\\ controller};
  \draw[->] (el.east) -- (12.9,2.2) node[lbl, right] {Motor\\ controller};
  \node[lbl, above right] at (10.45,4.2) {PWM signal (motor torque)\\ $\qty{500}{\hertz}<f<\qty{1000}{\hertz}$};
  \foreach \x in {6.3,7.5,8.7} {
    \node[sensor] at (\x,-0.2) {};
    \draw (\x,1.4) -- (\x,0.35);}
  \node[align=center, below] at (7.5,-0.8) {Speed sensors: $3\times$ Cherry GS100701,\\ sensor rings with 20 pulses per wheel revolution};
\end{tikzpicture}
\caption{Block diagram of the control electronics (planning status; the final implementation uses an Arduino Due with SAM3X8E, see Section~\ref{sec:implementation})\nocc}\label{fig:sysblock}
\end{figure}

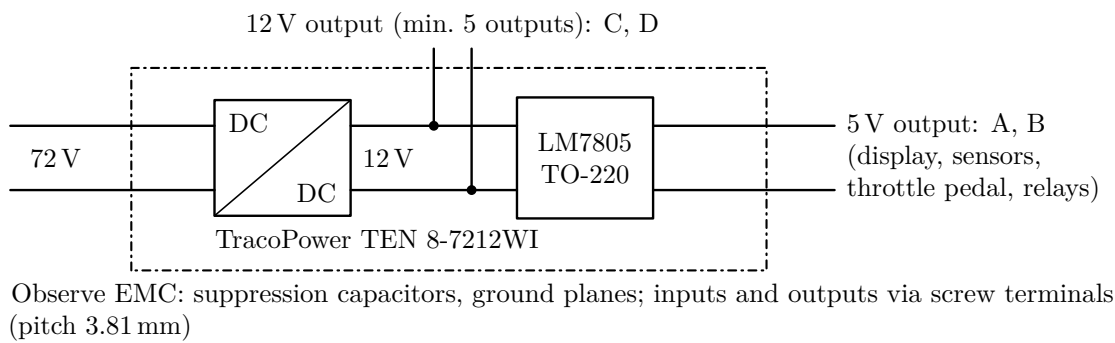
\begin{figure}[!ht]
\centering
\begin{tikzpicture}[schem, y=0.85cm, jnc/.style={circle, fill, minimum size=1.3mm, inner sep=0pt}]
  \draw[dash dot, line width=0.8pt] (1.6,-0.4) rectangle (10.0,2.75);
  \draw[part] (2.7,0.45) rectangle (4.5,2.25);
  \draw (2.7,0.45) -- (4.5,2.25);
  \node at (3.15,1.9) {DC};
  \node at (4.05,0.8) {DC};
  \node[below right, inner xsep=0pt] at (2.7,0.4) {TracoPower TEN 8-7212WI};
  \node[block, minimum width=1.8cm, minimum height=1.6cm] (reg) at (7.6,1.35) {LM7805\\ TO-220};
  \foreach \y in {1.85,0.85} {
    \draw[part] (0,\y) -- (2.7,\y);
    \draw[part] (4.5,\y) -- (6.7,\y);
    \draw[part] (8.5,\y) -- (10.9,\y);}
  \node at (0.6,1.35) {\qty{72}{\volt}};
  \node at (5.0,1.35) {\qty{12}{\volt}};
  \draw[part] (5.6,1.85) -- (5.6,3.05);
  \draw[part] (6.1,0.85) -- (6.1,3.05);
  \node[jnc] at (5.6,1.85) {};
  \node[jnc] at (6.1,0.85) {};
  \node[above] at (5.85,3.05) {\qty{12}{\volt} output (min.\ 5 outputs): C, D};
  \node[lbl, right] at (11.0,1.35) {\qty{5}{\volt} output: A, B\\ (display, sensors,\\ throttle pedal, relays)};
  \node[lbl, below right, inner sep=0pt, text width=14.6cm] at (0,-0.6) {Observe EMC: suppression capacitors, ground planes; inputs and outputs via screw terminals (pitch \qty{3.81}{\milli\metre})};
\end{tikzpicture}
\caption{Block diagram of the power supply\nocc}\label{fig:powersupply}
\end{figure}

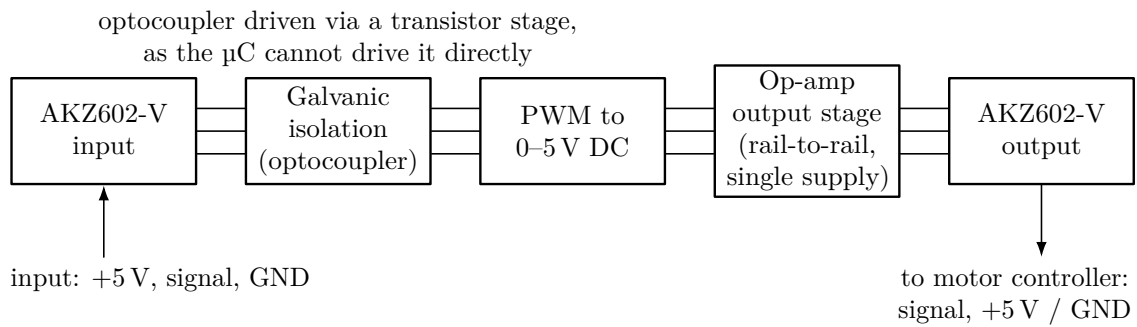
\begin{figure}[tbp]
\centering
\begin{tikzpicture}[schem, b/.style={block, text width=2.3cm, minimum height=1.4cm, inner sep=2pt}]
  \foreach \k/\t in {0/{AKZ602-V\\ input}, 1/{Galvanic isolation\\ (optocoupler)}, 2/{PWM to\\ 0--\qty{5}{\volt} DC},
      3/{Op-amp output stage (rail-to-rail, single supply)}, 4/{AKZ602-V\\ output}}
    \node[b] (n\k) at (\k*3.1,0) {\t};
  \foreach \k [evaluate=\k as \j using int(\k+1)] in {0,...,3}
    \foreach \dy in {-0.3,0,0.3} \draw ([yshift=\dy cm]n\k.east) -- ([yshift=\dy cm]n\j.west);
  \draw[<-] (n0.south) -- ++(0,-0.95) node[below, xshift=0.75cm] {input: $+\qty{5}{\volt}$, signal, GND};
  \draw[->] (n4.south) -- ++(0,-0.95) node[below, align=center, xshift=-0.35cm] {to motor controller:\\ signal, $+\qty{5}{\volt}$ / GND};
  \node[above, align=center] at (n1.north) {optocoupler driven via a transistor stage,\\ as the \textmu C cannot drive it directly};
\end{tikzpicture}
\caption{Block diagram of the motor controller interface\nocc}\label{fig:mcif}
\end{figure}

\begin{figure}[tbp]
\centering
\includegraphics[width=\textwidth]{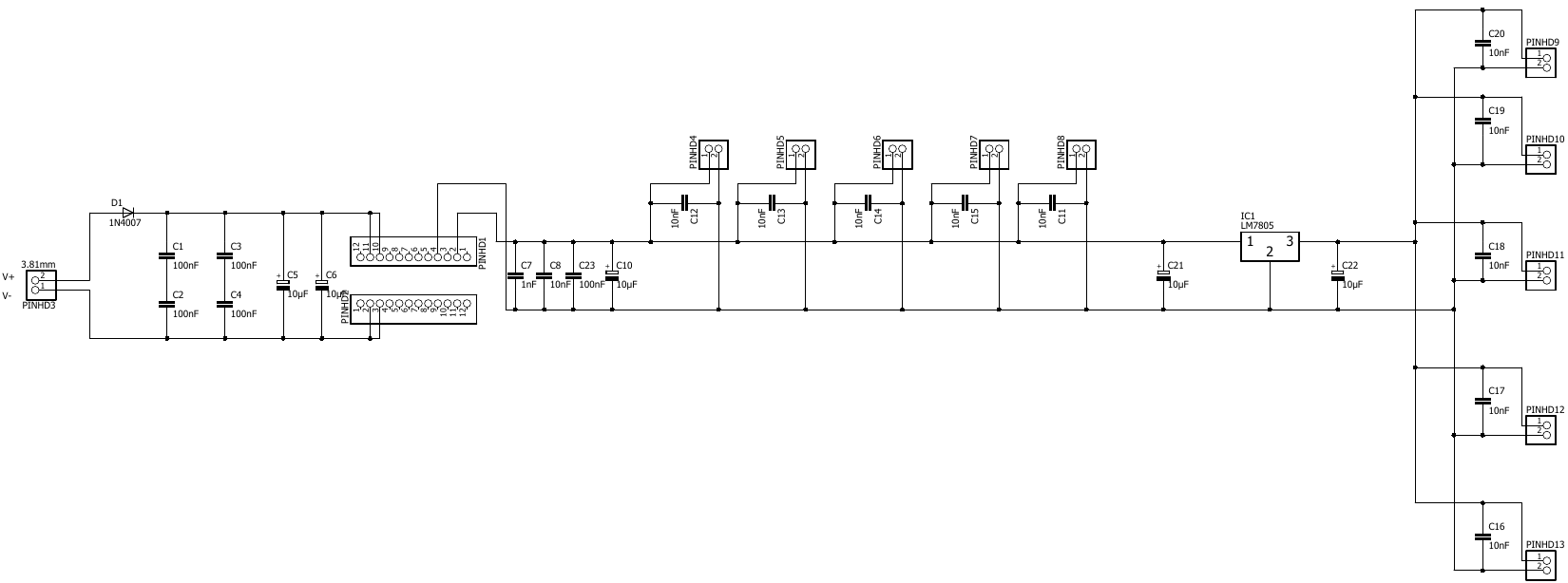}
\caption{Circuit diagram of the power supply (original)}\label{fig:image75}
\end{figure}

\begin{figure}[tbp]
\centering
\includegraphics[width=\textwidth]{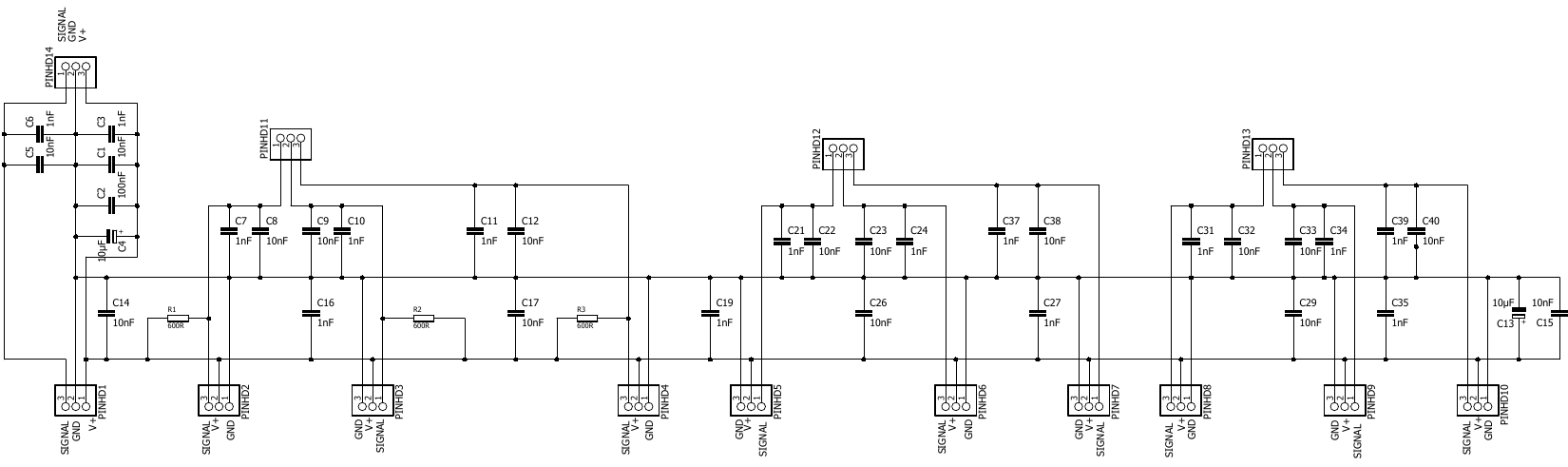}
\caption{Circuit diagram of the I/O board (original)}\label{fig:image74}
\end{figure}

\begin{figure}[tbp]
\centering
\includegraphics[width=\textwidth]{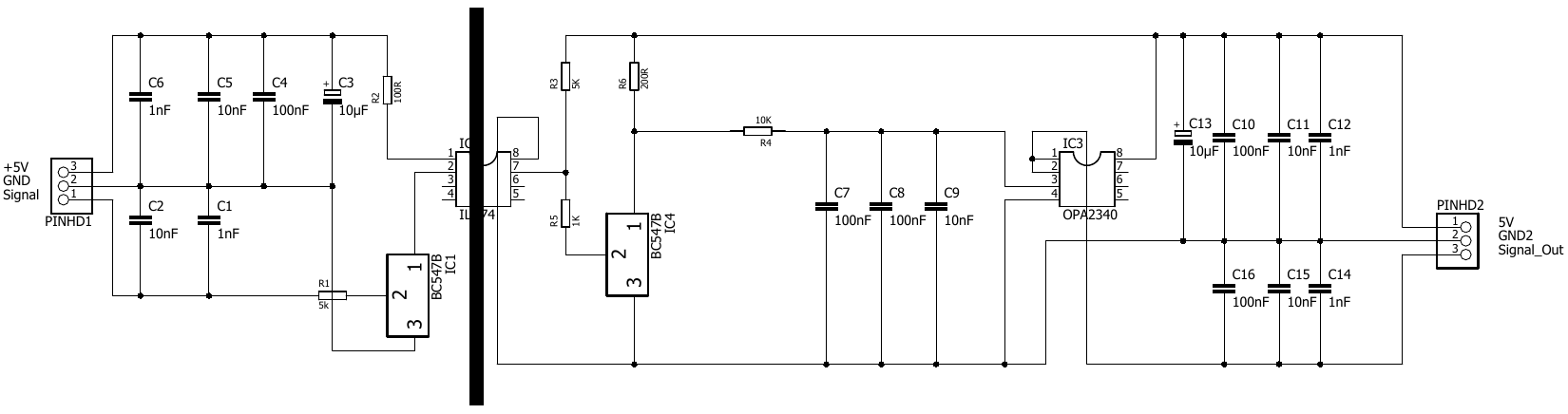}
\caption{Circuit diagram of the motor controller interface (original)}\label{fig:image76}
\end{figure}

The circuits were designed for electromagnetic compatibility and low power consumption. Figure~\ref{fig:sysblock} gives an overview of the control electronics. The block diagrams of the power supply, the I/O board and the motor controller interface follow in Figures~\ref{fig:powersupply} to~\ref{fig:mcif}, the corresponding circuit diagrams in Figures~\ref{fig:image75} to~\ref{fig:image76}. The block diagrams were redrawn in English. The circuit diagrams are reproduced unchanged and are also available in the project repository~\cite{Sto26}.

\begin{figure}[tbp]
\centering
\begin{tikzpicture}[schem, every node/.append style={execute at begin node={\hyphenpenalty=10000\exhyphenpenalty=10000}},
  b/.style={block, text width=1.72cm, inner ysep=1pt, inner xsep=2pt}]
  \draw[dash dot, line width=0.8pt] (0.9,-0.45) rectangle (15.45,4.5);
  \node at (8.2,4.17) {To microcontroller};
  \node[block, minimum width=0.6cm, minimum height=2.2cm] (sup) at (1.45,2.6) {};
  \node[rotate=90] at (sup) {AKZ602-V};
  \draw[->] (-0.3,2.6) -- (sup.west);
  \node[align=left, above right, inner sep=1.5pt] at (-0.3,2.6) {\qty{5}{\volt}\\ supply};
  \foreach \k/\t in {0/{Speed\\ sensor}, 1/{Speed\\ sensor}, 2/{Speed\\ sensor}, 3/{Switch 1}, 4/{Switch 2}, 5/{Switch 3}, 6/{Switch 4}} {
    \pgfmathsetmacro{\x}{2.75+\k*1.95}
    \node[b, minimum height=0.9cm] at (\x,3.35) {output\\ terminal};
    \node[b, minimum height=0.9cm] at (\x,2.35) {input\\ terminal};
    \node[b, minimum height=1.0cm] at (\x,1.2) {ground\\ plane};
    \node[align=center] at (\x,0.15) {\t};}
  \node[lbl, below right, inner sep=0pt, text width=14.5cm] at (0.9,-0.65) {All terminals: AKZ602-V. Speed sensors require a pull-up resistor (value see datasheet).\\
    EMC: all inputs and outputs with \qty{1}{\nano\farad} parallel \qty{100}{\nano\farad} filter capacitors;
    supply: \qty{10}{\nano\farad} ceramic parallel \qty{10}{\micro\farad} low-ESR electrolytic};
\end{tikzpicture}
\caption{Block diagram of the I/O board\nocc}\label{fig:ioboard}
\end{figure}

\chapter{Design Documentation}\label{app:cad-design-documentation}
\begin{figure}[!ht]
\centering
\begin{minipage}[t]{0.49\textwidth}
\textbf{a)}\par\medskip
\centering
\includegraphics[width=\linewidth]{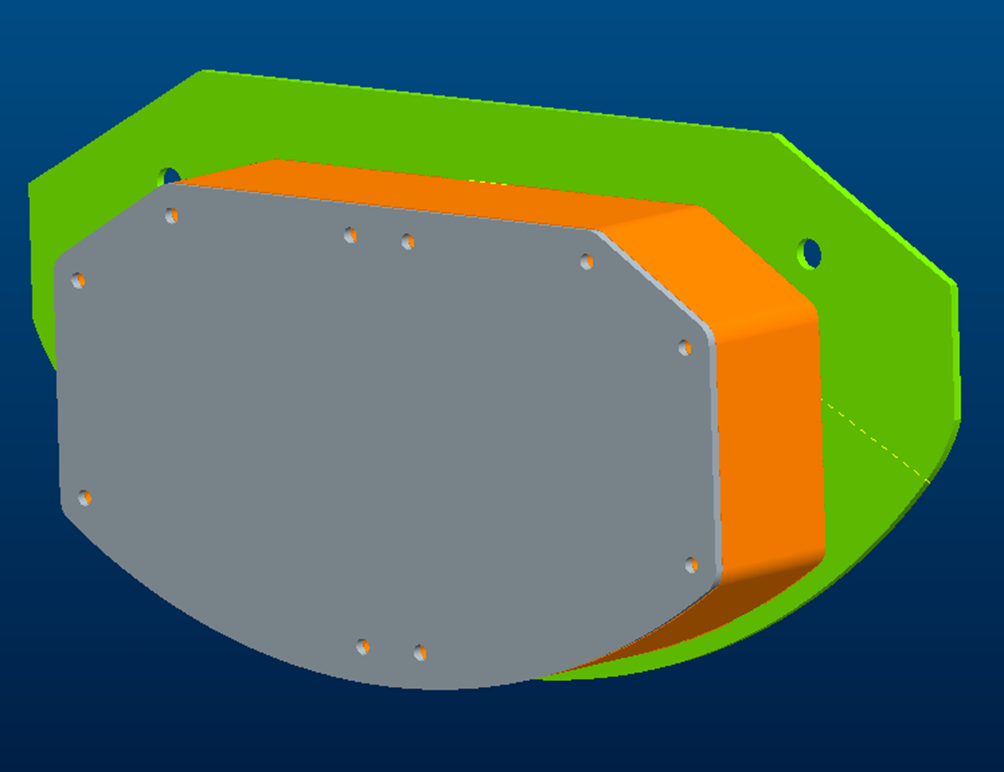}
\end{minipage}\hfill
\begin{minipage}[t]{0.49\textwidth}
\textbf{b)}\par\medskip
\centering
\includegraphics[width=\linewidth]{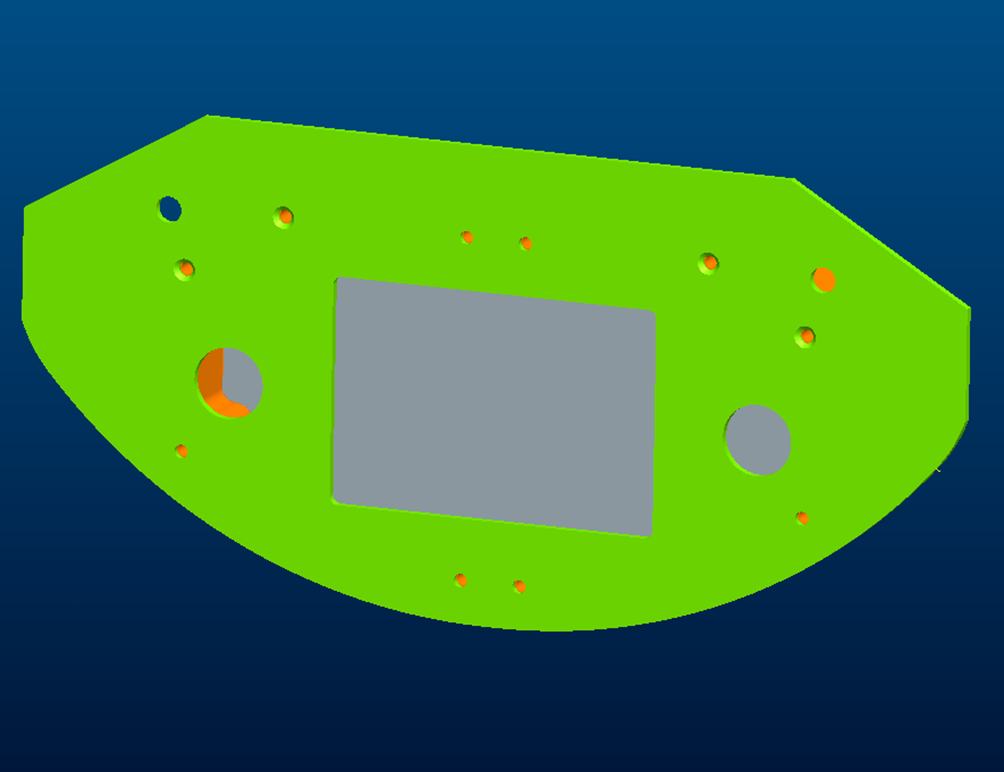}
\end{minipage}
\caption{3D views of the display holder: a) view 1, b) view 2}\label{fig:image82}
\end{figure}

\begin{figure}[!ht]
\centering
\begin{minipage}[t]{0.49\textwidth}
\textbf{a)}\par\medskip
\centering
\includegraphics[width=\linewidth]{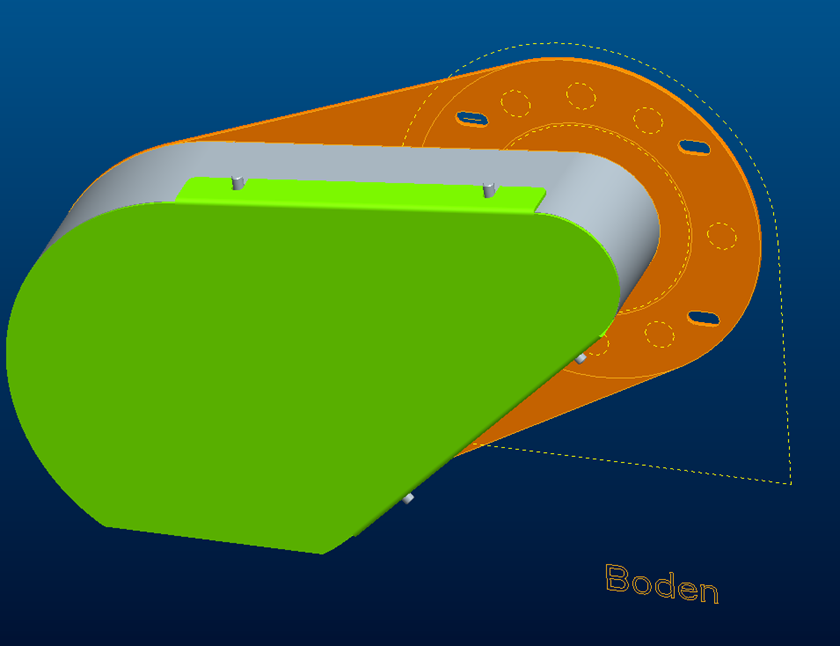}
\end{minipage}\hfill
\begin{minipage}[t]{0.49\textwidth}
\textbf{b)}\par\medskip
\centering
\includegraphics[width=\linewidth]{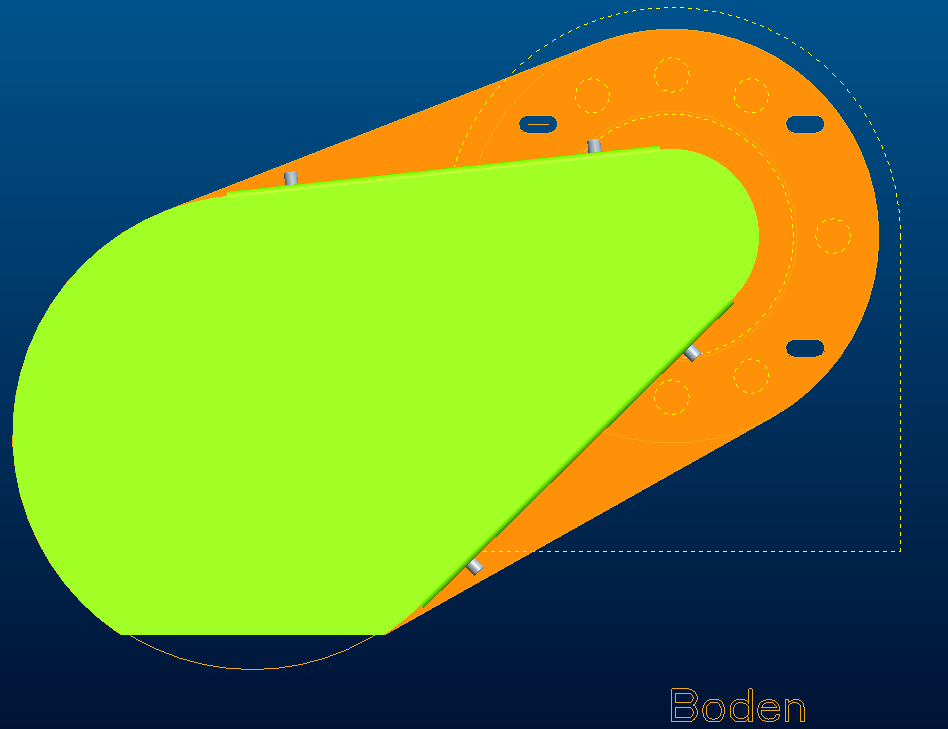}
\end{minipage}
\caption{3D views of the chain guard: a) view 1, b) view 2}\label{fig:image84}
\end{figure}

\begin{figure}[p]
\centering
\includegraphics[width=0.9\textwidth]{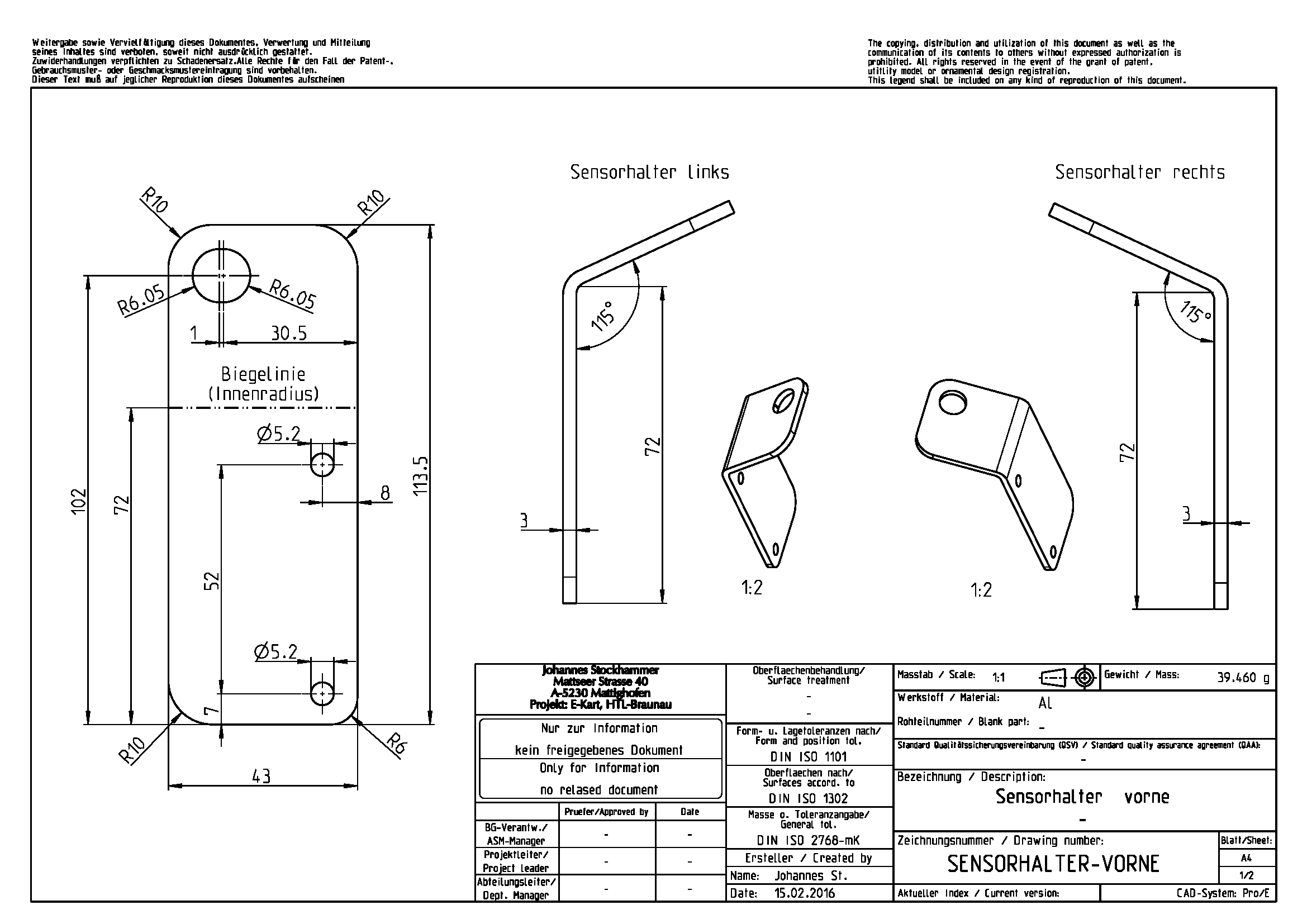}
\caption{Drawing of the front sensor holder}\label{fig:image77}
\vspace{\baselineskip}
\includegraphics[width=0.9\textwidth]{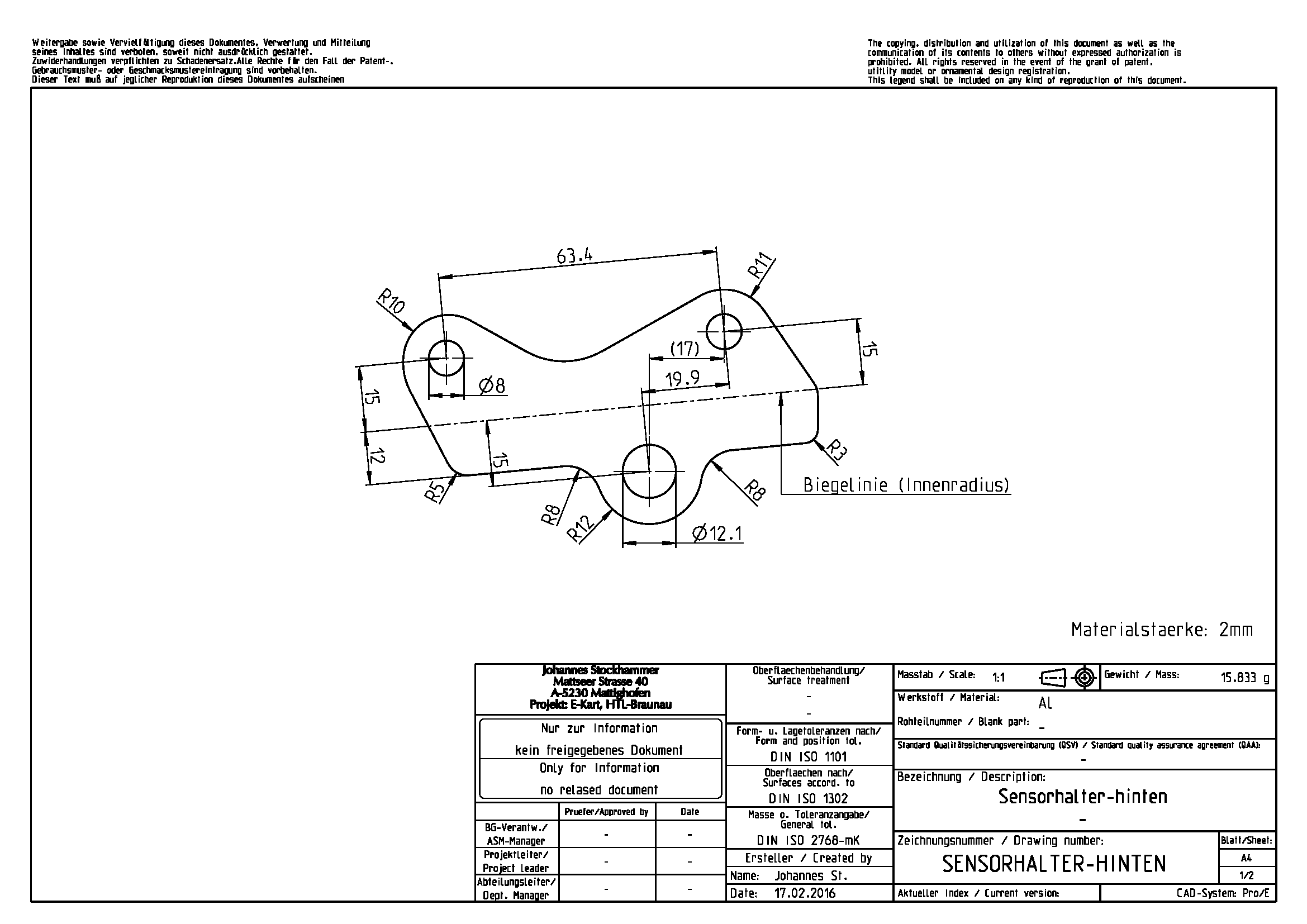}
\caption{Drawing of the rear sensor holder}\label{fig:image78}
\end{figure}

\begin{figure}[p]
\centering
\includegraphics[width=0.9\textwidth]{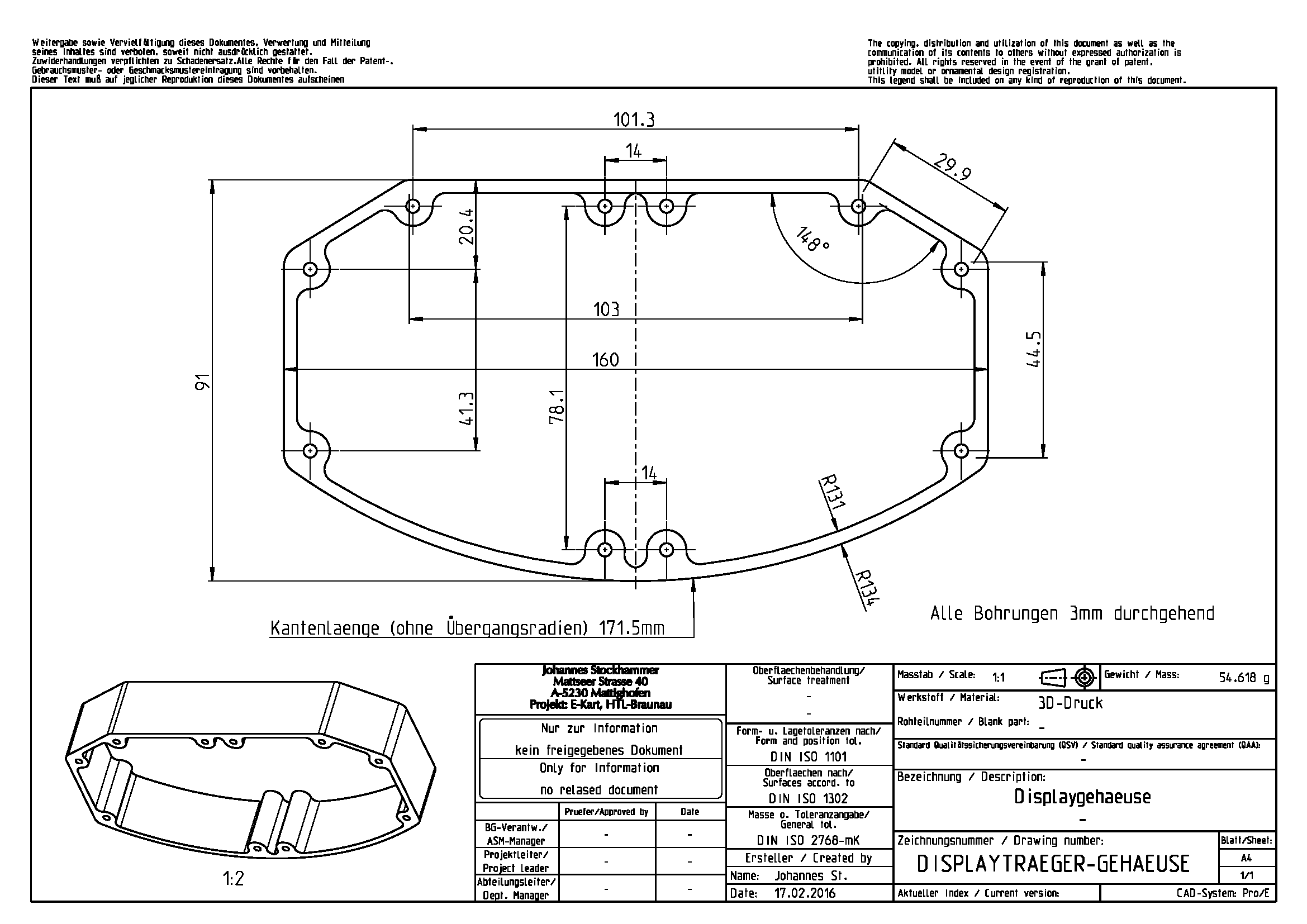}
\caption{Drawing of the display carrier housing}\label{fig:image79}
\vspace{\baselineskip}
\includegraphics[width=0.9\textwidth]{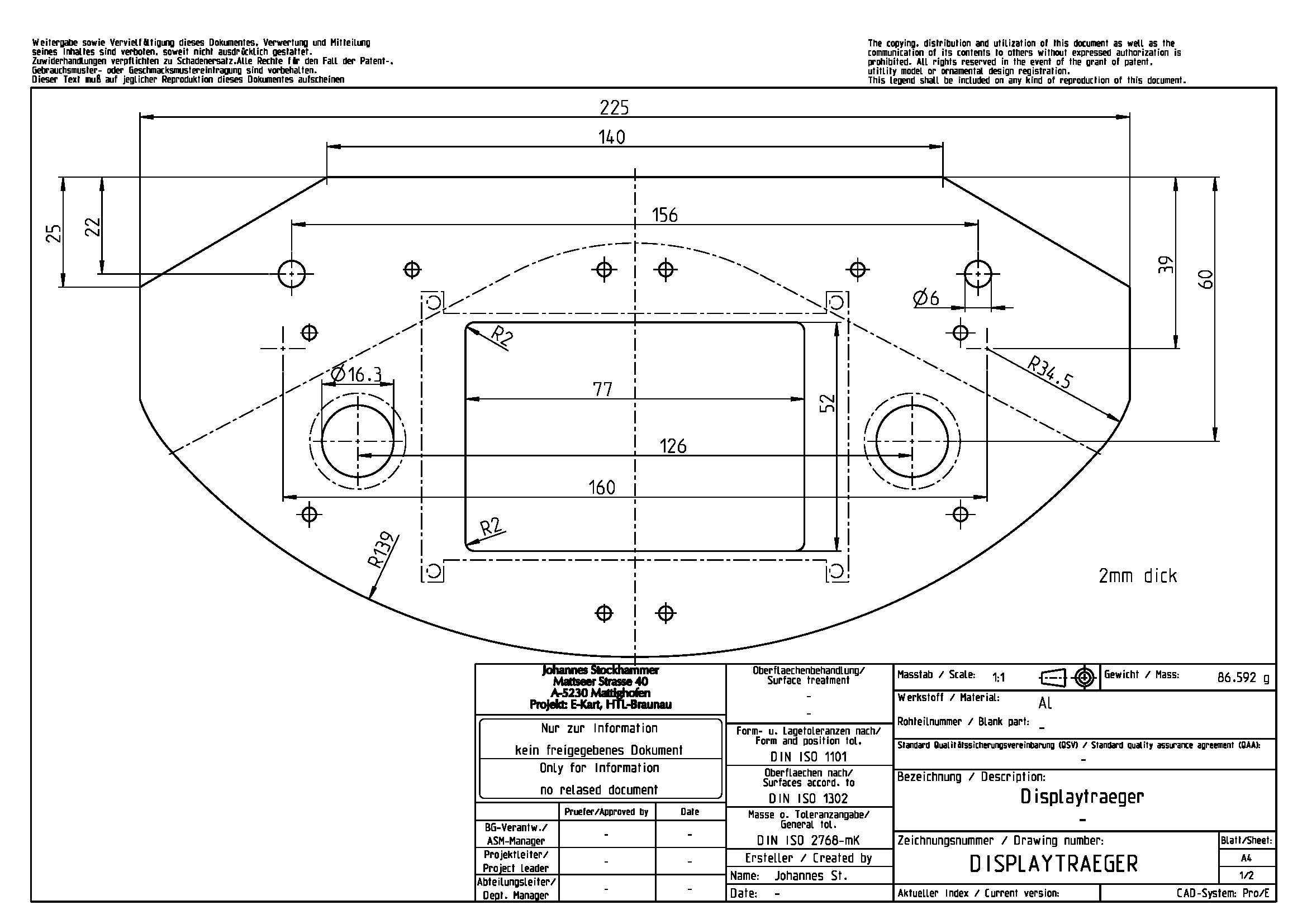}
\caption{Drawing of the display carrier}\label{fig:image80}
\end{figure}

\begin{figure}[p]
\centering
\includegraphics[width=0.9\textwidth]{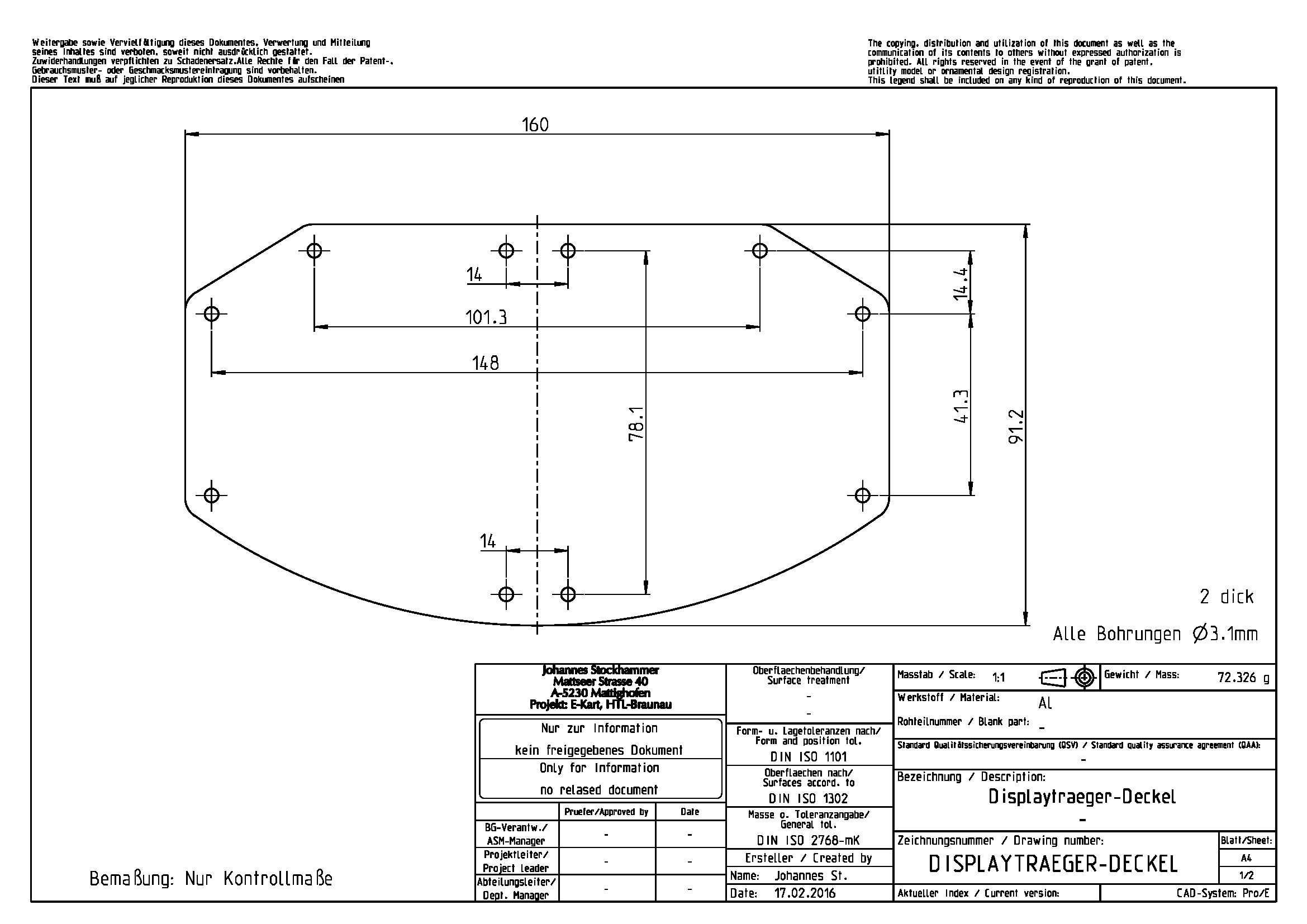}
\caption{Drawing of the display carrier cover}\label{fig:image81}
\end{figure}

The following 3D views and drawings document the parts designed in-house: the display holder and the chain guard (Figures~\ref{fig:image82} and~\ref{fig:image84}), the sensor holders (Figures~\ref{fig:image77} and~\ref{fig:image78}) and the display carrier (Figures~\ref{fig:image79} to~\ref{fig:image81}). The drawings are reproduced unchanged and contain German labels. Datasheets of the purchased components are not reproduced for copyright reasons; they are available from the respective manufacturers.

\clearpage
\begingroup
\let\clearpage\relax\let\cleardoublepage\relax
\RedeclareSectionCommand[beforeskip=2.2\baselineskip, afterskip=1.2\baselineskip]{chapter}
\renewcommand{\nocc}{}
\renewcommand{\noccpart}[1]{}
\listoffigures

\chapter*{List of Abbreviations}\addcontentsline{toc}{chapter}{List of Abbreviations}
\begin{longtable}{@{}p{3cm}p{\dimexpr\textwidth-3cm-2\tabcolsep\relax}@{}}
ADC & Analogue-to-digital converter \\
BLDC & Brushless DC motor \\
BMS & Battery management system \\
CAD & Computer-aided design \\
DAC & Digital-to-analogue converter \\
DOD & Depth of discharge \\
EMC & Electromagnetic compatibility \\
ERS & Energy Recovery System \\
HV & High voltage \\
I/O & Input/output \\
\LFP{} & Lithium iron phosphate (battery technology) \\
LiPo & Lithium polymer (battery technology) \\
LV & Low voltage \\
MS & Milestone \\
PM & Permanent magnet \\
PWM & Pulse-width modulation \\
TC & Traction control \\
WBS & Work breakdown structure \\
\textmu C & Microcontroller \\
\end{longtable}

\endgroup

\end{document}